\documentclass[usegraphicx,usenatbib]{mn2e}

\usepackage{times}
\usepackage{graphicx,epsf}

\newcommand{\apj}{ApJ}           
\newcommand{\apjs}{ApJS}
\newcommand{\mnras}{MNRAS}       
\newcommand{\aj}{AJ}
\newcommand{\araa}{ARAA}
\newcommand{\aap}{A\&A}
\newcommand{\aaps}{A\&AS}
\newcommand{\plotone}{\includegraphics[width=\columnwidth,clip=true]}

\title[Photometric characterization of isolated galaxies]{Photometric characterization of a well defined
sample of isolated galaxies in the context of the AMIGA project}

\author[Durbala et al.]{A. Durbala$^1$\thanks{E-mail: adriana.durbala@ua.edu}, J.~W.
Sulentic$^1$, R. Buta$^1$ and L. Verdes-Montenegro$^2$\\
$^1$Department of Physics and Astronomy, University of Alabama,
Box 870324, Tuscaloosa, AL 35487-0324, USA\\
$^2$Instituto de Astrof\'{\i}sica de Andaluc\'{\i}a, CSIC, Apdo.
3004, 18080 Granada, Spain\\}

\begin{document}

\maketitle

\begin{abstract}
We perform a detailed photometric analysis (bulge-disk-bar
decomposition and Concentration-Asymmetry-Clumpiness - CAS
parametrization) for a well defined sample of isolated galaxies,
extracted from the Catalog of Isolated Galaxies
\citep{karachentseva73} and reevaluated morphologically in the
context of the AMIGA project (\textbf{A}nalysis of the interstellar
\textbf{M}edium of \textbf{I}solated \textbf{GA}laxies). We focus on
Sb-Sc morphological types, as they are the most representative
population among the isolated spiral galaxies. Our analysis yields a
large number of important galactic parameters and various
correlation plots are used to seek relationships that might shed
light on the processes involved in determining those parameters.
Assuming that the bulge S\'{e}rsic index and/or Bulge/Total
luminosity ratios are reasonable diagnostics for pseudo- versus
classical bulges, we conclude that the majority of late-type
isolated disk galaxies likely host pseudobulges rather than
classical bulges. Our parametrization of galactic bulges and disks
suggests that the properties of the pseudobulges are strongly
connected to those of the disks. This may indicate that pseudobulges
are formed through internal processes within the disks (i.e. secular
evolution) and that bars may play an important role in their
formation. Although the sample under investigation covers a narrow
morphological range, a clear separation between Sb and Sbc-Sc types
is observed in various measures, e.g. the former are redder,
brighter, have larger disks and larger bars, more luminous bulges,
are more concentrated, more symmetric and clumpier than the latter.
A comparison with samples of spiral galaxies (within the same
morphological range) selected without isolation criteria reveals
that the isolated galaxies tend to host larger bars, are more
symmetric, less concentrated and less clumpy.

\end{abstract}


\section{Introduction}

The properties of galaxies and their evolution are thought to be
strongly related to their environment. The empirical quantification
of environmental influence (``nurture'') on morphology, structure,
nuclear activity, star formation properties, etc. requires a robust
definition of a sample of galaxies that are minimally perturbed by
other galaxies. Such a sample could serve as a ``pure nature''
baseline. In this sense, perhaps the best compilation of isolated
galaxies available at this time is the Catalog of Isolated Galaxies
\citep[CIG;][]{karachentseva73}. Both the size (n=1050 galaxies) and
the restrictive isolation criteria in the catalog contribute to its
statistical value. The definition of isolation requires that, for a
galaxy of diameter D, there is no companion/neighbor with a diameter
d in the range D/4 to 4D within a distance of 20d. The isolation
criteria used to construct the CIG suggest that a typical galaxy of
25 kpc diameter has not been visited by a similar mass perturber in
the past $\sim$ 3Gyr \citep[assuming a typical field velocity of
$\sim$ 150 km s$^{-1}$;][]{verdes05}. Thus, the evolution of such
isolated galaxies is mostly driven by internal processes and to a
much lesser degree by environment, at least for the last $\sim$ 3Gyr
of their existence.

A recent morphological reevaluation of the CIG galaxies in the
context of the AMIGA project (\textbf{A}nalysis of the interstellar
\textbf{M}edium of \textbf{I}solated \textbf{GA}laxies) revealed
that the bulk ($\sim$ 63\%) show morphological types in the range
Sb-Sc \citep{sulentic06}. In this study we present the results of a
photometric characterization for a representative subsample of n
$\sim$ 100 CIG galaxies classified as Sb-Sc in this latter
reference. We perform multicomponent decomposition (bulge/disk/bar)
using the BUDDA
code\footnote{http://www.mpa-garching.mpg.de/~dimitri/budda.html}
\citep{deSouza04}. Additionally, we evaluate CAS parameters
Concentration(C)-Asymmetry(A)-Clumpinesss(S)
\citep[e.g.][]{conselice00,bershady00,conselice03,taylor07}.
Assembling a set of parameters combining a model-dependent
description of the main components of galaxies (BUDDA) with global
structural measures (CAS) could provide valuable hints into the
formation and evolution of galaxies.

This is the first attempt to date to present a detailed examination
of this kind (bulge-disk-bar decomposition combined with CAS
parameters) for a well defined sample of isolated galaxies. This
study is an integral part of the AMIGA project, which is a dedicated
multiwavelength study of the revised CIG catalog. The goal of AMIGA
is to quantify the fundamental properties of a statistically
meaningful sample of isolated galaxies which can then be used as a
baseline for comparison and for estimation of the effects of
environment in other less isolated samples of galaxies. The CIG
catalog has recently been reevaluated in terms of galaxy positions
\citep{leon03}, isolation \citep{verley07a,verley07b} and morphology
\citep{sulentic06}. A series of studies were produced in the context
of the AMIGA project: 1) an optical characterization of the refined
sample \citep{verdes05}, 2) an analysis of mid- and far-infrared
properties \citep{lisenfeld07}, 3) a study of the neutral CO and HI
gas \citep{espada05,espada06}, 4) radio continuum emission
\citep{leon08} and 5) nuclear activity \citep{sabater08}. Another
recent study used a subsample of isolated AMIGA galaxies to
investigate the role of bars in star formation processes
\citep[e.g.][]{verley07c}. Our present study offers a detailed
photometric analysis of a representative sample of the core AMIGA
population of Sb-Sc morphological types. We should note that all
data produced within the AMIGA project are periodically updated and
made publicly available at http://www.iaa.es/AMIGA.html.

Theoretical models and numerical simulations exploring the formation
and evolution of galaxies rely on empirical results that could
separate and quantify the relative roles of internal secular
processes (that develop on time scales much longer than the galaxy
formation/collapse process itself) and slow or fast external
perturbations (environment) in defining the structural properties of
galaxies. In this sense, our present study has a twofold importance:
a) it explores a representative and well defined sample of
\textit{the most isolated galaxies} in the local Universe and b)
provides an \textit{extensive photometric structural analysis} of
these galaxies. Our main goal is to identify potential scaling
relations and correlations: i) between parameters describing the
same structural component (bulge, disk or bar), ii) between
components, iii) between components and global properties of the
galaxy (morphological type, color, luminosity, concentration,
asymmetry, clumpiness, etc.). With such correlations available one
could explore for example the nature of bulges in isolated galaxies
and how they are formed, the role of bars (if any) in the
formation/evolution of bulges, whether the isolated spiral galaxies
are different relative to spirals in richer environments in terms of
global properties and/or in terms of properties of their components
(bulge, bar, disks).

This paper is organized as follows: \S~2 presents the selection and
basic properties of the sample, \S~3 offers a concise view on data
reduction, \S~4 and \S~5 present the results of BUDDA decomposition
analysis and CAS parametrization, respectively. \S~6 combines
various measures obtained from the BUDDA code with CAS parameters.
\S~7 is dedicated to discussion and conclusions. Throughout the
paper we use H$_{o}$ = 75 km s$^{-1}$ Mpc$^{-1}$.

\section{The Sample}

\subsection{Sample Selection}

Galaxies of morphological types Sb-Sc were found to be the most
abundant (dominant) population in the AMIGA reanalysis of the CIG
\citep{sulentic06}, Sb-Sc galaxies represent 2/3 (n$\sim$637) of the
1018 galaxies with recession velocity V$_{R}$ $>$ 1000 km s$^{-1}$.
This motivated us to focus on the Sb-Sc morphological range since
earlier and later types are so rare that they cannot be considered
representative of an isolated sample. The sample adopted here was
drawn from that Sb-Sc  population \citep{sulentic06} after applying
the following constraints: (i) 1500 $<$ V$_{R}$ $<$ 10000 km
s$^{-1}$, (ii) blue corrected magnitudes \citep{verdes05}
m$_{Bcorr}$ $<$ 15, (iii) inclination $<$ 70$^\circ$ and (iv)
available images in SDSS (Data Release 6: DR6, \citealt{adelman08}).
We ended up with a representative sample of n = 101 Sb-Sc galaxies
all having SDSS (Sloan Digital Sky Survey) i-band magnitudes
brighter than 15.0. The lower limit to the V$_{R}$ range avoids
inclusion of local supercluster galaxies where the degree of
isolation is most uncertain. The upper limit ensures a large enough
SDSS overlap sample and at the same time adequate resolution to
permit evaluation of basic structural parameters for all of the
galaxies. We are preparing a complementary Fourier analysis of
spiral structure in the same sample considered here. The need for a
sufficiently accurate deprojection of galaxies in the context of the
Fourier analysis requires the third constraint on inclination for
the sample selection. The results of the Fourier analysis will be
reported in a later paper.

SDSS images are obtained with  a dedicated 2.5 m telescope
\citep{gunn06}. The imaging process is carried out under photometric
conditions \citep{hogg01} in five filters (ugriz)
\citep{fukugita96,smith02} employing a wide-field CCD camera
\citep{gunn98}. Data are processed by completely automated pipelines
that detect and measure photometric properties of objects and
astrometrically calibrate the data \citep{lupton01,pier03}.

Table 1 presents the sample of CIG galaxies that we analyze here in
terms of coordinates \citep{leon03}, recession velocity
\citep{verdes05}, morphological type from \citet{sulentic06} and
inclination, which was estimated using the formula $cos(i) = b/a$,
where $a$ and $b$ are the semimajor and semiminor axes of the disk,
respectively. Morphological reevaluation of all CIG galaxies
\citep{sulentic06} was based on the second Palomar Observatory Sky
Survey POSSII with some confirmation using  SDSS-DR3. DR6 provides a
much larger number of CIG galaxies with SDSS images. Therefore we
decided to do a SDSS based morphological classification of our
target sample of 101 galaxies in the framework presented in ``The de
Vaucouleurs Atlas of Galaxies'' \citep{buta07}. The last column of
Table 1 shows our revised and more complete (visual) classification.
Four galaxies (CIG: 250, 291, 308 and 392) have been excluded from
our sample because they are not in the range Sb-Sc according to our
revised classification (see the last four lines of Table 1).
Hereafter we consider the N=97 confirmed Sb-Sbc-Sc galaxies.

\subsection{Basic Properties of the Sample}

Figures 1a-d present some basic properties of our sample (the
measures shown in panels a, c and d are based upon photometric
estimates reported within the SDSS photo-pipeline): (a) distribution
of galactic size, as indicated by a$_{25}^{i}$, i.e. the semimajor
axis of the isophote where the disk surface brightness profile drops
to 25th mag arcsec$^{-2}$; (b) distribution of inclination; (c)
distribution of i-band absolute magnitudes M$_{i}$ and (d)
distribution of (g-i)$_{o}$ color. The size a$_{25}^{i}$ is
calculated from the SDSS photometric parameter ``isoA''. The
computation of (g-i)$_{o}$ is based on g- and i-band ``model'' SDSS
apparent magnitudes. The M$_{i}$ is obtained from the SDSS i-band
``cmodel'' magnitude (for more information on various types of
galaxy magnitudes reported within SDSS we direct the reader to:
http://www.sdss.org/dr6/algorithms/photometry.html). We applied
appropriate corrections to SDSS magnitudes and colors, i.e. Galactic
and internal extinction, K-correction using
YES\footnote{http://cadwww.hia.nrc.ca/yes} (York Extinction Solver;
\citealt{mccall04}) and a (1+z)$^{4}$ factor due to redshift
dimming. The color excess values E(B-V) required as input for YES
come from NED\footnote{This research has made use of the NASA/IPAC
Extragalactic Database (NED) which is operated by the Jet Propulsion
Laboratory, California Institute of Technology, under contract with
the National Aeronautics and Space Administration.}(NASA
Extragalactic Database) and are based on \citet{schlegel98}.

Each panel of Figure 1 indicates the mean ($\pm$ standard deviation)
and the median for the distribution. The galaxies in our sample show
semimajor axes covering a wide range between 4 and 28 kpc with the
majority concentrated between 8-20 kpc (panel a). We visually
examined galaxies in the first and last two bins of panel a (4-8 and
20-28kpc). At the large end we see galaxies with grand design spiral
structure - luminosity class I
\citep[]{vandenBergh60a,vandenBergh60b,sandage81}. At the small end
we find a majority with flocculent structure characteristic of
luminosity classes IV-V. One exception involves CIG522 which shows
surprisingly grand design structure given its small size.
Inclinations in our sample span a range from 0$^\circ$ to 70$^\circ$
with the bulk of the sample between 20$^\circ$ and 60$^\circ$ (panel
b). Our sample covers a range in i- band absolute magnitude from -19
to -23 (panel c), with an average absolute magnitude typical for an
L$^{\ast}$-galaxy \citep[e.g.][]{verdes05}. The faintest galaxies in
our sample are similar in luminosity to the Large Magellanic Cloud
(LMC). Galaxies in our sample show a wide spread in (g-i)$_{o}$
colors (0.1-1.2) with the bulk between 0.5 and 1 (panel d). Table 2
presents basic photometric measures based on SDSS photo-pipeline:
(g-i)$_{o}$ colors, absolute global i-band magnitudes M$_{i}$ and
the a$_{25}$ galactic semimajor axes both in i- and g-band. We
checked our sample for biases driven by inclination effects and we
found that SDSS photometric measures and BUDDA derived parameters
seem to be insensitive to inclination. Because our sample covers a
wide range in redshift, it is affected by Malmquist bias. Lowest
redshift favors the lowest luminosity galaxies and highest redshift
the most luminous objects.

We also marked on all the plots in the present paper (not shown in
the paper) the galaxies classified in \citet{sulentic06} as I/A
(interacting). The interaction code was either ``y'' (KIG 446, 712)
or ``?'' (KIG 11, 33, 282, 328, 339, 366, 386, 466, 508, 640, 645,
743, 912, 943). A ``y'' indicates a morphologically distorted system
and/or almost certain interacting system while ``?'' indicates
evidence for interaction/asymmetry with/without certain detection of
a companion. We find no trends for the galaxies flagged as I/A.

Table 3 presents average values (mean and median) for various
interesting photometric measures, some of which having been employed
in Figure 2. Figure 2a shows the correlation between disk size
a$_{25}^{i}$ and total i-band absolute magnitude M$_{i}$. The three
morphological types (Sb-Sbc-Sc) are shown with different symbols.
The solid line represents the best fit linear
regression\footnote{All correlation coefficients we report in this
paper refer to an ordinary least square linear regression of Y on X
or OLS(Y|X), e.g. \citet{isobe90}. The error bars of individual data
points are not taken into account for linear regression fits.}
(correlation coefficient R=0.89). Galaxies classified Sb and Sc
favor opposite ends of the correlation with means (15.5kpc; -21.5)
and (12.2kpc; -20.9) respectively. A similar correlation is seen
when g-band absolute magnitudes are used. Figure 2b is a
color-magnitude diagram and indicates that the more luminous
galaxies are also redder \citep[e.g.][]{tully82,wyse82,gildepaz07},
with linear regression correlation coefficient R=0.76. Figure 2c
shows the relation between the galaxy color (g-i)$_{o}$ and disk
size a$_{25}^{i}$ (linear regression correlation coefficient
R=0.75). We see that galaxies with larger disks tend to be redder
than those with smaller disks. Correlations 2b and 2c appear to be
largely driven by the location of Sb galaxies which tend to be
larger, redder and more luminous than the other subclasses. Visual
classification of Hubble subtypes, while rather subjective, appears
to retain some utility for isolating galaxies according to first
order physical properties. All of the trends involving  size,
luminosity and color are consistent with previous studies
\citep[e.g.][]{roberts94,shimasaku01}. Table 3 suggests that i-band
a$_{25}$ disk measures are systematically larger than corresponding
g-band measures. This effect is likely caused by the lower s/n of
the g-band images because the i-band filter is more
sensitive\footnote
{http://www.sdss.org/dr6/instruments/imager/index.html}.

The purpose of Figures 2a-c is twofold: to reveal correlations
between basic properties and to identify  outliers. The former
correlations are expected to be better defined in a sample with
minimal effects of nurture while, by the same reasoning, outliers
are likely to indicate problematic data or remaining galaxies
affected by interactions (that were not previously suspected as
showing signs of interaction). The initial correlations based only
on SDSS data revealed a small number of extreme outliers. We
corrected all the measurements of the outlier galaxies (e.g. KIG:
397, 406, 502, 716, 928). It became clear that the automated
photometric SDSS pipeline cannot deal properly with galaxies that
are strongly contaminated by nearby bright stars. One galaxy (KIG
924) didn't have any photometric measurements and two more galaxies
(KIG 491, 712) didn't have any isoA measurements in the SDSS
pipeline. For a few other galaxies SDSS provided measures that fell
away from the correlations well described by the rest of our sample.
The SDSS magnitudes of the galaxies that fell on the correlations
agree within 0.1-0.2 magnitudes with our new measurements. Two
galaxies (KIG 502 and KIG 716) show an a$_{25}^{i}$ too large for
their absolute magnitudes. These two galaxies show peripheral
structures that raise the possibility they were affected by an
interaction or accretion event. We conclude that the natural sizes
of the galaxies KIG 502 and KIG 716 are much smaller than suggested
by a$_{25}^{i}$. They are excluded from Figures 1a, 2a and 2c. One
galaxy (KIG 322) was excluded from the sample because the
contamination of the nearby bright star makes it impossible to
obtain reliable photometric measurements.

We derive the following best fit regressions for the panels of
Figure 2: (a) $a_{25}^{i}=-87.27-4.74*M_{i}$; (b)
$(g-i)_{o}=-3.03-0.18*M_{i}$; (c)
$(g-i)_{o}=-0.28+0.96*log(a_{25}^{i})$.

\subsection{The Choice of SDSS i-Band Images}

The BUDDA-based decomposition and evaluation of CAS parameters
reported in later sections are performed on SDSS i-band images. The
choice of i-band was motivated by several considerations: 1) the
internal extinction in i-band is significantly less than in a bluer
filter, e.g. is $\sim$ 60\% of that in g-band and $\sim$ 80\% of
that in r-band (based on the sample of galaxies discussed here using
YES extinction solver), 2) the presence of star forming regions
within spiral arms would be associated with H$\alpha$ emission,
which is almost exclusively contained within the r-filter for the
range of V$_{R}$ we consider here and 3) the BUDDA code models the
galaxy stellar background, including bars (best revealed by a redder
filter) and does not fit the spiral structure (best traced by a
bluer filter). The typical surface brightness zero-point for the
i-band images we used is $\sim$ 26 mag arcsec$^{-2}$.

\section{Data Reduction}

We used the i-band frames that are flat-field, bias, cosmic-ray, and
pixel-defect corrected within the SDSS photometric pipeline
\citep{stoughton02}. In a few cases more than one frame was needed
in order to fully reconstruct the image of a galaxy. Frames were
combined using IRAF\footnote{Image Reduction and Analysis Facility
(IRAF) is distributed by the National Optical Astronomy
Observatories, which are operated by the Association of Universities
for Research in Astronomy, Inc., under cooperative agreement with
the National Science Foundation - http://iraf.noao.edu/} task
IMCOMBINE. We cleaned images removing contaminating stars using the
IRAF task IMEDIT. The sky was fitted with a two-dimensional 2nd
order polynomial and subtracted from the image using IRAF tasks
IMSURFIT. Photometric
calibration\footnote{http://www.sdss.org/dr6/algorithms/fluxcal.html}
was accomplished with \emph{aa}, \emph{kk} and \emph{airmass}
coefficients (zeropoint, extinction coefficient and airmass) from
the SDSS TsField files. We computed the zeropoint for the surface
brightness using
$2.5\times\log(exptime\times0.396^{2})-2.5\times0.4\times(aa+kk\times
airmass)$, considering the exposure time \emph{exptime} 53.907456
seconds and the pixel size 0\farcs396.

After performing these preliminary steps, we followed two different
(but complementary) approaches toward describing quantitatively the
galaxies structure and morphology:

1) \textit{BULGE/DISK/BAR/AGN Decomposition}: We used BUDDA
\citep{deSouza04} code version 2.1 to perform bulge/disk/bar/AGN
decomposition. The program can fit simultaneously multiple
components: a S\'{e}rsic bulge, two exponential disks, a S\'{e}rsic
bar and a Moffat central source (Active Galactic Nucleus -AGN).

The S\'{e}rsic surface brightness profile \citep{sersic68} is
described by:
$$\mu(r)=\mu_{e}+c_{n}[(r/r_{e})^{1/n}-1]$$
where r$_{e}$ is the effective radius (half-light radius), $\mu_{e}$
is the effective surface brightness (surface brightness at r$_{e}$),
n is the S\'{e}rsic index - a parameter describing the shape of the
profile and $c_{n}=2.5(0.868n-0.142)$. A S\'{e}rsic model is most
suitable to describe the shape of luminosity profiles in bulges of
galaxies \citep{andredakis95}. A pure deVaucouleurs profile
\citep{deVaucouleurs48} is characterized by a S\'{e}rsic index of 4
and a pure exponential profile is described by a S\'{e}rsic index of
1. The S\'{e}rsic index n$_{bulge}$ ranges from about 1 for late
type spiral galaxies (exponential profile, e.g.
\citealt{andredakis94,deJong96}) to about 6 for elliptical galaxies.

The exponential surface brightness profile of the disk
\citep{freeman70} is given by:
$$\mu(r)=\mu_{o}+1.086 r/h_{R}$$
where $\mu_{o}$ is the central surface brightness of the disk and
h$_{R}$ is the radial scalelength of the disk.

For a full description of the analytical functions used to fit each
component see section \S\S~3.1 of \cite{gadotti08}. IRAF task
ELLIPSE was used to get an initial guess for position angle (PA) and
ellipticity ($\epsilon$) for the bulge, disk and bar.

2) \textit{CAS Parametrization}:

The concentration index appears to be an integral part of any
morphological classification of galaxies \citep[e.g.][]{bershady00}.
The asymmetry and clumpiness indices are more sensitive to
environmental (i.e. external) influences and are reasonable
interaction diagnostics. Customarily, the three parameters are
described quantitatively as follows:

Concentration $C = 5 \times log(r_{80\%}/r_{20\%})$, where
r$_{80\%}$ and r$_{20\%}$ are the radii that include 80\% and 20\%
of the total light respectively \citep[c.f.][]{conselice03}.

Asymmetry $A_{abs} = \frac{\sum|I_{0}-I_{180}|}{\sum|I_{0}|} -
\frac{\sum|B_{0}-B_{180}|}{\sum|I_{0}|}$, where I$_{0}$ and
I$_{180}$ represent the pixel light intensity in the initial and the
180$^\circ$ rotated image. The letter ``B'' in this context refers
to background and has a similar meaning. The summation is done over
all pixels. The IRAF task IMCNTR identifies the center of the galaxy
(maximum intensity). The image is rotated 180$^\circ$ about that
center using the IRAF task ROTATE via linear interpolation. The
standard procedure for computation of the asymmetry index involves
also a minimization of A. (see section 3.3 in
\citealt{conselice00}). This method is effective for irregular and
edge-on galaxies where the centroid is most uncertain. We note that
Sb-Sc galaxies in our sample have a range of inclinations that allow
less ambiguous determination of their centers, i.e. the brightest
central grid point, whose coordinates are real numbers. In our cases
the uncertainty in identifying the center of the galaxy using IMCNTR
task is less than 1\% of a pixel. Therefore, we did not minimize the
asymmetry index. Nonetheless, we imposed the condition that the
center of the galaxy (initial estimate) does not shift upon
rotation. We used IRAF task IMSHIFT to correct for any displacements
that occurred.

Clumpiness $S = 10 \times [\frac{\sum(I_{0}-I_{\sigma})}{\sum I_{0}}
- \frac{\sum(B_{0}-B_{\sigma})}{\sum I_{0}}$], where I$_{0}$ and
B$_{0}$ have the same meaning as in definition of A. The subscript
``$\sigma$'' refers to the image that is smoothed with a boxcar of
size $\sigma = 0.3 \times r(\eta=0.2)$, where $r(\eta=0.2)$ is the
inverted Petrosian radius. We note that all central pixels (within
1/20 of the defined total radius of the galaxy) are set to nil
value. Our S-definition is adapted from \citet{taylor07}. We should
also add that CAS are calculated within the total radius of the
galaxy, defined as 1.5 $\times$ $r(\eta=0.2)$. $\eta (r)=
\frac{I(r)}{\langle I(r)\rangle}$, where in practice $I(r)$ is the
(mean) pixel-intensity at radius $r$ from the galactic center and
$\langle I(r)\rangle$ is the average intensity within $r$ \citep[see
also][]{takamiya99}.

\section{Galaxy Decomposition Using BUDDA}

Figure 3 shows examples of the BUDDA-based decomposition for the
first four galaxies listed in Table 1 (ordered by CIG/KIG name). The
left panel displays, from left to right, the initial image, the
fitted model and the enhanced residual image (initial image
normalized by the model). The residual image shows high spatial
frequency structure associated with the spiral arms and emission
regions in each galaxy. The BUDDA code does not fit such high
spatial frequency  structure. The right panel of Figure 3 shows the
surface brightness profile of the galaxy (black), fitted components
(green=bulge, blue=disk and bar=turquoise) and total model (red; sum
of all components). Data products for the entire sample are
available at www.iaa.es/AMIGA.html. Table 4 provides a full set of
parameters that describe the bulge-disk-bar components of the
galaxies obtained using BUDDA\footnote{The surface brightness
profiles shown in Figure 3 and the numbers reported in columns (7)
and (10) of Table 4 do not include Galactic extinction, (1+z)$^{4}$
redshift dimming or K- corrections. However, everywhere else
hereafter the averages (mean/median) are calculated after such
corrections were applied. Moreover, all plots involving surface
brightness $\mu_{o}$ and $\mu_{e}$ include the corrected values.}.
We checked if our BUDDA parameters are affected by any biases and it
turns out that the parameters derived by the code are insensitive to
galaxy inclination and recession velocity. We conclude that the
BUDDA code decomposition provides reasonable parametrization for all
94 galaxies. Two galaxies show clear shell structure (KIG 600 and
KIG 754) suggesting that despite all of our efforts a few nurtured
galaxies remain in the sample. The code was unable to model such
complex structures, clearly overestimating the contribution of a
bulge component (KIG 600) or not being able to isolate a bulge
component at all (KIG 754). The sample consists of 25, 34 and 35
galaxies classified as Sb, Sbc and Sc, respectively according to our
reclassification using the SDSS images. We included an AGN component
for two galaxies (KIG: 671 and 719) because they are classified as
Seyfert 1 in NED. All statistical analysis will be based upon
results for the  94 galaxies suitable for BUDDA decomposition.

\subsection{Properties of Bulges}

Spiral galaxies show a wide range of bulge sizes and Bulge/Disk
luminosity ratios. It has been suggested that ``bulge building'' via
nurture processes (external acquisitions/accretion of companions)
may be responsible for many or all large bulge systems
\citep{carlberg99}. Alternatively, other studies propose that
dissipative processes in disks (internal secular evolution) are
responsible for building up bulges in most spirals
\citep[e.g.][]{hunt04}.

As expected, the fraction of CIG spirals with large bulges has
decreased as classifications have evolved from the original low
resolution POSS to higher resolution POSS2 images and finally SDSS.
If small bulge spirals represent some kind of primordial unnurtured
spiral population then AMIGA/CIG represents the best sample to study
the population statistically. It is especially interesting to
quantify their properties in order to see how much of the known
morphological and structural diversity is likely due to pure nature
rather than nurture.

In the past 15-20 years, the concept of exponential bulges has been
systematically investigated and nowadays it is accepted that there
exist two general types of bulges: classical and pseudobulges. A
number of criteria \citep{kormendy04} have been proposed to identify
pseudobulges based upon: 1) morphological analysis of high
resolution Hubble images (they show flattened geometry, associated
with nuclear spirals, rings or bars; e.g. \citealt[][]{carollo99,
fisher08}), 2) kinematics (they are described by low velocity
dispersion, which makes them outliers relative to the Faber-Jackson
relation; \citealt{faber76}), 3) photometric analysis of the surface
brightness profile (they show nearly exponential profiles typical of
disks) and 4) color (age) of bulge stellar population (they may be
dominated by Population I material, without obvious signs of
mergers). In what follows, we evaluate the nature of bulges in our
isolated sample by exploiting the photometric decomposition of the
light profiles and we adopt a few simple criteria involving
n$_{bulge}$ and Bulge/Total luminosity ratio to distinguish between
pseudo- from classical bulges.

It has been proposed that all galaxies with a bulge contribution
(relative to the total galaxy luminosity) of 10\% or less are
pseudobulges, i.e. disk-like structures formed by secular evolution
\citep[e.g.][]{kormendy04,laurikainen07}. About 68\% of our sample
shows Bulge/Total ratios B/T$<$ 0.1. Based on BUDDA parameters
$\sim$ 45\% of the Sb galaxies and 75-80\% of the Sbc-Sc galaxies
show B/T $<$ 0.1. A more extreme view (i.e. less restrictive)
involves the proposal that all bulges characterized by n$_{bulge}$
$<$ 2.5 and B/T $<$ 0.45 are pseudobulges
\citep[e.g.][]{kormendy04,drory07}. These considerations raise the
possibility that $\sim$ 94\% of our sample contain pseudobulges.
Applying similar criteria to the sample of n=95 Sb-Sc galaxies from
\citet{laurikainen04a} (based on Ohio State University Bright Spiral
Galaxy Survey; OSUBSGS - \citealt{eskridge02}) we found a similar
fraction $\sim$ 92\% pseudobulges. The latter sample was not
selected using an isolation criterion, so a higher degree of
nurtured galaxies might be expected. In this context we note that a
smaller fraction (59\% of the OSU sample) show B/T$<$0.1.

We note that the largest Bulge/Total ratio in our sample is B/T
$\sim$ 0.35 and the largest n$_{bulge}$ value we find is close to
3.5. The typical uncertainty in n$_{bulge}$ is $\sigma\approx$ 0.5
(confirming the result of \citealt{gadotti08}). Figure 4 shows the
distribution of BUDDA-derived S\'{e}rsic bulge indices n$_{bulge}$
vs. the B/T ratio. The pseudo-/classical bulge proposed boundaries
mentioned above are indicated by dotted lines. Figure 4 shows the
strong concentration of much of our sample within the extreme
pseudobulge domain. Although Bulge/Total is a rather robust
empirical measure \citep{gadotti08}, it is a very challenging task
to quantify its uncertainties, which are not provided by the BUDDA
code. We attempted to calculate the uncertainty of Bulge/Total
varying the input parameters (giving BUDDA a range of reasonable
starting values). We estimate (2-3$\sigma$) uncertainties generally
smaller than 15\%.

In Figure 4 the three morphological types Sb, Sbc and Sc are
indicated with distinct symbols (see figure's legend). There are six
galaxies (4 Sb and 2 Sbc) with n$_{bulge}$ $>$ 2.5 and Bulge/Total
$>$ 0.1 that could be interpreted as classical bulges because they
lie outside the extreme suggested pseudo-/classical bulge
boundaries. Thirty galaxies lie outside of the more restrictive
pseudobulge domain with Sb again showing the largest fraction. This
would still leave 2/3 of our sample (and most Sbc-Sc) as pseudobulge
systems. The subset of Sb galaxies show an apparent trend or linear
correlation in Figure 4. Best fit regression line (solid) and
bisector (dashed) are indicated for the Sb population. In either
case, the correlation coefficient is R$\approx$0.7. For the Sbc and
Sc subsets we find no evidence for a correlation (correlation
coefficients R$\sim$ 0.3-0.4 for Sbc types and 0.01-0.07 for Sc
types). Clearly the scatter increases with lateness of type. As a
check for hidden luminosity trends in Figure 4 we compared the
location of the five least and most luminous galaxies (not shown
here). Those points scatter everywhere on the plots suggesting no
systematic effects on the distribution of measures.

Table 5 presents the average values (mean/median) of the structural
parameters of bulges, disks and bars for the entire sample. The
table is organized as follows: Column 1 - Morphological Type, Column
2 - Bulge/Total luminosity ratio, Column 3 - S\'{e}rsic index of the
bulge n$_{bulge}$, Column 4 - bulge effective radius r$_{e}$, Column
5 - effective surface brightness of the bulge $\mu_{e}$, Column 6 -
disk scalelength h$_{R}$, Column 7 - central surface brightness of
the disk $\mu_{o}$. Table 5 shows a decreasing trend for mean and
median Bulge/Total and S\'{e}rsic index n$_{bulge}$ measures from
earlier to later types, with a larger gradient between Sb and Sbc
than between Sbc and Sc. The S\'{e}rsic index n$_{bulge}$ appears to
be more sensitive to differences in Hubble type. We also point out
that the bulge effective radius r$_{e}$ shows no trend among Hubble
subtypes (Tables 5). Mean/median effective surface brightness
increases from Sc to Sb by about 1.3 magnitudes/arcsec$^2$.

Figure 5a shows a 2D projection within the fundamental plane defined
by $\mu_{e}$ (surface brightness at r$_{e}$) and r$_{e}$
\citep[Hamabe-Kormedy relation;][]{hamabe87} for bulges in our
sample - (log r$_{e}$, $\mu_{e}$) plane. The bulge effective surface
brightness $\mu_{e}$ is corrected for Galactic extinction,
(1+z)$^{4}$ and K-corrected. There is no clear trend, as previously
reported for late-type spiral bulges \citep{capaccioli92,carollo99}.
We again see a separation between the earliest and latest types in
our sample where Sb galaxies show the highest and Sc the lowest
effective surface brightness (see also Table 5). The two outliers
with lowest effective radii may be Sd galaxies. The segregation of
morphological types is driven along the ordinate by lines of
constant luminosity $L_{T}^{bulge}\propto I_{e} r_{e}^2$  in the
(log r$_{e}$, $\mu_{e}$) plane. Isolated Sb galaxies tend to be more
luminous and have larger bulges compared to isolated Sc galaxies
(see Tables 3 and 5). Galaxies with luminosities differing as much
as 4 magnitudes share the same range in r$_{e}$. Bulges in our Sb -
Sc sample do not grow larger than r$_{e}$$\simeq$2.5 kpc regardless
of their luminosity. These results are consistent with those
reported by \citet{capaccioli92} (see their Figure 4).

Figure 5b shows the distribution of the bulge effective radius as a
function of bulge absolute magnitude in i-band, which can be
regarded as a surrogate (log r$_{e}$, $\mu_{e}$) plane. The
M$_{i}^{bulge}$ is obtained from the SDSS cmodel i-band magnitude of
the galaxy (\S~2) taking into account the Bulge/Total luminosity
ratio from the BUDDA code. The more luminous bulges show larger
effective radii, although the correlation between the two parameters
shows large scatter, with a linear regression correlation
coefficient R=0.65 (solid line). Actually we see three parallel
sequences of galaxies that we classify as Sb, Sbc and Sc,
respectively. The r$_e$ ranges are similar but Sc galaxies are
displaced between $\Delta$M$_i$= 1-2 magnitudes lower than Sb
galaxies. Bulge effective radius is relatively insensitive to Hubble
subtypes while bulge luminosity is useful in this context. This
interpretation allows one to identify visually misclassified objects
(or galaxies for which BUDDA derived bulge parameters are
unreliable).

Figure 5c plots $\mu_{e}$ versus M$_{i}^{bulge}$. It suggests that
brighter bulges are also characterized by higher effective surface
brightness, yet again the scatter is quite large. Figure 5d displays
the distribution of the bulge S\'{e}rsic indices as a function of
M$_{i}^{bulge}$. Fainter bulges tend to have lower values for
n$_{bulge}$.

\subsection{Properties of Bars}

Visual inspection of SDSS i-band images suggests that 57\% of our
sample (55 out of 96) could be classified as SB or SAB, showing bars
or ovals. This fraction is consistent with that reported in other
studies (in near-IR or r-band) with no restriction on morphological
type
\citep{knapen99,eskridge00,menendez07,marinova07,verley07c,barazza08}.
The BUDDA code identified a bar component in 51\% of the sample (48
out of 94). There is a slight discrepancy between our visual
estimate and BUDDA results (55 versus 48 barred galaxies,
respectively). Six out of seven galaxies for which BUDDA did not
identify a bar component were visually classified as SAB, i.e.
transitional or intermediate between barred SB and non-barred SA.
The most intriguing case is KIG 689, which visually could be
classified as SB, yet the code cannot separate bulge-bar components.

We find that 34 galaxies have a Bar/Total luminosity ratio smaller
than 10\%. The fraction of barred galaxies decreases from 84\% (for
Sb type) to $\sim$ 40-50\% for each of the later types Sbc and Sc.
The BUDDA code provides a parameter called ``maximum radius of the
bar'', which we tabulate as l$_{bar}$ and use as an estimate for the
length of the bar \citep{gadotti08}. Figure 6 shows the distribution
of l$_{bar}$ (semimajor axis of the bar). Barred galaxies show a
wide range l$_{bar}$= 1-12 kpc with a large concentration in the
range $\sim$ 2-6 kpc.

Tables 6ab present average values (mean/median) for the most
important structural parameters for bulge, bar and disk components
as estimated by BUDDA. We present numbers for barred (6a) and
non-barred (6b) galaxies separately and by morphological subtype.
Tables 6ab are organized as follows: Column 1 - Morphological Type,
Column 2 - Bulge/Total luminosity ratio, Column 3 - n$_{bulge}$,
Column 4 - bulge effective radius, Column 5 - effective surface
brightness of the bulge, Column 6 - disk scalelength, Column 7 -
central surface brightness of the disk, Column 8 - semimajor axis of
the bar.

Table 6a indicates that l$_{bar}$ decreases by a factor of two from
Sb through Sc (qualitatively consistent with \citealt{erwin05}; see
also \citealt{combes93,zhang07}).

\subsection{Properties of Disks}

Figure 7 illustrates the relation between the two parameters
describing the disk exponential profile: $\mu_{o}$ and h$_{R}$. The
surface brightness $\mu_{o}$ was corrected for Galactic extinction.
We also applied a (1+z)$^{4}$ and a K-correction. In Figure 7a the
three morphological types are indicated with different symbols
Sb-Sbc-Sc. In Figure 7b the ten most luminous and the ten least
luminous galaxy disks in the sample are shown with solid circles and
solid triangles, respectively. The ten most luminous disks have
absolute magnitudes in the range -22.8 to -22 and the ten least
luminous have absolute magnitudes in the range -19.9 to -19. The
disk central surface brightness $\mu_{o}$ and disk scalelength
h$_{R}$ are strongly correlated, with linear regression correlation
coefficient R=0.88 \citep[see
also][]{grosbol85,kent85,khosroshahi00,grahamblok01,mendez08}. The
slope of the linear regression fit is 3.0$\pm$0.2, well below a
constant luminosity disk, which could be described by a slope of 5
(based on the approximation $L_{T}^{disk}\approx2\pi
I_{o}h_{R}^{2}$). This scaling relation seems to hold for all spiral
types and it is observed for low surface brightness galaxies as well
\citep{beijersbergen99}. Other studies reported a slope in the range
1.5-3.0 (see \citealt{graham01a} and references therein). The
($\mu_{o}$, log h$_{R}$) plane is part of what some label as ``the
fundamental plane'' of spiral galaxy disks
\citep[e.g.][]{graham02,shen02} described by v$_{max}$, $\mu_{o}$,
h$_{R}$, where v$_{max}$ is the maximum rotation velocity. For a
constant velocity, virial theorem expressed as $v_{max}^{2} \propto
I_{o} h_{R}$ predicts a slope of 2.5, assuming a constant M/L
(mass-to-light luminosity ratio). We find no morphological
separation Sb-Sbc-Sc in the ($\mu_{o}$, log h$_{R}$) plane in our
sample (Figure 7a). A separation is seen when one compares more
extreme morphological types (S0/Sa-Sb versus Scd-Sm/Irr;
\citealt{grahamblok01}). In Figure 7b the ten most luminous disks
seem to define the upper envelope and the ten least luminous the
lower envelope of the $\mu_{o}$ - log h$_{R}$ correlation. If we
divide our sample in luminosity bins we would see parallel lines of
constant luminosity in Figure 7b.

Most galaxies ($\sim$ 90\%) have h$_{R}$ $<$ 10 kpc. We observe that
the Sb barred galaxies have a disk scalelength larger by a factor of
two compared to the nonbarred Sb galaxies. For Sbc and Sc galaxies
the presence/absence of bars does not appear to affect h$_{R}$
(Tables 6 a-b). We also note that the barred galaxies exhibit a
significant change in the disk scalelength between Sb and Sbc-Sc
(Table 6a), the earlier types Sb showing the largest values. In the
case of non-barred galaxies the trend is rather reversed (Table 6b).

In \S~2 we found that the size of the disk decreases from Sb to Sc
morphological type (Table 3). When a$_{25}^{i}$ is normalized to the
disk scalelength h$_{R}$ we get a rather different picture (Table
7). Sbc types are characterized by the largest values (also noted in
\citealt{erwin05}; see their Figure 7).

\subsection{Bar-Bulge-Disk Scaling Relations}

\subsubsection{Bar-Bulge Interplay}

Figure 8a shows n$_{bulge}$ versus Bulge/Total luminosity ratio for
the barred galaxies in the sample. Most barred galaxies with
Bulge/Total $>$ 0.1 are morphological type Sb ($\sim$81\%); 67\% of
all barred Sb-Sc fall in the high probability pseudobulge space
(Bulge/Total $<$ 0.1). Sb galaxies show a correlation even when we
restrict the plot only to barred galaxies. The correlation
coefficients are R$\approx$0.7 for both the regression line and the
bisector fit, shown with a solid and a dashed line, respectively.
Figure 8b shows n$_{bulge}$ versus Bulge/Total luminosity ratio for
the non-barred galaxies in the sample; 70\% of all non-barred Sb-Sc
fall in the high probability pseudobulge space (Bulge/Total $<$
0.1). Non-barred galaxies appear to concentrate at lower values of
n$_{bulge}$ compared to the barred galaxies. One can notice a scarce
occupation for n$_{bulge}$ $>$ 1.7 for non-barred galaxies.

While for the barred galaxies we see a tendency to get lower values
for both Bulge/Total and the S\'{e}rsic index n$_{bulge}$ from Sb to
Sc, the non-barred galaxies show rather unchanged numbers (Tables
6ab). We also point out that the bulge effective radius r$_{e}$
doesn't appear sensitive to the presence/absence of bars (Tables
6ab). While barred galaxies show lower effective surface brightness
from Sb through Sc, non-barred galaxies do not appear to change in
bulge effective surface brightness with morphological type (Tables 6
ab).

\subsubsection{Bar-Disk Interplay}

Figure 9a shows a robust correlation between the size of the bar
l$_{bar}$ and the disk scalelength h$_{R}$, regression line
correlation coefficient R=0.84. Larger bars are hosted by larger
disks. Figure 9b suggests that larger bars are found in disks with
lower central surface brightness $\mu_{o}$. Figure 9c indicates that
the bar size is correlated with the absolute magnitude of the
galaxy, linear regression correlation coefficient R=0.69, the more
luminous galaxies harboring the largest bars \citep{kormendy79}.

Figure 9d displays a relation between the galaxy color (g-i)$_{o}$
and the size of the bar l$_{bar}$. We observe that larger bars are
hosted by redder galaxies.

In panel a we presented a tight linear correlation between l$_{bar}$
and the disk scalelength h$_{R}$. At this time we would like to see
if a similar correlation holds between bar length l$_{bar}$ and the
size of the disk a$_{25}^{i}$ (Figure 9e). Although a larger scatter
is evident, the trend is consistent with the previously reported
result (Figure 9a), namely larger bars are hosted in larger disks.

As we reported in \S\S~4.2 the size of the bar l$_{bar}$ appears
dependent upon the morphological type, Sb hosting the largest bars
and Sc the smallest ones. Even when $l_{bar}$ is normalized to
a$_{25}^{i}$ the trend is still preserved. However, if l$_{bar}$ is
normalized by h$_{R}$ we notice a similarity between Sb and Sbc
galaxies and a rather large drop for Sc types (Table 7).

In all trends seen in Figures 9a-e one can notice a morphological
separation Sb-Sbc-Sc. At one end Sb galaxies tend to have the
largest bars, the largest disks, the lowest central surface
brightness, being the most luminous and the reddest. At the other
end lie the Sc galaxies.

\subsubsection{Bulge-Disk Interplay}

Figure 10a plots the bulge effective radius r$_{e}$ versus the disk
scalelength h$_{R}$ (linear regression correlation coefficient
R=0.55). Larger bulges are associated with larger disks. Figure 10b
shows the bulge effective radius normalized to the disk scalelength
as a function of Bulge/Total ratio. This later panel should be
considered in conjunction with Tables 8ab. Table 8a presents the
average values (mean/median) of $r_{e}/h_{R}$ for each morphological
type (considering also barred vs. non-barred galaxies) and Table 8b
displays those averages for galaxies with Bulge/Total $<$ 0.1 and
Bulge/Total $>$ 0.1, respectively. We found a similar proportion
(half-half) of barred/non-barred among galaxies with Bulge/Total $<$
0.1. The same relative distribution of barred/non-barred we find for
galaxies with Bulge/Total $>$ 0.1. We observe that non-barred
galaxies have on average larger $r_{e}/h_{R}$ than barred galaxies
for all morphological types (Table 8a). This seems to be the case
even when we divide the sample about Bulge/Total=0.1 (Table 8b).

Three main conclusions emerge from Figure 10b: a) for galaxies with
Bulge/Total $<$ 0.1 the points appear evenly distributed about
$r_{e}/h_{R} \sim 0.16-0.18$, b) the dispersion of $r_{e}/h_{R}$
values increases as Bulge/Total gets larger and c) Sb galaxies seem
to detach themselves from a clear correlation described by Sbc and
Sc galaxies (see also \citealt{laurikainen07}).

In Figure 10c we test whether Bulge/Disk luminosity ratio scales
with the bulge effective radius r$_{e}$ normalized to the size of
the disk a$_{25}^{i}$. The scatter is rather large, with an
increasing dispersion for larger Bulge/Disk values. Nonetheless, a
global trend is evident, with $r_{e}/a_{25}^{i}$ getting larger as
Bulge/Disk gets larger. We see again, just like in panel b, a
separation of Sb galaxies from the rest of the sample, Sbc-Sc
showing a correlation between the two parameters.

\section{CAS parametrization - Data Analysis}

CAS parameters are a useful diagnostic indicating possible
interacting processes. Our sample is particularly useful in defining
the natural levels of these parameters. We have considered for this
part of the analysis our sample of n=96 galaxies (including KIG 600
and KIG 754 for which we could not perform reliable BUDDA
decomposition - see \S~4). The results are summarized in Table 9.
Table 10 presents CAS averages (mean/median) measures. Concentration
parameter C shows a clear decrement with morphological type from Sb
to Sc. We also note that Sb galaxies appear more symmetric than
Sbc-Sc types.

The three panels of Figure 11 show the CS-CA-SA parameter planes for
our isolated galaxies. The most significant correlation appears
between the Clumpiness index S and the Concentration index C (linear
regression correlation coefficient R=0.66). The more concentrated
galaxies appear clumpier as well. We note that Sb galaxies show a
behavior rather different relative to Sbc-Sc types. The former show
a wide range of concentration indices (C) and are clustered in the
region of very low values of asymmetry index (A), while the latter
show a smaller extent in C, yet a much wider range of asymmetry.

\section{Combining BUDDA-based and CAS measures}

It is useful to look for trends between physical parameters that
describe the larger scale (low frequency) components of galaxies
(i.e. BUDDA-based measures) and morphological parameters (i.e. CAS)
that are sensitive to the higher frequency structures.

Figure 12 (a-d) presents relationships between the Concentration
index C and parameters that describe the bulge. As reported in
previous sections all measures (C, n$_{bulge}$, Bulge/Total,
$\mu_{e}$) appear dependent upon the morphological type. Therefore,
in general terms, the trends shown in the panels of Figure 12 are
somehow predictable. Yet, it is important to emphasize a few
aspects: 1) for Bulge/Total $<$ 0.1, galaxies show a clear linear
correlation between Bulge/Total and C (linear regression correlation
coefficient R=0.68), but for Bulge/Total $>$ 0.1 there is a large
scatter in that plot (panels a-b); 2) Concentration index C scales
with the bulge S\'{e}rsic index n$_{bulge}$ with an increasing
dispersion as n$_{bulge}$ gets larger (panel c); 3) larger
concentration indices C are found only in galaxies characterized by
brighter bulge effective surface brightness $\mu_{e}$ (panel d)
\citep[see also][]{graham01b}. We also find that more asymmetric
galaxies (larger A) are restricted to brighter central surface
brightness disks ($\mu_{o}$) (Figure 13). Barred and non-barred
galaxies show very similar behavior in the A-$\mu_{o}$ plane (not
shown in the paper). The meaning of the curved trends illustrated in
Figures 12d and 13 are not completely clear at this time. Extending
the morphological range in both directions outside the Sb-Sc
morphological range may be relevant in this regard.

\section{Discussion and Conclusions}

We have presented a detailed structural analysis for a well defined
sample of $\sim$ 100 late-type isolated galaxies. If a {\it bona
fide} isolated (pure ``nature'') population of galaxies exists then
our previous work \citep{sulentic06} suggests that it is dominated
by systems with spiral morphology ($\sim$ 84\%) with the bulk in the
range Sb-Sc (63\%). We assume that the galaxies we investigate here
are best described as ``minimal nurture and maximal nature'' systems
because they are as isolated as individual galaxies can be. This
hypothesis does not imply that these isolated galaxies have
undergone no merger activity since their epoch of formation but
rather that major mergers are probably absent from their past $\sim$
3Gyr history. We do note that the AMIGA sample includes 14\%
early-type galaxies and those are systems of such low luminosity as
to suspect little or no major merger activity over their entire
history \citep{sulentic06}.

One might reasonably expect the tightest correlations between
various intrinsic properties from a sample of isolated galaxies,
where it is assumed that nurture (i.e. interactions) would increase
the scatter (e.g. UBV-colors; \citealt{larson78}). The strength of
this study is manifold: the large size of the sample, the uniformity
of the SDSS data, the robustness of the BUDDA code and the stringent
isolation criteria underlying the definition of the parent AMIGA
sample. In this study we have retained subjective morphological
classifications and investigate morphological type dependence of
various properties even though the typical range is narrow. This
narrowness coupled with our ``nurture-free'' assumption raises the
possibility that Hubble type T=4$\pm$1 may represent the seed
population for all spiral galaxies.

\subsection{Pseudobulges in Isolated Galaxies}

We present evidence favoring the hypothesis that most or all
late-type isolated galaxies host pseudobulges (\S\S~4.1) rather
than classical bulges:

A. A large majority of our isolated systems host relatively
``unevolved'' bulge structures (as hypothesized by
\citealt{hunt04}); most S\'{e}rsic indices (n$_{bulge}$) are smaller
than 2.0-2.5 (see Table 5 and Figure 4) with the largest
concentration around n$_{bulge}$ $\sim$ 1.3-1.4. Such bulges are
probably not as relaxed as larger bulges in earlier spiral types.
They are likely dominated by rotation unlike higher S\'{e}rsic index
bulges (for a detailed discussion on this subject see section 4.6 in
\citealt{kormendy04}, section 4.2 in \citealt{laurikainen07} and
references therein).

B. We observe a large range of effective surface brightness
$\mu_{e}$ for a rather narrow range of r$_{e}$ (these two last
parameters defining in part the fundamental plane) - see Figure 5a.
The locus occupied by the bulges of our Sb-Sc galaxies in this plane
is similar to that of disky bulges of galaxies at the end of
dissipative collapse \citep{capaccioli92}. The lack of correlation
between $\mu_{e}$ and r$_{e}$ supports the case of ``pseudobulges''
for isolated spiral galaxies in our sample. As pointed out in
\citet{macarthur03}, these results support an ``iceberg'' scenario,
i.e. late-type spiral bulges are ``more deeply embedded in their
host galaxy disk than earlier type bulges''. This idea is further
complemented by the fact that the size of the bulge (r$_{e}$) scales
with the scalelength of the disk (Figure 10a) \citep[see also
e.g.][]{khosroshahi00,macarthur03,mendez08,fisher08}.

We observe a larger dispersion in r$_{e}$/h$_{R}$ for larger
Bulge/Total luminosity ratios (Figure 10b). However, in Figures
10b-c one can clearly see that Sbc-Sc galaxies do show a clear
increasing trend for r$_{e}$/h$_{R}$ and r$_{e}$/a$_{25}^{i}$ with
Bulge/Total and Bulge/Disk luminosity ratio, respectively. In
contrast, Sb galaxies appear detached from the Sbc-Sc population.
Assuming that the bulge S\'{e}rsic index and/or Bulge/Total
luminosity ratios are reasonable discriminators of pseudo- versus
classical bulges (\S\S~4.1), then amongst our sample Sb galaxies
have the greatest chance of hosting classical bulges. Thus, in
Figures 10b-c we may have yet another indication that the
pseudobulges and the galactic disks are clearly connected, while the
classical bulges do not show similar scaling relations.

Some studies (e.g. \citealt{thomas06}) argue that ``secular
evolution through the disk and the phenomenon of pseudobulge
formation are most likely restricted to spirals of types Sc and
later''. Our results (but see also \citealt{laurikainen07}) find a
large fraction of pseudobulges among spiral types earlier than Sc
(see \S\S~4.1). This may be telling us that the formation of
pseudobulges does not appear exclusively restricted to Sc types or
later. Our results suggest that if one considers only morphological
types later than Sc, one may identify an almost pure pseudobulge
population of galaxies. A fundamental question mentioned earlier is
whether the isolated Sb-Sc spiral galaxies constitute the seed
population of unnurtured spirals? If so then isolated galaxies might
be expected to host a pure pseudobulge population. In this context
Sb types in our sample have the greatest chance of bulge building
via nurture and may involve a mixed classical and psuedobulge
population. In Figure 4 it is interesting that a linear correlation
emerges only for galaxies of Sb type which bridges the classical and
pseudobulges domains. Alternatively, the trend may be telling us
that all/most Sb galaxies contain a real (classical) bulge. This
would suggest that some large bulges are natural or that all Sb
spirals in the sample are a product of nurture. The latter
interpretation is disfavored by the extreme isolation of our sample.

\subsection{The Role of Bars in the Formation of Pseudobulges}

The results of the present study could set constraints for various
galaxy formation and evolution models. Two important galaxy
formation scenarios have been proposed and advocated: 1) spheroidal
component (bulge) forms prior to the disk component in a monolithic
collapse or via early mergers (so called ``inside out'' formation,
e.g.
\citealt{eggen62,baugh96,kauffmann96,vandenBosch98,cole00,merlin06})
and 2) bulges form after the disk component as a result of secular
dynamics/evolution driven by a disk instability
\citep[e.g.][]{courteau96,zhang04} possibly triggered by external
satellite accretion \citep[e.g][]{aguerri01,eliche06}. The former
mechanism may be dominant for elliptical galaxies and in early
spiral galaxies with large bulges (as they all appear to share
similar properties and scaling relations within the fundamental
plane; e.g. \citealt{kormendy85,djorgovski87,faber87}). The latter
mechanism may be more plausible for late type spiral systems
\citep[e.g.][]{carollo99,hunt04,debattista04}, as they largely
harbor pseudobulges.

Some authors proposed that bulges of late type spiral galaxies are
formed primarily through secular evolution of bars
\citep[e.g.][]{kormendy79,kormendy93,norman96,hasan98,fathi03,kormendy04,athanassoula05,jogee05,debattista06}.
Others have suggested that bars can help the process of
``pseudobulge'' formation (making it faster and more efficient), but
is not a necessary requirement for that process \citep[e.g.][and
references therein]{laurikainen07}. Bars can transport gas inward
\citep[e.g.][]{sakamoto99, sheth05} potentially contributing to the
formation of a bulge. On the other hand it has been proposed that
even without a bar the stellar disk component could be redistributed
due to a secular torque action \citep[e.g.][]{zhang07}.

We find a larger fraction of barred galaxies among Sb types relative
to Sbc-Sc types (\S\S~4.2). Sb galaxies also appear to host the
largest bars (Table 6a) within the morphological sequence Sb-Sbc-Sc.
If bars are assumed as necessary precursors of all pseudobulges,
then the smaller bars in later type galaxies ``dissolve'' more
efficiently in the process of bulge formation. It is interesting to
mention that for Sb and Sbc types in our sample of isolated galaxies
we find systematically larger values of the index n$_{bulge}$ for
barred galaxies compared to the non-barred galaxies (Tables 6a-b).
The difference almost vanishes for Sc barred and non-barred.
\citet{laurikainen07} report a rather opposite result for Sb type
(see their Figure 3). If n$_{bulge}$ is one of the empirical
discriminators between classical and pseudobulges then any
connection with the presence/absence of bars merits further
attention. In this context it is relevant to review our Figures
8a-b. We note the ``disappearance'' of objects with n$_{bulge}$
above 1.7 for non-barred galaxies (Figure 8b in contrast to Figure
8a). We tested whether this may be caused by the resolution
limitation in SDSS images, thus the BUDDA code's inability to
identify the presence of a bar. First, we analyzed the distribution
of n$_{bulge}$ values of non-barred galaxies with V$_{R}$ lower and
higher than the median V$_{R}$ of the full non-barred sample ($\sim$
5700 km s$^{-1}$), respectively. We found no significant difference.
Secondly, for our galaxies, the typical seeing FWHM is better than
1\arcsec, with very few cases at 1.5\arcsec. Considering the most
extreme case, for a galaxy showing V$_{R}$ $\simeq$ 10,000 km
s$^{-1}$ a 1.5\arcsec seeing would translate into a spatial
resolution of $\sim$ 1.0 kpc, which is well within the capability of
the BUDDA code to provide reliable structural measures
\citep{gadotti08}. The scarcity of non-barred galaxies with
n$_{bulge}$ above 1.7 is consistent with the scenario that bars
could transform by dissolution into pseudobulges. The presence of
bars may influence the degree of relaxation of bulges in the sense
that n$_{bulge}$ decreases from Sb through Sc only for barred
galaxies, but not for non-barred spirals (\S\S~4.2).

The formation and lifetime of bars may be sensitive to environment
\citep[e.g.][]{gerin90}. It has been suggested that bars in early
type spiral galaxies are formed by tidal interactions with other
galaxies and those in late types have intrinsic origin
\citep{noguchi96}. The connection bars-environment may be different
for early and late type spirals \citep{noguchi96, noguchi00}, being
proposed a ``bimodality'' of bars in this sense. Moreover, numerical
simulations have shown that for Sb-Sc galaxies bars are transient
features and dissolve progressively in $\sim$ 1-2 Gyr
\citep{bournaud05}. As we pointed out earlier, the AMIGA/CIG
isolated galaxies have been basically nurture-free for at least a
comparable time. We find that $\sim$ 50-60\% of our present sample
are barred galaxies. The conclusion here could be that the bars we
observe in these late type isolated spiral galaxies have been likely
renewed or reformed through internal processes and not by external
accretion or interactions \citep[e.g.][]{block02,berentzen04}. It is
also interesting to mention we find that the largest bars lie in
disks with the lowest central surface brightness $\mu_{o}$ (Figure
9b). This is consistent with the idea that bars build up from the
material in the central parts of disks and they are products of
secular dynamical evolution within the disk.

We find that our isolated galaxies tend to host large
bars\footnote{However, one must be aware that there is no standard
definition for the length of a bar in a galaxy \citep{erwin05} and
scaling parameters for galaxy components may be sensitive to the
filter that is used for photometry.}. Our Figure 6 shows that most
bar radii are clustered in the range 2-6 kpc. \citealt{erwin05}
(based on \citealt{martin95}) reports typical bar sizes in the range
1-3 kpc (B-band) for morphological types Sb-Sc having absolute
magnitudes similar to our sample. A more recent study
\citep{marinova07} presents a characterization of bars in optical
(B-band) and near-IR (H-band) for the OSUBSGS sample of galaxies. In
order to compare the bar sizes with their estimates, we restrict
their sample to Sb-Sc morphological range, based on the RC3 catalog
\citep{deVaucouleurs91}. The OSUBSGS-based sample has a similar
distribution of absolute magnitudes as our sample. In terms of
l$_{bar}$, our sample of barred galaxies (n=48) is characterized by
a mean $\sim$ 5.0 kpc and a median $\sim$ 4.8 kpc. For The OSUBSGS
sample of n=49 barred galaxies, the mean and median values (H-band)
of l$_{bar}$ $\sim$ 3.8 kpc and $\sim$ 3.4 kpc, respectively. The
conclusion is that the size of bars may be related to the
environment, isolation favoring larger bars.

Moreover, this conclusion seems to be consistent with reports that
the disk scalelength h$_{R}$ of spiral galaxies in rich environments
is typically smaller than that of field (i.e. isolated) galaxies
\citep[e.g.][]{aguerri04}. We find that the bar size scales with the
disk scalelength h$_{R}$ (our Figure 9; see also \citealt{laine02}).
In extreme environments (e.g. compact groups) spiral galaxies tend
to lose their disk components by dissolution into a stellar halo.
The size of the disks in what we assumed were initially late type
spiral galaxies in Seyfert's Sextet for example (estimated by the
last concentric isophote) is less than 10 kpc diameter, comparable
to the smallest disks in our present sample \citep{durbala08}.
Comparing our $l_{bar}/a_{25}^{i}$ estimates with the similar
quantities reported in \citet{erwin05} we observe the same declining
trend from Sb through Sc; for our sample we do obtain systematically
larger $l_{bar}/a_{25}^{i}$ ratios relative to that study, although
we note that \citet{erwin05} measures are based on B-band data from
\citet{martin95}.

\subsection{CAS Structural Measures in Isolated Galaxies}

The minimal environmental influence on AMIGA/CIG galaxies
investigated here is revealed also by an analysis of the structural
properties in terms of CAS parameters (\S~5). Due to the narrow
morphological range represented in our sample of isolated galaxies,
any attempt at comparison with other studies must be cautiously
explored. Nonetheless, the size of the sample examined in the
present study allows a meaningful comparison of the 96 galaxies as a
whole (i.e. the full set of Sb-Sc galaxies) with galaxies of same
morphological types selected without isolation constraints. Table 11
offers such a comparison with the subsample of Sb-Sc galaxies (n=49)
examined in \citet{conselice03}, extracted from the \citet{frei96}
sample, assumed representative for the population of nearby normal
galaxies. The general conclusion is that the isolated galaxies are
less concentrated, less asymmetric and less clumpy than other
galaxies of same morphological type selected without isolation
criteria. Thus, we may have clear indications of environmental
influence on the structure of galaxies. This may be telling us that
the formation of large central concentrations and large clumps
within disks are disfavored in the absence of comparable sized
neighbors.

\subsection{Describing the Morphological Classification}

Although our study involves a narrow range of morphological types
all plots that involve exclusively bulge measures show clear
morphological separation (e.g. Figure 5). When we combine disk
measures (e.g. Figure 7a), the morphological segregation is less
clear or absent suggesting some commonality among disk properties
over the Sb-Sc range. Thus, it appears that the morphological
separation may be associated with a change in the luminosity profile
of bulges as indicated by their S\'{e}rsic indices. \citet{hunt04}
proposed that spiral galaxies may begin with low bulge S\'{e}rsic
index. As they age they change into structurally more evolved
systems (toward n$_{bulge}$ = 4 or higher) also  characterized by
higher surface brightness (see Figures 5a, 12d) and an increased
absolute magnitude (see Figures 5b-d). However, \citet{carollo99}
argue that pseudobulges cannot evolve into denser r$^{1/4}$ (i.e.,
n$_{bulge}$=4) bulges just by repeated cycles of bar
formation/disruption.

At the same time one should keep in mind that the bulge S\'{e}rsic
index is associated with rather large uncertainties (\S\S~4.1),
which complicates its use for a quantitative morphological
classification \citep{gadotti08}. It appears that the concentration
index C, the Bulge/Total (or Bulge/Disk) luminosity ratio and the
bulge S\'{e}rsic index are relevant parameters when one describes
the morphological sequence of spiral galaxies from earlier to later
types. Nonetheless, Figure 12 suggests that the morphological
diversity of spiral galaxies is deeply connected to the structure of
their bulges. The concentration index C is not a good tracer of
Bulge/Total ratio for Bulge/Total $>$ 0.1. This is true because the
bulge light is no longer concentrated within the radius that
includes 20\% of the total light (section \S~3; see also
\citealt{graham01b}). It is not obvious why the bulge surface
brightness shows a plateau in its trend versus C (Figure 12d).
However, one may speculate that the fact that some of the Sb
galaxies curve away from the main trend (described largely by Sbc
and Sc types) toward larger C values could be due to a different
type of bulges they host.

\subsection{Final Remarks}

This present study could be complemented by an extension of a
similar type of analysis to the whole set of isolated spiral
galaxies, which would include the whole sequence of Hubble
morphological types. This would provide a more general and a more
clear picture on the morphological type dependence of various
structural properties and scaling relations presented and discussed
here. Measures of bulge colors and kinematics would both provide
strong tests of our hypothesis that most isolated spirals involve
pseudobulges. Another complementary approach is a Fourier analysis
of our images, which would provide a quantitative description of the
spiral structure, intimately connected to galactic morphology as
well. This is part of an ongoing project we are working on at this
time and the results will be presented in a future paper.

\section*{Acknowledgments}

We thank Dr. Dimitri Alexei Gadotti for answering many questions
about the BUDDA code. His prompt and clear replies are sincerely
appreciated. LVM is partially supported by DGI Grant AYA
2005-07516-C02-01 and Junta de Andaluc\'{i}a (Spain).

This study has made use of SDSS Data Release 6. Funding for the SDSS
and SDSS-II has been provided by the Alfred P. Sloan Foundation, the
Participating Institutions, the National Science Foundation, the
U.S. Department of Energy, the National Aeronautics and Space
Administration, the Japanese Monbukagakusho, the Max Planck Society,
and the Higher Education Funding Council for England. The SDSS Web
Site is http://www.sdss.org/. The SDSS is managed by the
Astrophysical Research Consortium for the Participating
Institutions. The Participating Institutions are the American Museum
of Natural History, Astrophysical Institute Potsdam, University of
Basel, University of Cambridge, Case Western Reserve University,
University of Chicago, Drexel University, Fermilab, the Institute
for Advanced Study, the Japan Participation Group, Johns Hopkins
University, the Joint Institute for Nuclear Astrophysics, the Kavli
Institute for Particle Astrophysics and Cosmology, the Korean
Scientist Group, the Chinese Academy of Sciences (LAMOST), Los
Alamos National Laboratory, the Max-Planck-Institute for Astronomy
(MPIA), the Max-Planck-Institute for Astrophysics (MPA), New Mexico
State University, Ohio State University, University of Pittsburgh,
University of Portsmouth, Princeton University, the United States
Naval Observatory, and the University of Washington.

\clearpage

\begin{table}
\centering \caption{CIG/KIG Galaxies in our Sample}
\begin{tabular}{llccccll}

\hline\hline

Galaxy & & R.A.(J2000) & Dec.(J2000) & v$_r$ & i &
\multicolumn{2}{c}{Morphological Type} \\
KIG name & UGC/NGC name & (hh mm ss.ss) & (+dd mm ss.s) & (km
s$^{-1}$) & ($^{o}$) & old$^{i}$ & revised \\

\hline

KIG 11      &      UGC 00139                &     00 14 31.88   &     -00 44 10.4   &   3963    &   66  &    Sbc    &     SA(s)c    \\
KIG 33      &      UGC 00461 / NGC 0237     &     00 43 27.81   &     -00 07 26.9   &   4175    &   55  &    Sb     &     SA(s)c    \\
KIG 56      &      UGC 01143 / NGC 0622     &     01 36 00.13   &     +00 39 48.8   &   5155    &   43  &    Sb     &     SB(rs)b       \\
KIG 187     &      UGC 03825                &     07 23 33.16   &     +41 26 05.6   &   8281    &   25  &    Sc     &     SAB(s)bc      \\
KIG 198     &      UGC 03935                &     07 37 49.45   &     +46 23 52.4   &   9628    &   32  &    Sc     &     SAB(s)c   \\
KIG 203     &                               &     07 44 36.40   &     +38 02 39.6   &   7998    &   65  &    Sc     &     SA(s)c    \\
KIG 217     &      UGC 04107                &     07 57 01.84   &     +49 34 02.1   &   3504    &   25  &    Sc     &     SA(rs)bc      \\
KIG 222     &      UGC 04158 / NGC 2503     &     08 00 36.75   &     +22 24 00.8   &   5506    &   28  &    Sb     &     SAB(s)b   \\
KIG 232     &      UGC 04256 / NGC 2532     &     08 10 15.20   &     +33 57 22.5   &   5260    &   30  &    Sc     &     SA(rs)c   \\
KIG 238     &      UGC 04283                &     08 14 22.05   &     +39 15 05.3   &   8295    &   18  &    Sb     &     (R$_{1}$$^{\prime}$)SB(s)b    \\
KIG 241     &                               &     08 19 15.77   &     +19 18 48.0   &   5681    &   46  &    Sc     &     SA(s)c    \\
KIG 242     &                               &     08 19 22.31   &     +23 44 50.0   &   4603    &   54  &    Sb     &     SA(s)bc   \\
KIG 258     &                               &     08 31 49.40   &     +28 32 11.0   &   6047    &   52  &    Sb     &     SAB(rs)b      \\
KIG 260     &      UGC 04456                &     08 32 03.53   &     +24 00 38.5   &   5488    &   11  &    Sc     &     SAB(r)c   \\
KIG 271     &      UGC 04512                &     08 39 39.91   &     +60 58 07.7   &   7911    &   58  &    Sc     &     SAB(s)c   \\
KIG 281     &      UGC 04555 / NGC 2649     &     08 44 08.30   &     +34 43 01.8   &   4244    &   32  &    Sc     &     SA(rs)bc      \\
KIG 282     &~~~~~~~~~~~~~~~~~~~~~~~NGC 2651&     08 43 55.12   &     +11 46 15.5   &   8696    &   34  &    Sc     &     SAB(s)bc      \\
KIG 287     &      UGC 04624                &     08 50 23.58   &     +25 57 14.5   &   8297    &   58  &    Sbc    &     SAB(rs)bc     \\
KIG 292     &      UGC 04708 / NGC 2712     &     08 59 30.53   &     +44 54 51.5   &   1818    &   58  &    Sb     &     SA(s)b    \\
KIG 298     &      UGC 04770 / NGC 2746     &     09 05 59.41   &     +35 22 38.3   &   7065    &   29  &    Sb     &     SB(rs)b   \\
KIG 302     &~~~~~~~~~~~~~~~~~~~~~~ NGC 2761&     09 07 30.76   &     +18 26 05.2   &   8728    &   55  &    Sc     &     SA(s)c    \\
KIG 314     &      UGC 04838 / NGC 2776     &     09 12 14.37   &     +44 57 17.8   &   2626    &   8   &    Sc     &     SA(rs)c   \\
KIG 322     &                               &     09 21 10.80   &     +45 53 17.4   &   1858    &   22  &    Sc     &     SA(rs)c   \\
KIG 325     &      UGC 04973                &     09 22 38.12   &     +60 51 55.6   &   7823    &   21  &    Sc     &     SAB(s)bc      \\
KIG 328     &      UGC 05002                &     09 24 22.90   &     +28 17 34.9   &   6518    &   42  &    Sbc    &     SAB(s)bc      \\
KIG 330     &                               &     09 25 30.87   &     +45 31 57.2   &   4276    &   43  &    Sbc    &     SAB(s)bc      \\
KIG 336     &      UGC 05038                &     09 27 23.45   &     +30 26 26.5   &   8070    &   32  &    Sb     &     (R$_{1}$$^{\prime}$)SB(r)b    \\
KIG 339     &      UGC 05055                &     09 30 11.77   &     +55 51 07.4   &   7540    &   32  &    Sbc    &     (R$_{2}$$^{\prime}$)SB(s)bc   \\
KIG 351     &                               &     09 35 21.80   &     +13 32 55.9   &   5920    &   42  &    Sb     &     SB(rs)b   \\
KIG 365     &                               &     09 42 31.53   &     +07 05 57.1   &   8627    &   27  &    Sc     &     SAB(rs)b      \\
KIG 366     &      UGC 05184                &     09 43 02.20   &     +37 49 22.3   &   6581    &   53  &    Sb     &     SB(s)b    \\
KIG 367     &      UGC 05201                &     09 44 34.94   &     +55 45 48.3   &   7627    &   51  &    Sc     &     SA(s)c    \\
KIG 368     &                               &     09 44 47.09   &     +51 41 19.5   &   9959    &   25  &    Sc     &     SA(s)c    \\
KIG 386     &                               &     09 57 48.95   &     +51 49 16.7   &   7490    &   49  &    Sc     &     SAB(s)c   \\
KIG 397     &      UGC 05425 / NGC 3107     &     10 04 22.53   &     +13 37 18.1   &   2791    &   44  &    Sb     &     SAB(s)bc      \\
KIG 399     &                               &     10 07 50.53   &     +34 18 56.5   &   6106    &   40  &    Sc     &     SA(s)bc   \\
KIG 401     &      UGC 05472                &     10 08 32.01   &     -00 39 57.2   &   6436    &   55  &    Sbc    &     SB(rs)bc      \\
KIG 405     &      UGC 05521                &     10 13 52.60   &     +00 33 03.0   &   6232    &   28  &    Sc     &     SAB(s)c   \\
KIG 406     &                               &     10 14 08.03   &     +10 08 54.5   &   8355    &   49  &    Sc     &     SA(s)c    \\
KIG 409     &                               &     10 22 00.37   &     +25 52 20.1   &   6371    &   40  &    Sbc    &     SAB(s)bc      \\
KIG 410     &      UGC 05606                &     10 22 24.06   &     +01 11 55.1   &   6492    &   36  &    Sbc    &     SA(s)bc   \\
KIG 429     &                               &     10 39 45.32   &     +11 38 49.8   &   8991    &   42  &    Sc     &     SAB(s)c   \\
KIG 444     &      UGC 05956                &     10 50 58.37   &     -02 09 00.6   &   4469    &   13  &    Sb     &     SA\underline{B}(s)c   \\
KIG 446     &                               &     10 52 25.05   &     +59 41 08.9   &   8417    &   44  &    Sc     &     SA(rs)bc      \\
KIG 460     &                               &     11 05 00.56   &     +14 53 43.8   &   6186    &   52  &    Sb     &     SAB(s)bc      \\
KIG 466     &      UGC 06194                &     11 09 00.70   &     +22 55 45.4   &   2643    &   26  &    Sc     &     (R$_{2}$$^{\prime}$)SAB(s)c   \\
KIG 489     &      UGC 06568                &     11 35 36.88   &     +00 07 38.6   &   5910    &   56  &    Sbc    &     SA(s)c    \\
KIG 491     &      UGC 06608                &     11 38 33.25   &     -01 11 05.2   &   6251    &   48  &    Sc     &     SA(rs)bc      \\
KIG 494     &                               &     11 40 39.23   &     +28 51 49.3   &   6825    &   24  &    Sbc    &     SAB(s)c   \\
KIG 499     &      UGC 06769                &     11 47 43.71   &     +01 49 34.1   &   8537    &   62  &    Sbc    &     SAB(r)b       \\
KIG 502     &      UGC 06780                &     11 48 50.37   &     -02 01 57.5   &   1729    &   70  &    Sbc    &     SA(s)c    \\
KIG 508     &      UGC 06854                &     11 52 43.45   &     +01 44 27.0   &   6128    &   29  &    Sbc    &     SAB(s)c   \\
KIG 512     &      UGC 06903                &     11 55 36.90   &     +01 14 13.5   &   1892    &   28  &    Sc     &     SB(s)c    \\
KIG 515     &                               &     11 58 36.22   &     +18 51 47.6   &   6876    &   60  &    Sbc    &     SAB(s)c   \\
KIG 520     &      UGC 07144                &     12 09 45.08   &     +56 31 26.8   &   7864    &   33  &    Sc     &     SAB(rs)bc     \\
KIG 522     &                               &     12 15 40.77   &     +61 53 22.6   &   6102    &   34  &    Sb     &     SB(rs)b   \\
KIG 525     &      UGC 07416                &     12 21 39.18   &     +40 50 54.9   &   6901    &   22  &    Sb     &     SB(r)b    \\
KIG 532     &                               &     12 31 34.55   &     +37 58 47.9   &   7129    &   39  &    Sb     &     SAB(rs)c      \\
KIG 550     &      UGC 07917 / NGC 4662     &     12 44 26.17   &     +37 07 17.1   &   6985    &   25  &    Sbc    &     SB(r)b    \\
KIG 553     &      UGC 07987 / NGC 4719     &     12 50 08.69   &     +33 09 33.0   &   7091    &   28  &    Sb     &     SB(r)b    \\
KIG 560     &                               &     12 56 53.52   &     +22 22 24.7   &   6529    &   38  &    Sc     &     (R$_{1}$$^{\prime}$)S\underline{A}B(s)c   \\
KIG 571     &      UGC 08184 / NGC 4964     &     13 05 24.69   &     +56 19 24.7   &   2520    &   54  &    Sc     &     SA(s)c    \\
KIG 575     &      UGC 08279 / NGC 5016     &     13 12 06.63   &     +24 05 42.2   &   2612    &   42  &    Sb     &     SA(rs)c   \\
\end{tabular}

\end{table}

\clearpage

\begin{table}
\textbf{Table 1.}--continued


\centering
\begin{tabular}{llccccll}

\hline\hline

Galaxy & & R.A.(J2000) & Dec.(J2000) & v$_r$ & i &
\multicolumn{2}{c}{Morphological Type} \\
KIG name & UGC/NGC name & (hh mm ss.ss) & (+dd mm ss.s) & (km
s$^{-1}$) & ($^{o}$) & old$^{i}$ & revised \\

\hline
KIG 580     &                               &     13 19 01.90   &     +14 47 28.0   &   6643    &   57  &    Sbc    &     SA(s)c    \\
KIG 598     &      UGC 08705                &     13 46 32.29   &     +20 50 51.3   &   6938    &   58  &    Sc     &     SAB(s)bc      \\
KIG 600     &                               &     13 49 28.89   &     +13 52 36.5   &   7228    &   41  &    Sc     &     (R$_{2}$$^{\prime}$)SA(rs)c   \\
KIG 612     &      UGC 09035                &     14 07 55.41   &     +29 52 22.2   &   8244    &   28  &    Sbc    &     SB(rs)b   \\
KIG 626     &      UGC 09201 / NGC 5584     &     14 22 23.67   &     -00 23 14.1   &   1640    &   45  &    Sc     &     SAB(s)c   \\
KIG 630     &      UGC 09248 / NGC 5622     &     14 26 12.18   &     +48 33 50.4   &   3861    &   58  &    Sb     &     SA(s)b    \\
KIG 633     &                               &     14 32 27.42   &     +27 25 38.3   &   4298    &   29  &    Sbc    &     SA(s)bc   \\
KIG 639     &                               &     14 37 49.61   &     +06 44 54.1   &   8659    &   52  &    Sc     &     SA(s)c    \\
KIG 640     &                               &     14 38 38.22   &     +54 16 40.4   &   8790    &   24  &    Sbc    &     SA(s)bc   \\
KIG 641     &      UGC 09461                &     14 39 33.01   &     +62 00 10.5   &   6728    &   45  &    Sb     &     SB(r)b    \\
KIG 645     &      UGC 09516                &     14 45 48.85   &     +50 23 38.5   &   4027    &   35  &    Sc     &     (R$_{2}$)SA(s)c   \\
KIG 652     &      UGC 09564 / NGC 5768     &     14 52 08.05   &     -02 31 47.9   &   1962    &   27  &    Sc     &     SAB(s)bc      \\
KIG 665     &                               &     15 12 24.98   &     +18 38 47.7   &   6408    &   54  &    Sb     &     SA(s)b    \\
KIG 671     &      UGC 09826                &     15 21 33.05   &     +39 12 04.3   &   8822    &   11  &    Sb     &     SAB(rs)b      \\
KIG 689     &                               &     15 36 36.23   &     +17 20 17.5   &   4292    &   58  &    Sbc    &     SB(s)c    \\
KIG 712     &      UGC 10083 / NGC 6012     &     15 54 13.74   &     +14 36 06.9   &   1854    &   55  &    Sbc    &     (R$_{2}$)SA(r)b   \\
KIG 716     &      UGC 10104                &     15 57 27.86   &     +30 03 34.6   &   9841    &   13  &    Sc     &     SA(rs)bc      \\
KIG 719     &      UGC 10120                &     15 59 09.56   &     +35 01 47.2   &   9438    &   22  &    Sb     &     (R$_{1}$$^{\prime}$)SB(r)b    \\
KIG 731     &                               &     16 17 39.45   &     +10 21 45.7   &   9817    &   46  &    Sb     &     SAB(rs)bc     \\
KIG 743     &      UGC 10435                &     16 31 21.62   &     +22 41 49.3   &   7297    &   43  &    Sb     &     SB(rs)b   \\
KIG 754     &      UGC 10490                &     16 38 49.56   &     +17 21 11.6   &   4594    &   45  &    Sc     &     SA(rs)b   \\
KIG 757     &                               &     16 39 30.75   &     +21 19 02.2   &   9338    &   48  &    Sbc    &     SAB(s)bc      \\
KIG 795     &      UGC 10774                &     17 14 08.93   &     +58 49 06.3   &   8873    &   53  &    Sc     &     SAB(rs)bc     \\
KIG 805     &      UGC 10829                &     17 23 47.31   &     +26 29 11.6   &   4730    &   51  &    Sbc    &     SA(rs)bc      \\
KIG 807     &                               &     17 23 09.59   &     +63 54 28.4   &   8228    &   57  &    Sbc    &     SA(s)bc   \\
KIG 839     &                               &     17 56 03.62   &     +49 01 41.7   &   9458    &   45  &    Sbc    &     SAB(s)c   \\
KIG 892     &                               &     20 52 22.38   &     +00 04 32.3   &   9087    &   36  &    Sc     &     SA(rs)bc      \\
KIG 907     &                               &     21 20 21.01   &     +10 19 13.6   &   5257    &   55  &    Sbc    &     SA(s)bc   \\
KIG 912     &                               &     21 23 22.14   &     +10 07 59.9   &   5122    &   53  &    Sb     &     SA(rs)c   \\
KIG 924     &      UGC 11790                &     21 41 29.92   &     +00 53 40.8   &   4540    &   39  &    Sc     &     SA(s)bc   \\
KIG 928     &                               &     21 45 54.72   &     +11 40 41.5   &   6985    &   19  &    Sc     &     SA(s)bc   \\
KIG 931     &      UGC 11816                &     21 49 07.30   &     +00 26 50.5   &   4750    &   0   &    Sbc    &     (R$_{2}$)SA(s)bc      \\
KIG 932     &      UGC 11817 / NGC 7138     &     21 49 01.10   &     +12 30 51.9   &   8406    &   59  &    Sbc    &     SB(r)b    \\
KIG 943     &                               &     22 04 12.67   &     -00 01 52.5   &   9778    &   46  &    Sb     &     (R$_{1}$$^{\prime}$)SB(rs)b   \\
\hline
KIG 250     &      UGC 04393                &     08 26 04.51   &     +45 58 06.0   &   2125    &   52  &    Sc     &     SB(s)dm   \\
KIG 291     &      UGC 04684                &     08 56 40.68   &     +00 22 29.6   &   2521    &   23  &    Sc     &     SAB(rs)d      \\
KIG 308     &                               &     09 09 34.93   &     +18 36 56.9   &   8487    &   12  &    Sc     &     SA(s)ab   \\
KIG 392     &                               &     10 03 23.21   &     +48 21 56.6   &   7413    &   40  &    Sbc    &     SAB(r)ab      \\
\hline \\[1ex]   

\end{tabular}

\begin{minipage}{13cm}

Col.(1): KIG Name. Col.(2): UGC/NGC Name. Col.(3): Right Ascension
(J2000). Col.(4): Declination(J2000). Col.(5): recession velocity
(km s$^{-1}$). Col.(6): Inclination in degrees. Col.(7):
Morphological Type -- \citet{sulentic06}. Col.(8): Morphological
Type - revised classification. The last four galaxies in this table
(between horizontal lines) were excluded from our sample, because
according to our revised classification they are not in the range
Sb-Sc.

References.-- (i) \citet{sulentic06}

\end{minipage}

\end{table}

\clearpage
\begin{table}
\setcounter{table}{1} \centering \caption{Photometric Measures based
on SDSS Photo-pipeline}
\begin{tabular}{lcccc}

\hline\hline

Galaxy & (g-i)$_{o}$ & M$_{i}$ & a$_{25}^{i}$ (kpc) & a$_{25}^{g}$
(kpc) \\

\hline

KIG 11	&	0.61	&	-20.80	&	14.7	&	15.6	\\
KIG 33	&	0.71	&	-21.50	&	13.8	&	13.1	\\
KIG 56	&	0.99	&	-21.53	&	17.3	&	14.7	\\
KIG 187	&	1.00	&	-22.18	&	19.5	&	17.3	\\
KIG 198	&	1.15	&	-22.09	&	19.2	&	15.6	\\
KIG 203	&	0.70	&	-21.09	&	13.9	&	12.0	\\
KIG 217	&	0.85	&	-20.61	&	10.0	&	9.3	\\
KIG 222	&	1.01	&	-21.40	&	13.3	&	11.4	\\
KIG 232	&	1.09	&	-21.98	&	19.6	&	17.7	\\
KIG 238	&	0.93	&	-21.30	&	17.2	&	13.0	\\
KIG 241	&	0.52	&	-20.16	&	7.3	&	6.8	\\
KIG 242	&	0.41	&	-19.36	&	5.9	&	5.7	\\
KIG 258	&	0.76	&	-20.53	&	9.8	&	8.2	\\
KIG 260	&	0.77	&	-21.57	&	18.7	&	17.7	\\
KIG 271	&	0.79	&	-22.16	&	20.5	&	18.3	\\
KIG 281	&	0.91	&	-21.32	&	15.5	&	14.3	\\
KIG 282	&	0.93	&	-21.20	&	11.2	&	8.9	\\
KIG 287	&	0.90	&	-21.79	&	16.8	&	16.0	\\
KIG 292	&	0.72	&	-20.79	&	10.9	&	10.7	\\
KIG 298	&	1.14	&	-22.55	&	18.4	&	17.9	\\
KIG 302	&	0.87	&	-22.34	&	17.4	&	14.3	\\
KIG 314	&	0.68	&	-21.60	&	16.3	&	14.5	\\
KIG 325	&	0.82	&	-21.67	&	13.9	&	13.4	\\
KIG 328	&	0.83	&	-21.53	&	13.6	&	12.2	\\
KIG 330	&	0.66	&	-19.92	&	7.0	&	6.3	\\
KIG 336	&	1.07	&	-22.21	&	18.9	&	17.4	\\
KIG 339	&	0.94	&	-21.78	&	18.8	&	21.6	\\
KIG 351	&	0.92	&	-20.74	&	12.7	&	11.5	\\
KIG 365	&	0.73	&	-21.44	&	13.0	&	11.8	\\
KIG 366	&	1.11	&	-21.88	&	16.2	&	14.4	\\
KIG 367	&	0.69	&	-20.89	&	13.6	&	13.3	\\
KIG 368	&	0.69	&	-21.90	&	16.0	&	15.2	\\
KIG 386	&	0.82	&	-21.38	&	11.5	&	10.1	\\
KIG 397	&	0.82	&	-20.53	&	8.5	&	5.9	\\
KIG 399	&	0.77	&	-21.35	&	13.5	&	11.8	\\
KIG 401	&	0.87	&	-21.57	&	15.4	&	14.3	\\
KIG 405	&	0.73	&	-20.62	&	11.9	&	10.7	\\
KIG 406	&	0.94	&	-21.61	&	15.4	&	11.6	\\
KIG 409	&	0.52	&	-20.18	&	7.6	&	7.1	\\
KIG 410	&	0.52	&	-21.01	&	10.1	&	9.2	\\
KIG 429	&	0.88	&	-21.68	&	15.1	&	13.7	\\
KIG 444	&	0.82	&	-21.29	&	8.5	&	7.7	\\
KIG 446	&	0.80	&	-21.97	&	15.1	&	13.6	\\
KIG 460	&	0.37	&	-20.12	&	7.9	&	8.5	\\
KIG 466	&	0.50	&	-19.37	&	6.4	&	6.6	\\
KIG 489	&	0.58	&	-21.07	&	10.8	&	10.4	\\
KIG 491	&	0.85	&	-21.72	&	13.9	&	11.5	\\
KIG 494	&	0.55	&	-20.74	&	10.3	&	10.4	\\
KIG 499	&	0.94	&	-22.40	&	21.9	&	20.1	\\
KIG 502	&	0.18	&	-19.24	&	11.2	&	11.0	\\
KIG 508	&	0.56	&	-20.84	&	9.4	&	9.4	\\
KIG 512	&	0.65	&	-19.70	&	9.6	&	9.0	\\
KIG 515	&	0.68	&	-21.39	&	12.3	&	11.6	\\
KIG 520	&	1.02	&	-22.20	&	17.7	&	16.7	\\
KIG 522	&	0.56	&	-20.73	&	7.6	&	6.1	\\
KIG 525	&	1.06	&	-22.01	&	20.8	&	18.0	\\
KIG 532	&	0.59	&	-20.66	&	9.6	&	9.3	\\
KIG 550	&	1.13	&	-22.97	&	26.2	&	23.6	\\
KIG 553	&	0.88	&	-22.07	&	18.8	&	17.0	\\
KIG 560	&	0.30	&	-20.08	&	7.0	&	6.6	\\
KIG 571	&	0.65	&	-19.70	&	6.5	&	5.7	\\
KIG 575	&	0.78	&	-20.77	&	9.7	&	9.0	\\

\end{tabular}

\end{table}

\begin{table}
\textbf{Table 2.}--continued

\centering
\begin{tabular}{lcccc}

\hline\hline

Galaxy & (g-i)$_{o}$ & M$_{i}$ & a$_{25}^{i}$ (kpc) & a$_{25}^{g}$
(kpc) \\

\hline

KIG 580	&	0.63	&	-20.93	&	10.9	&	9.5	\\
KIG 598	&	0.72	&	-22.18	&	13.5	&	12.7	\\
KIG 600	&	0.57	&	-19.94	&	8.4	&	8.3	\\
KIG 612	&	1.05	&	-21.64	&	16.8	&	15.2	\\
KIG 626	&	0.38	&	-20.53	&	11.9	&	11.3	\\
KIG 630	&	0.76	&	-21.04	&	12.1	&	11.7	\\
KIG 633	&	0.49	&	-20.02	&	5.1	&	4.5	\\
KIG 639	&	0.63	&	-21.07	&	12.2	&	11.0	\\
KIG 640	&	0.68	&	-21.13	&	9.2	&	7.1	\\
KIG 641	&	1.03	&	-21.78	&	14.1	&	12.4	\\
KIG 645	&	0.76	&	-20.44	&	8.9	&	8.5	\\
KIG 652	&	0.67	&	-20.35	&	8.0	&	7.5	\\
KIG 665	&	0.71	&	-21.03	&	11.9	&	11.0	\\
KIG 671	&	0.92	&	-21.41	&	16.4	&	14.6	\\
KIG 689	&	0.39	&	-19.90	&	8.5	&	7.1	\\
KIG 712	&	0.58	&	-21.04	&	14.3	&	11.7	\\
KIG 716	&	0.74	&	-23.00	&	44.1	&	31.7	\\
KIG 719	&	0.79	&	-21.91	&	15.3	&	14.9	\\
KIG 731	&	0.76	&	-21.41	&	13.2	&	12.4	\\
KIG 743	&	0.94	&	-21.59	&	12.7	&	11.8	\\
KIG 754	&	0.59	&	-19.98	&	9.6	&	8.5	\\
KIG 757	&	0.69	&	-22.16	&	16.1	&	15.4	\\
KIG 795	&	0.90	&	-21.85	&	15.7	&	15.1	\\
KIG 805	&	0.83	&	-21.33	&	13.7	&	13.2	\\
KIG 807	&	0.76	&	-21.54	&	12.3	&	10.5	\\
KIG 839	&	0.75	&	-21.50	&	11.8	&	10.8	\\
KIG 892	&	0.99	&	-22.59	&	17.1	&	14.7	\\
KIG 907	&	0.40	&	-19.66	&	7.7	&	7.3	\\
KIG 912	&	0.65	&	-20.40	&	8.2	&	8.0	\\
KIG 924	&	0.76	&	-21.03	&	14.6	&	13.0	\\
KIG 928	&	0.66	&	-20.30	&	6.6	&	5.5	\\
KIG 931	&	0.88	&	-20.67	&	10.2	&	8.8	\\
KIG 932	&	1.00	&	-22.81	&	22.0	&	19.4	\\
KIG 943	&	0.70	&	-21.81	&	14.2	&	12.3	\\

\hline
\end{tabular}

\begin{minipage}{6.5cm}

Col.(1): Galaxy Name. Col.(2): (g-i) color corrected for Galactic,
internal extinction as well as K-corrected. Col.(3): Absolute
magnitude in i-band. Col.(4): Semimajor axis of $\mu_{i}$=25 mag
arcsec$^{-2}$ isophote. Col.(5): Semimajor axis of $\mu_{g}$=25 mag
arcsec$^{-2}$ isophote.
\end{minipage}

\end{table}

\clearpage
\begin{table}

\setcounter{table}{2}

\caption{Mean/Median for Some Photometric Measures}

\centering
\begin{tabular}{l cccc}
\hline\hline
Type & M$_{i}$ & (g-i)$_{o}$ & a$_{25}^{i}$ (kpc) & a$_{25}^{g}$ (kpc) \\
 & mean$\pm$SE median & mean$\pm$SE median & mean$\pm$SE median & mean$\pm$SE median \\

\hline
Sb    & -21.52$\pm$0.14~~-21.49 & 0.89$\pm$0.03~~0.92 & 15.5$\pm$0.9~~14.8 & 13.8$\pm$0.8~~12.7 \\
Sbc   & -21.25$\pm$0.15~~-21.38 & 0.76$\pm$0.03~~0.80 & 12.3$\pm$0.7~~13.5 & 11.2$\pm$0.7~~11.8 \\
Sc    & -20.92$\pm$0.13~~-20.91 & 0.67$\pm$0.03~~0.68 & 12.2$\pm$0.7~~11.8 & 11.1$\pm$0.6~~10.7 \\
Sb-Sc & -21.20$\pm$0.08~~-21.33 & 0.76$\pm$0.02~~0.76 & 13.1$\pm$0.4~~13.2 & 11.9$\pm$0.4~~11.7 \\
\hline
\end{tabular}

\begin{minipage}{13.5cm}

Note: \emph{SE} is standard deviation of the mean.

Col.(1): Morphological Type. Col.(2): Absolute Magnitude in i-band.
Col.(3): (g-i) color corrected for Galactic, internal extinction as
well as K-corrected. Col.(4): Semimajor axis
of $\mu_{i}$=25 mag arcsec$^{-2}$ isophote. Col.(5): Semimajor axis
of $\mu_{g}$=25 mag arcsec$^{-2}$ isophote.

\end{minipage}

\end{table}

\clearpage
\begin{table}

\begin{minipage}{155mm}
\caption{Structural Parameters Obtained with BUDDA for CIG/KIG Galaxies in our Sample in i-band} 
\end{minipage}

\centering
\begin{tabular}{lcccccccrccc}
\hline\hline

(1) & (2) & (3) & (4) & (5) & (6) & (7) & (8) & (9) & (10) & (11) & (12) \\
Galaxy & $\frac{Bulge}{Total}$ & $\frac{Disk}{Total}$ &
$\frac{Bar}{Total}$  & $\frac{Bulge}{Disk}$ & r$_{e}$ & $\mu_{e}$ &
n$_{bulge}$ & h$_{R}$ & $\mu_{o}$ & l$_{bar}$ & n$_{bar}$ \\
 &  &  &  &  &
(\arcsec) & ($\frac{mag}{arcsec^{2}}$) &  & (\arcsec) &
($\frac{mag}{arcsec^{2}}$) & (\arcsec) &  \\

\hline
KIG 11  &   0.294   &   0.706   &           &   0.416   &   9.66    &   21.05   &   1.14    &   142.12  &   23.15   &           &       \\
KIG 33  &   0.349   &   0.651   &           &   0.536   &   5.71    &   20.04   &   1.17    &   16.08   &   20.33   &           &       \\
KIG 56  &   0.160   &   0.714   &   0.126   &   0.224   &   1.69    &   18.32   &   1.75    &   16.68   &   20.62   &   19.01   &   0.60    \\
KIG 187 &   0.072   &   0.865   &   0.063   &   0.083   &   1.66    &   20.22   &   2.33    &   13.54   &   20.80   &   9.11    &   0.68    \\
KIG 198 &   0.038   &   0.671   &   0.292   &   0.056   &   1.31    &   19.75   &   0.77    &   37.94   &   22.84   &   15.92   &   0.65    \\
KIG 203 &   0.012   &   0.988   &           &   0.013   &   0.72    &   21.00   &   1.00    &   6.94    &   19.88   &           &       \\
KIG 217 &   0.040   &   0.960   &           &   0.042   &   2.09    &   20.54   &   2.55    &   15.77   &   20.36   &           &       \\
KIG 222 &   0.038   &   0.927   &   0.035   &   0.041   &   1.51    &   19.91   &   0.76    &   11.81   &   20.31   &   13.46   &   0.57    \\
KIG 232 &   0.096   &   0.904   &           &   0.106   &   2.53    &   19.33   &   2.76    &   15.81   &   19.89   &           &       \\
KIG 238 &   0.150   &   0.688   &   0.163   &   0.217   &   1.48    &   19.44   &   1.97    &   19.87   &   22.19   &   11.88   &   0.59    \\
KIG 241 &   0.010   &   0.990   &           &   0.010   &   1.04    &   20.92   &   0.37    &   5.18    &   19.66   &           &       \\
KIG 242 &   0.165   &   0.835   &           &   0.197   &   1.97    &   20.26   &   0.70    &   4.13    &   19.72   &           &       \\
KIG 258 &   0.204   &   0.766   &   0.030   &   0.267   &   1.32    &   19.66   &   2.90    &   8.26    &   20.55   &   8.71    &   1.05    \\
KIG 260 &   0.020   &   0.949   &   0.031   &   0.021   &   1.84    &   21.08   &   1.93    &   14.92   &   20.77   &   9.90    &   0.60    \\
KIG 271 &   0.020   &   0.964   &   0.016   &   0.020   &   1.36    &   20.85   &   2.09    &   11.45   &   20.01   &   7.92    &   0.62    \\
KIG 281 &   0.024   &   0.976   &           &   0.024   &   1.80    &   20.04   &   1.07    &   15.86   &   20.04   &           &       \\
KIG 282 &   0.047   &   0.907   &   0.046   &   0.051   &   1.32    &   20.48   &   1.55    &   4.42    &   19.48   &   8.51    &   0.76    \\
KIG 287 &   0.153   &   0.804   &   0.044   &   0.190   &   2.33    &   20.34   &   1.66    &   7.80    &   19.99   &   7.79    &   0.78    \\
KIG 292 &   0.173   &   0.827   &           &   0.209   &   4.91    &   19.73   &   3.24    &   21.41   &   19.66   &           &       \\
KIG 298 &   0.123   &   0.805   &   0.071   &   0.153   &   2.04    &   18.81   &   1.78    &   19.38   &   20.52   &   19.80   &   0.70    \\
KIG 302 &   0.103   &   0.897   &           &   0.115   &   1.93    &   19.78   &   1.03    &   6.33    &   19.05   &           &       \\
KIG 314 &   0.075   &   0.925   &           &   0.081   &   3.31    &   19.46   &   1.64    &   17.79   &   19.52   &           &       \\
KIG 325 &   0.083   &   0.917   &           &   0.091   &   1.78    &   20.07   &   1.10    &   8.69    &   20.15   &           &       \\
KIG 328 &   0.045   &   0.950   &   0.005   &   0.047   &   1.21    &   19.42   &   0.62    &   7.27    &   19.34   &   12.28   &   0.53    \\
KIG 330 &   0.026   &   0.974   &           &   0.026   &   1.37    &   20.35   &   0.43    &   7.12    &   19.93   &           &       \\
KIG 336 &   0.213   &   0.662   &   0.125   &   0.322   &   2.87    &   19.47   &   2.09    &   26.79   &   21.79   &   22.18   &   0.50    \\
KIG 339 &   0.299   &   0.564   &   0.137   &   0.531   &   2.37    &   18.92   &   2.60    &   54.07   &   22.91   &   21.78   &   0.48    \\
KIG 351 &   0.031   &   0.867   &   0.102   &   0.036   &   1.16    &   20.11   &   0.34    &   11.83   &   20.97   &   11.09   &   0.53    \\
KIG 365 &   0.051   &   0.898   &   0.051   &   0.057   &   1.11    &   19.52   &   0.90    &   7.81    &   20.37   &   10.30   &   0.76    \\
KIG 366 &   0.081   &   0.779   &   0.140   &   0.104   &   1.50    &   19.44   &   1.90    &   9.53    &   19.70   &   16.24   &   0.82    \\
KIG 367 &   0.216   &   0.784   &           &   0.276   &   7.08    &   22.04   &   1.55    &   58.30   &   23.19   &           &       \\
KIG 368 &   0.061   &   0.939   &           &   0.065   &   1.24    &   19.91   &   1.10    &   7.42    &   20.11   &           &       \\
KIG 386 &   0.025   &   0.962   &   0.014   &   0.026   &   0.84    &   19.75   &   0.66    &   5.73    &   19.24   &   6.53    &   0.47    \\
KIG 397 &   0.060   &   0.930   &   0.010   &   0.064   &   1.92    &   19.61   &   1.56    &   7.65    &   18.76   &   5.94    &   0.60    \\
KIG 399 &   0.042   &   0.958   &           &   0.044   &   1.50    &   20.14   &   1.46    &   9.25    &   19.83   &           &     \\
KIG 401 &   0.007   &   0.969   &   0.024   &   0.008   &   0.83    &   20.59   &   0.60    &   9.69    &   19.77   &   4.99    &   0.70    \\
KIG 405 &   0.056   &   0.944   &           &   0.059   &   1.99    &   20.79   &   1.13    &   19.29   &   21.74   &           &       \\
KIG 406 &   0.209   &   0.791   &           &   0.263   &   3.41    &   21.25   &   1.08    &   7.53    &   20.47   &           &       \\
KIG 409 &   0.004   &   0.962   &   0.034   &   0.004   &   1.00    &   21.85   &   0.40    &   4.54    &   19.62   &   4.75    &   0.62    \\
KIG 410 &   0.065   &   0.935   &           &   0.070   &   1.22    &   19.36   &   0.63    &   6.06    &   19.63   &           &        \\
KIG 429 &   0.014   &   0.982   &   0.004   &   0.014   &   1.07    &   21.14   &   0.73    &   7.81    &   20.04   &   4.95    &   0.58  \\
KIG 444 &   0.024   &   0.961   &   0.015   &   0.025   &   1.57    &   19.82   &   1.23    &   6.23    &   18.52   &   10.10   &   0.65    \\
KIG 446 &   0.064   &   0.936   &           &   0.069   &   1.16    &   19.64   &   0.84    &   5.10    &   18.83   &           &       \\
KIG 460 &   0.053   &   0.947   &           &   0.056   &   1.83    &   20.83   &   0.54    &   5.00    &   19.48   &           &       \\
KIG 466 &   0.243   &   0.726   &   0.032   &   0.334   &   8.54    &   21.64   &   0.66    &   29.30   &   22.07   &   14.14   &   0.77    \\
KIG 489 &   0.017   &   0.983   &           &   0.017   &   1.19    &   20.20   &   0.48    &   7.25    &   19.35   &           &       \\
KIG 491 &   0.076   &   0.924   &           &   0.083   &   1.60    &   19.76   &   1.66    &   7.45    &   19.28   &           &       \\
KIG 494 &   0.040   &   0.916   &   0.044   &   0.044   &   2.07    &   21.35   &   0.69    &   7.28    &   20.25   &   6.66    &   0.50    \\
KIG 499 &   0.045   &   0.840   &   0.115   &   0.054   &   0.99    &   19.61   &   2.10    &   12.81   &   20.23   &   7.07    &   0.69    \\
KIG 502 &   0.085   &   0.915   &           &   0.093   &   8.98    &   21.87   &   1.45    &   337.49  &   23.34   &           &       \\
KIG 508 &   0.031   &   0.912   &   0.057   &   0.034   &   1.19    &   19.70   &   1.41    &   11.00   &   20.61   &   5.94    &   0.70    \\
KIG 512 &   0.016   &   0.951   &   0.033   &   0.017   &   3.93    &   21.45   &   1.50    &   38.30   &   21.74   &   19.80   &   0.62    \\
KIG 515 &   0.044   &   0.952   &   0.003   &   0.046   &   2.43    &   20.96   &   0.88    &   7.13    &   19.42   &   5.15    &   1.37    \\
KIG 520 &   0.094   &   0.866   &   0.040   &   0.109   &   1.94    &   19.87   &   1.97    &   8.38    &   19.60   &   8.17    &   0.71    \\
KIG 522 &   0.123   &   0.742   &   0.135   &   0.166   &   0.77    &   18.76   &   2.50    &   4.02    &   19.16   &   7.92    &   0.50    \\
KIG 525 &   0.150   &   0.795   &   0.055   &   0.189   &   2.04    &   19.08   &   2.69    &   14.67   &   20.50   &   13.07   &   0.91    \\
KIG 532 &   0.040   &   0.844   &   0.116   &   0.048   &   1.54    &   19.80   &   0.51    &   5.89    &   20.02   &   6.34    &   0.85    \\
KIG 550 &   0.066   &   0.879   &   0.055   &   0.075   &   1.96    &   18.98   &   1.32    &   19.67   &   20.43   &   15.84   &   0.63    \\
KIG 553 &   0.331   &   0.557   &   0.112   &   0.593   &   2.82    &   18.76   &   3.38    &   68.00   &   22.99   &   24.95   &   0.45    \\
KIG 560 &   0.143   &   0.857   &           &   0.166   &   1.76    &   20.24   &   0.65    &   4.04    &   19.79   &           &       \\
KIG 571 &   0.031   &   0.969   &           &   0.032   &   2.53    &   21.50   &   1.22    &   9.06    &   19.41   &           &       \\

\end{tabular}

\end{table}

\clearpage
\begin{table}
\textbf{Table 4.}--continued


\centering
\begin{tabular}{lcccccccrccc}
\hline\hline

(1) & (2) & (3) & (4) & (5) & (6) & (7) & (8) & (9) & (10) & (11) & (12) \\
Galaxy & $\frac{Bulge}{Total}$ & $\frac{Disk}{Total}$ &
$\frac{Bar}{Total}$  & $\frac{Bulge}{Disk}$ & r$_{e}$ & $\mu_{e}$ &
n$_{bulge}$ & h$_{R}$ & $\mu_{o}$ & l$_{bar}$ & n$_{bar}$ \\
 &  &  &  &  &
(\arcsec) & ($\frac{mag}{arcsec^{2}}$) &  & (\arcsec) &
($\frac{mag}{arcsec^{2}}$) & (\arcsec) &  \\

\hline
KIG 575 &   0.038   &   0.962   &           &   0.040   &   2.57    &   20.42   &   1.50    &   11.73   &   19.08   &           &       \\
KIG 580 &   0.052   &   0.948   &           &   0.055   &   1.13    &   20.36   &   1.45    &   5.71    &   19.42   &           &       \\
KIG 598 &   0.052   &   0.937   &   0.011   &   0.055   &   1.36    &   19.50   &   1.96    &   10.33   &   19.48   &   11.88   &   0.74    \\
KIG 612 &   0.153   &   0.768   &   0.079   &   0.199   &   1.63    &   19.35   &   2.22    &   15.76   &   21.32   &   13.88   &   0.44    \\
KIG 626 &   0.004   &   0.982   &   0.014   &   0.004   &   1.47    &   20.71   &   1.60    &   34.42   &   20.64   &   9.90    &   0.67    \\
KIG 630 &   0.061   &   0.939   &           &   0.065   &   2.56    &   20.24   &   0.86    &   12.71   &   19.69   &           &       \\
KIG 633 &   0.052   &   0.948   &           &   0.055   &   1.03    &   19.31   &   0.39    &   3.41    &   18.48   &           &       \\
KIG 639 &   0.033   &   0.967   &           &   0.034   &   0.80    &   20.27   &   1.02    &   5.25    &   19.71   &           &       \\
KIG 640 &   0.099   &   0.901   &           &   0.110   &   0.93    &   19.24   &   0.97    &   2.14    &   18.05   &           &       \\
KIG 641 &   0.141   &   0.801   &   0.058   &   0.176   &   1.50    &   18.76   &   1.74    &   10.33   &   19.97   &   11.88   &   0.72    \\
KIG 645 &   0.111   &   0.889   &           &   0.124   &   3.23    &   20.81   &   1.49    &   14.35   &   20.71   &           &       \\
KIG 652 &   0.029   &   0.940   &   0.031   &   0.031   &   1.17    &   19.03   &   1.84    &   12.09   &   19.39   &   8.32    &   1.19    \\
KIG 665 &   0.086   &   0.914   &           &   0.094   &   1.90    &   20.56   &   0.98    &   6.73    &   19.70   &           &       \\
KIG 671 &   0.110   &   0.648   &   0.218   &   0.169   &   1.27    &   18.65   &   0.76    &   34.38   &   22.74   &   14.54   &   0.50    \\
KIG 689 &   0.067   &   0.933   &           &   0.072   &   4.04    &   21.64   &   1.02    &   8.51    &   20.22   &           &       \\
KIG 712 &   0.059   &   0.941   &           &   0.062   &   5.06    &   20.72   &   1.05    &   22.06   &   19.65   &           &       \\
KIG 716 &   0.175   &   0.825   &           &   0.212   &   3.51    &   20.28   &   3.16    &   39.34   &   22.28   &           &       \\
KIG 719 &   0.121   &   0.679   &   0.153   &   0.186   &   1.27    &   18.84   &   1.28    &   14.88   &   21.42   &   14.26   &   0.63    \\
KIG 731 &   0.040   &   0.885   &   0.075   &   0.045   &   0.81    &   19.91   &   0.94    &   5.96    &   20.16   &   7.92    &   0.31    \\
KIG 743 &   0.056   &   0.865   &   0.079   &   0.064   &   1.32    &   20.23   &   2.11    &   14.10   &   21.10   &   13.46   &   0.30    \\
KIG 757 &   0.074   &   0.847   &   0.079   &   0.088   &   1.16    &   19.48   &   2.24    &   7.29    &   19.61   &   11.29   &   0.79    \\
KIG 795 &   0.046   &   0.954   &           &   0.048   &   1.45    &   20.61   &   1.53    &   9.12    &   19.99   &           &       \\
KIG 805 &   0.041   &   0.959   &           &   0.042   &   1.39    &   19.99   &   2.09    &   11.90   &   19.77   &           &       \\
KIG 807 &   0.022   &   0.978   &           &   0.023   &   1.02    &   20.03   &   0.70    &   5.07    &   19.01   &           &       \\
KIG 839 &   0.011   &   0.979   &   0.010   &   0.012   &   0.57    &   20.63   &   1.00    &   4.74    &   19.70   &   3.96    &   0.60    \\
KIG 892 &   0.161   &   0.839   &           &   0.192   &   2.39    &   19.72   &   1.53    &   8.23    &   19.75   &           &       \\
KIG 907 &   0.303   &   0.697   &           &   0.435   &   4.93    &   21.49   &   0.87    &   71.95   &   23.46   &           &       \\
KIG 912 &   0.039   &   0.961   &           &   0.041   &   1.26    &   20.89   &   1.39    &   5.18    &   19.22   &           &       \\
KIG 924 &   0.018   &   0.982   &           &   0.018   &   1.63    &   20.63   &   0.60    &   17.81   &   20.95   &           &       \\
KIG 928 &   0.289   &   0.711   &           &   0.406   &   1.63    &   19.93   &   1.17    &   2.45    &   19.14   &           &       \\
KIG 931 &   0.120   &   0.880   &           &   0.137   &   5.12    &   21.33   &   1.75    &   16.36   &   21.29   &           &       \\
KIG 932 &   0.075   &   0.886   &   0.039   &   0.084   &   1.87    &   19.83   &   2.05    &   8.93    &   19.13   &   10.27   &   0.78    \\
KIG 943 &   0.102   &   0.475   &   0.423   &   0.215   &   0.64    &   18.46   &   2.00    &   9.43    &   21.22   &   7.52    &   1.25    \\

\hline\\[1ex]
\end{tabular}

\begin{minipage}{14.5cm}

Col.(1): Galaxy Name. Col.(2): Bulge/Total luminosity ratio.
Col.(3): Disk/Total luminosity ratio. Col.(4): Bar/Total luminosity
ratio. Col.(5): Bulge/Disk luminosity ratio. Col.(6): effective
radius of the bulge in arcsec. Col.(7): effective surface brightness
of the bulge in mag arcsec$^{-2}$. Col.(8): S\'{e}rsic index of the
bulge. Col.(9): disk scalelength in arcsec. Col.(10): central
surface brightness of the disk in mag arcsec$^{-2}$. Col.(11): bar
length, i.e. semimajor axis of the bar in arcsec. Col.(12):
S\'{e}rsic index of the bar.
\end{minipage}

\end{table}

\clearpage
\begin{table}

\setcounter{table}{4}
\begin{minipage}{155mm}
\caption{Mean/Median for Structural Parameters of Bulges, Disks and
Bars of \textbf{All} Galaxies}
\end{minipage}
\begin{tabular}{lcccccc}
\hline\hline

Type (N) & Bulge/Total & n$_{bulge}$ & r$_{e}$ (kpc) & $\mu_{e}$ & h$_{R}$ (kpc) & $\mu_{o}$\\
     & mean$\pm$SE median & mean$\pm$SE median & mean$\pm$SE median & mean$\pm$SE median & mean$\pm$SE median & mean$\pm$SE median \\
\hline
Sb (25)    & 0.12$\pm$0.01~~0.11 & 1.79$\pm$0.17~~1.90 & 0.73$\pm$0.05~~0.64 & 19.24$\pm$0.13~~19.30 & 7.40$\pm$1.27~~5.56 & 20.47$\pm$0.21~~20.34\\
Sbc (34)   & 0.09$\pm$0.01~~0.06 & 1.35$\pm$0.13~~1.32 & 0.74$\pm$0.08~~0.60 & 19.89$\pm$0.12~~19.85 & 5.23$\pm$1.11~~3.65 & 19.77$\pm$0.19~~19.52\\
Sc (35)    & 0.08$\pm$0.01~~0.04 & 1.18$\pm$0.08~~1.13 & 0.79$\pm$0.10~~0.65 & 20.48$\pm$0.12~~20.60 & 6.74$\pm$1.58~~3.59 & 20.21$\pm$0.21~~19.84\\
Sb-Sc (94) & 0.09$\pm$0.01~~0.06 & 1.40$\pm$0.07~~1.30 & 0.76$\pm$0.05~~0.64 & 19.94$\pm$0.09~~19.90 & 6.37$\pm$0.79~~4.07 & 20.11$\pm$0.12~~19.80\\
\hline
\end{tabular}

\begin{minipage}{155mm}

Col.(1): Galaxy Name. Col.(2): Bulge/Total luminosity ratio.
Col.(3): S\'{e}rsic index of the bulge. Col.(4): effective radius of
the bulge in kpc. Col.(5): effective surface brightness of the bulge
in mag arcsec$^{-2}$. Col.(6): disk scalelength in kpc. Col.(7):
central surface brightness of the disk in mag arcsec$^{-2}$.

Note: N=number of galaxies; \emph{SE} is standard deviation of the
mean.
\end{minipage}

\end{table}

\begin{table}
\begin{minipage}{155mm}
\textbf{Table 6a.} Mean/Median for Structural Parameters of Bulges,
Disks and Bars of \textbf{Barred} Galaxies
\end{minipage}
\centering
\begin{tabular}{lccccccc}
\hline\hline
Type (N) & Bulge/Total & n$_{bulge}$ & r$_{e}$ (kpc) & $\mu_{e}$ & h$_{R}$ (kpc) & $\mu_{o}$ & l$_{bar}$ (kpc)\\
 & mean$\pm$SE median & mean$\pm$SE median & mean$\pm$SE median & mean$\pm$SE median & mean$\pm$SE median & mean$\pm$SE median & mean$\pm$SE median\\

\hline
Sb (21) & 0.12$\pm$0.02~~0.12 & 1.84$\pm$0.16~~1.97 & 0.74$\pm$0.06~~0.65 & 19.06$\pm$0.12~~19.19 & 8.29$\pm$1.44~~6.55 & 20.64$\pm$0.23~~20.38 & 6.48$\pm$0.51~~6.33\\
Sbc (13) & 0.08$\pm$0.02~~0.05 & 1.56$\pm$0.20~~1.66 & 0.66$\pm$0.09~~0.61 & 19.75$\pm$0.22~~19.66 & 5.33$\pm$1.81~~4.03 & 19.72$\pm$0.28~~19.45 & 4.39$\pm$0.72~~4.79\\
Sc (14) & 0.04$\pm$0.02~~0.02 & 1.12$\pm$0.14~~0.94 & 0.66$\pm$0.09~~0.64 & 20.45$\pm$0.20~~20.60 & 5.26$\pm$1.44~~4.00 & 20.25$\pm$0.31~~19.99 & 3.23$\pm$0.55~~2.90\\
Sb-Sc (48) & 0.09$\pm$0.01~~0.05 & 1.55$\pm$0.10~~1.63 & 0.70$\pm$0.04~~0.65 & 19.65$\pm$0.13~~19.51 & 6.60$\pm$0.91~~4.54 & 20.28$\pm$0.16~~20.09 & 4.97$\pm$0.39~~4.78\\
\hline
\end{tabular}

\begin{minipage}{175mm}

Col.(1): Galaxy Name. Col.(2): Bulge/Total luminosity ratio.
Col.(3): S\'{e}rsic index of the bulge. Col.(4): effective radius of
the bulge in kpc. Col.(5): effective surface brightness of the bulge
in mag arcsec$^{-2}$. Col.(6): disk scalelength in kpc. Col.(7):
central surface brightness of the disk in mag arcsec$^{-2}$.
Col.(8): bar length, i.e. semimajor axis of the bar in kpc.

Note: N=number of galaxies; \emph{SE} is standard deviation of the
mean; Barred galaxies are those galaxies for which BUDDA returned a
non-zero bar contribution.

\end{minipage}

\end{table}

\begin{table}

\begin{minipage}{155mm}
\textbf{Table 6b.} Mean/Median for Structural Parameters of Bulges,
Disks and Bars of \textbf{Non-Barred} Galaxies
\end{minipage}


\centering
\begin{tabular}{lcccccc}

\hline\hline

Type (N) & Bulge/Total & n$_{bulge}$ & r$_{e}$ (kpc) & $\mu_{e}$ & h$_{R}$ (kpc) & $\mu_{o}$\\
     & mean$\pm$SE median & mean$\pm$SE median & mean$\pm$SE median & mean$\pm$SE median & mean$\pm$SE median & mean$\pm$SE median \\

\hline
Sb (4)    & 0.10$\pm$0.03~~0.07 & 1.53$\pm$0.57~~1.01 & 0.66$\pm$0.05~~0.63 & 20.20$\pm$0.20~~20.25 & 2.78$\pm$0.14~~2.72 & 19.56$\pm$0.02~~19.57\\
Sbc (21)  & 0.09$\pm$0.02~~0.06 & 1.23$\pm$0.16~~1.07 & 0.79$\pm$0.11~~0.59 & 19.98$\pm$0.13~~19.91 & 5.18$\pm$1.45~~3.57 & 19.80$\pm$0.27~~19.60\\
Sc (21)   & 0.10$\pm$0.02~~0.07 & 1.22$\pm$0.10~~1.14 & 0.88$\pm$0.16~~0.74 & 20.50$\pm$0.16~~20.63 & 7.74$\pm$2.47~~3.57 & 20.19$\pm$0.30~~19.65\\
Sb-Sc (46)& 0.10$\pm$0.01~~0.06 & 1.25$\pm$0.10~~1.10 & 0.82$\pm$0.09~~0.62 & 20.24$\pm$0.10~~20.12 & 6.14$\pm$1.31~~3.09 & 19.96$\pm$0.18~~19.59\\
\hline
\end{tabular}

\begin{minipage}{155mm}

Note: N=number of galaxies; \emph{SE} is standard deviation of the
mean. Columns have the same designations like in Table 6a.

\end{minipage}

\end{table}

\begin{table}

\setcounter{table}{6}

\caption{Mean/Median for Some Photometric Measures}

\centering
\begin{tabular}{l ccc}
\hline\hline
Type & a$_{25}^{i}$/h$_{R}$ & l$_{bar}$/a$_{25}^{i}$ & l$_{bar}$/h$_{R}$\\
 & mean$\pm$SE median & mean$\pm$SE median & mean$\pm$SE median\\

\hline
Sb    & 2.9$\pm$0.3~~3.0 & 0.40$\pm$0.03~~0.37 & 0.98$\pm$0.08~~0.96\\
Sbc   & 3.9$\pm$0.3~~3.8 & 0.30$\pm$0.04~~0.26 & 1.06$\pm$0.13~~1.00\\
Sc    & 3.3$\pm$0.2~~3.6 & 0.26$\pm$0.03~~0.25 & 0.75$\pm$0.09~~0.68\\
Sb-Sc & 3.4$\pm$0.1~~3.5 & 0.34$\pm$0.02~~0.34 & 0.93$\pm$0.06~~0.90\\
\hline
\end{tabular}

\begin{minipage}{8.5cm}

Note: \emph{SE} is standard deviation of the mean.

Col.(1): Morphological Type. Col.(2): Semimajor axis of $\mu_{i}$=25
mag arcsec$^{-2}$ isophote normalized by the disk radial scalelength
h$_{R}$. Col(3): semimajor axis of the bar normalized by the
semimajor axis of $\mu_{i}$=25 mag arcsec$^{-2}$ isophote. Col.(4):
semimajor axis of the bar normalized by disk radial scalelength
h$_{R}$.

\end{minipage}

\end{table}

\clearpage
\begin{table}

\begin{minipage}{155mm}
\textbf{Table 8a.} Mean/Median for \textbf{r$_{e}$/h$_{R}$}
\end{minipage}


\centering
\begin{tabular}{lccc}

\hline\hline

Type & all & barred & non-barred \\
     & mean$\pm$SE median & mean$\pm$SE median & mean$\pm$SE median \\

\hline
Sb    & 0.13$\pm$0.01~~0.11 & 0.11$\pm$0.01~~0.10 & 0.24$\pm$0.02~~0.23 \\
Sbc   & 0.22$\pm$0.02~~0.20 & 0.17$\pm$0.02~~0.16 & 0.24$\pm$0.04~~0.20 \\
Sc    & 0.20$\pm$0.02~~0.17 & 0.17$\pm$0.03~~0.13 & 0.22$\pm$0.03~~0.20 \\
Sb-Sc & 0.19$\pm$0.01~~0.16 & 0.15$\pm$0.01~~0.13 & 0.23$\pm$0.02~~0.20 \\
\hline
\end{tabular}

\begin{minipage}{155mm}

Note: \emph{SE} is standard deviation of the mean.

\end{minipage}

\end{table}

\begin{table}

\begin{minipage}{155mm}
\textbf{Table 8b.} Mean/Median for \textbf{r$_{e}$/h$_{R}$ for
Bulge/Total less than and larger than 0.1}
\end{minipage}


\centering
\begin{tabular}{ l | ccc c ccc }

\hline\hline

Type & \multicolumn{3}{c}{Bulge/Total$<$0.1} & \vline & \multicolumn{3}{c}{Bulge/Total$>$0.1} \\
     & all & barred & non-barred & \vline & all & barred & non-barred  \\
     & mean$\pm$SE median & mean$\pm$SE median & mean$\pm$SE median & \vline & mean$\pm$SE median & mean$\pm$SE median & mean$\pm$SE median \\

\hline
Sb    & 0.16$\pm$0.02~~0.14 & 0.13$\pm$0.02~~0.11 & 0.24$\pm$0.02~~0.23 & \vline & 0.11$\pm$0.02~~0.10 & 0.10$\pm$0.01~~0.10 & 0.23~~~~~~~~~~~~~~~~~~~ \\
Sbc   & 0.19$\pm$0.02~~0.18 & 0.17$\pm$0.02~~0.16 & 0.21$\pm$0.03~~0.20 & \vline & 0.28$\pm$0.08~~0.29 & 0.17$\pm$0.13~~0.17 & 0.32$\pm$0.01~~0.30\\
Sc    & 0.18$\pm$0.02~~0.16 & 0.16$\pm$0.03~~0.12 & 0.19$\pm$0.03~~0.18 & \vline & 0.28$\pm$0.05~~0.30 & 0.29~~~~~~~~~~~~~~~~~~~~& 0.28$\pm$0.06~~0.31\\
Sb-Sc & 0.18$\pm$0.01~~0.16 & 0.16$\pm$0.01~~0.13 & 0.20$\pm$0.02~~0.20 & \vline & 0.20$\pm$0.03~~0.14 & 0.13$\pm$0.02~~0.10 & 0.29$\pm$0.05~~0.30\\
\hline
\end{tabular}

\begin{minipage}{155mm}

Note: \emph{SE} is standard deviation of the mean.

\end{minipage}

\end{table}

\clearpage
\begin{table}

\setcounter{table}{8}

\caption{CAS Parameters}

\begin{tabular}{lccc}
\hline\hline

Galaxy  &    C   &   A   &   S \\

\hline

KIG 11  &   3.39    &   0.05    &   0.27    \\
KIG 33  &   3.04    &   0.12    &   0.22    \\
KIG 56  &   4.18    &   0.07    &   0.26    \\
KIG 187 &   3.22    &   0.04    &   0.20    \\
KIG 198 &   3.19    &   0.08    &   0.21    \\
KIG 203 &   2.63    &   0.02    &   0.20    \\
KIG 217 &   2.52    &   0.08    &   0.15    \\
KIG 222 &   2.57    &   0.06    &   0.15    \\
KIG 232 &   3.06    &   0.28    &   0.23    \\
KIG 238 &   4.25    &   0.07    &   0.28    \\
KIG 241 &   2.44    &   0.11    &   0.15    \\
KIG 242 &   3.19    &   0.05    &   0.20    \\
KIG 258 &   3.77    &   0.12    &   0.19    \\
KIG 260 &   2.86    &   0.11    &   0.22    \\
KIG 271 &   2.71    &   0.08    &   0.18    \\
KIG 281 &   2.42    &   0.05    &   0.14    \\
KIG 282 &   2.95    &   0.17    &   0.23    \\
KIG 287 &   3.53    &   0.07    &   0.30    \\
KIG 292 &   3.18    &   0.08    &   0.19    \\
KIG 298 &   3.69    &   0.07    &   0.22    \\
KIG 302 &   2.89    &   0.26    &   0.17    \\
KIG 314 &   3.22    &   0.09    &   0.26    \\
KIG 325 &   2.94    &   0.08    &   0.21    \\
KIG 328 &   2.61    &   0.11    &   0.16    \\
KIG 330 &   2.45    &   0.13    &   0.15    \\
KIG 336 &   3.96    &   0.04    &   0.24    \\
KIG 339 &   5.70    &   0.04    &   0.29    \\
KIG 351 &   2.79    &   0.05    &   0.21    \\
KIG 365 &   2.58    &   0.14    &   0.16    \\
KIG 366 &   3.45    &   0.05    &   0.26    \\
KIG 367 &   3.10    &   0.06    &   0.22    \\
KIG 368 &   2.79    &   0.10    &   0.21    \\
KIG 386 &   2.56    &   0.18    &   0.20    \\
KIG 397 &   2.83    &   0.19    &   0.21    \\
KIG 399 &   2.68    &   0.13    &   0.20    \\
KIG 401 &   2.61    &   0.03    &   0.16    \\
KIG 405 &   2.51    &   0.03    &   0.15    \\
KIG 406 &   2.61    &   0.10    &   0.14    \\
KIG 409 &   2.47    &   0.11    &   0.17    \\
KIG 410 &   2.76    &   0.28    &   0.20    \\
KIG 429 &   2.41    &   0.06    &   0.16    \\
KIG 444 &   2.62    &   0.28    &   0.15    \\
KIG 446 &   2.82    &   0.07    &   0.25    \\
KIG 460 &   2.96    &   0.07    &   0.18    \\
KIG 466 &   3.01    &   0.23    &   0.12    \\
KIG 489 &   2.65    &   0.17    &   0.18    \\
KIG 491 &   2.95    &   0.07    &   0.26    \\
KIG 494 &   2.82    &   0.16    &   0.20    \\
KIG 499 &   3.70    &   0.05    &   0.22    \\
KIG 502 &   3.71    &   0.01    &   0.27    \\
KIG 508 &   2.88    &   0.19    &   0.17    \\
KIG 512 &   2.47    &   0.00    &   0.16    \\
KIG 515 &   2.84    &   0.08    &   0.22    \\
KIG 520 &   3.34    &   0.07    &   0.24    \\
KIG 522 &   3.06    &   0.04    &   0.14    \\
KIG 525 &   3.88    &   0.08    &   0.24    \\
KIG 532 &   3.04    &   0.18    &   0.18    \\
KIG 550 &   3.14    &   0.07    &   0.19    \\
KIG 553 &   4.90    &   0.03    &   0.34    \\
KIG 560 &   3.20    &   0.09    &   0.25    \\
KIG 571 &   2.70    &   0.10    &   0.19    \\

\end{tabular}

\end{table}

\begin{table}

\textbf{Table 9.}--continued

\begin{tabular}{lccc}
\hline\hline

Galaxy  &    C   &   A   &   S \\

\hline

KIG 575 &   3.00    &   0.08    &   0.25    \\
KIG 580 &   2.79    &   0.08    &   0.16    \\
KIG 598 &   2.80    &   0.12    &   0.18    \\
KIG 600 &   2.94    &   0.04    &   0.20    \\
KIG 612 &   4.04    &   0.05    &   0.24    \\
KIG 626 &   2.42    &   0.08    &   0.17    \\
KIG 630 &   3.03    &   0.10    &   0.26    \\
KIG 633 &   2.74    &   0.20    &   0.23    \\
KIG 639 &   2.66    &   0.06    &   0.18    \\
KIG 640 &   2.97    &   0.10    &   0.22    \\
KIG 641 &   3.74    &   0.05    &   0.25    \\
KIG 645 &   2.85    &   0.06    &   0.20    \\
KIG 652 &   2.86    &   0.13    &   0.17    \\
KIG 665 &   3.06    &   0.05    &   0.22    \\
KIG 671 &   4.75    &   0.08    &   0.29    \\
KIG 689 &   2.66    &   0.14    &   0.14    \\
KIG 712 &   2.99    &   0.07    &   0.23    \\
KIG 716 &   3.90    &   0.03    &   0.22    \\
KIG 719 &   5.07    &   0.06    &   0.17    \\
KIG 731 &   2.89    &   0.06    &   0.21    \\
KIG 743 &   2.72    &   0.05    &   0.15    \\
KIG 754 &   2.46    &   0.02    &   0.15    \\
KIG 757 &   3.13    &   0.13    &   0.23    \\
KIG 795 &   2.57    &   0.19    &   0.15    \\
KIG 805 &   2.89    &   0.06    &   0.21    \\
KIG 807 &   2.50    &   0.15    &   0.16    \\
KIG 839 &   2.55    &   0.07    &   0.19    \\
KIG 892 &   3.21    &   0.10    &   0.23    \\
KIG 907 &   2.84    &   0.04    &   0.19    \\
KIG 912 &   2.72    &   0.07    &   0.20    \\
KIG 924 &   2.31    &   0.04    &   0.15    \\
KIG 928 &   2.83    &   0.11    &   0.21    \\
KIG 931 &   2.70    &   0.02    &   0.18    \\
KIG 932 &   3.39    &   0.04    &   0.26    \\
KIG 943 &   3.86    &   0.09    &   0.18    \\

\hline
\end{tabular}

\begin{minipage}{40mm}

C-Concentration \\
A-Asymmetry \\
S-Clumpiness.

\end{minipage}

\end{table}

\clearpage
\begin{table}

\setcounter{table}{9}

\begin{minipage}{155mm}
\caption{Mean/Median for CAS Parameters of \textbf{All} Galaxies}
\end{minipage}

\begin{tabular}{lccc}
\hline\hline

Type & C & A & S \\
     & mean$\pm$SE median & mean$\pm$SE median & mean$\pm$SE median \\
\hline
Sb       & 3.55$\pm$0.14~~3.57 & 0.07$\pm$0.01~~0.06 & 0.22$\pm$0.01~~0.22 \\
Sbc      & 2.94$\pm$0.10~~2.84 & 0.10$\pm$0.01~~0.08 & 0.20$\pm$0.01~~0.20 \\
Sc       & 2.83$\pm$0.05~~2.80 & 0.11$\pm$0.01~~0.09 & 0.19$\pm$0.01~~0.19 \\
Sb-Sc    & 3.06$\pm$0.06~~2.89 & 0.09$\pm$0.01~~0.08 & 0.20$\pm$0.01~~0.20 \\
\hline
\end{tabular}

\begin{minipage}{8.5cm}
Note: \emph{SE} is standard deviation of the mean.

Col.(1): Morphological Type. Col.(2): C-Concentration. Col.(3)
A-Asymmetry. Col.(4) S-Clumpiness.
\end{minipage}

\end{table}

\begin{table}

\setcounter{table}{10}

\begin{minipage}{95mm}
\caption{Mean/Median for CAS Parameters of \textbf{All} Sb-Sc
Galaxies in Our Sample versus the Sb-Sc Nearby Normal Galaxies from
Frei Sample \citep{conselice03}}
\end{minipage}

\begin{tabular}{lccc}
\hline\hline

     & C & A & S \\
     & mean$\pm$SE median & mean$\pm$SE median & mean$\pm$SE median \\
\hline
This study           & 3.06$\pm$0.06~~2.89 & 0.09$\pm$0.01~~0.08 & 0.20$\pm$0.01~~0.20 \\
\citet{conselice03}  & 3.47$\pm$0.08~~3.44 & 0.14$\pm$0.01~~0.13 & 0.28$\pm$0.02~~0.25 \\
\hline
\end{tabular}

\begin{minipage}{9cm}
Note: \emph{SE} is standard deviation of the mean.

Col.(1): Morphological Type. Col.(2): C-Concentration. Col.(3)
A-Asymmetry. Col.(4) S-Clumpiness.
\end{minipage}

\end{table}

\begin{figure*}
\centering
\includegraphics[width=\columnwidth,clip=true]{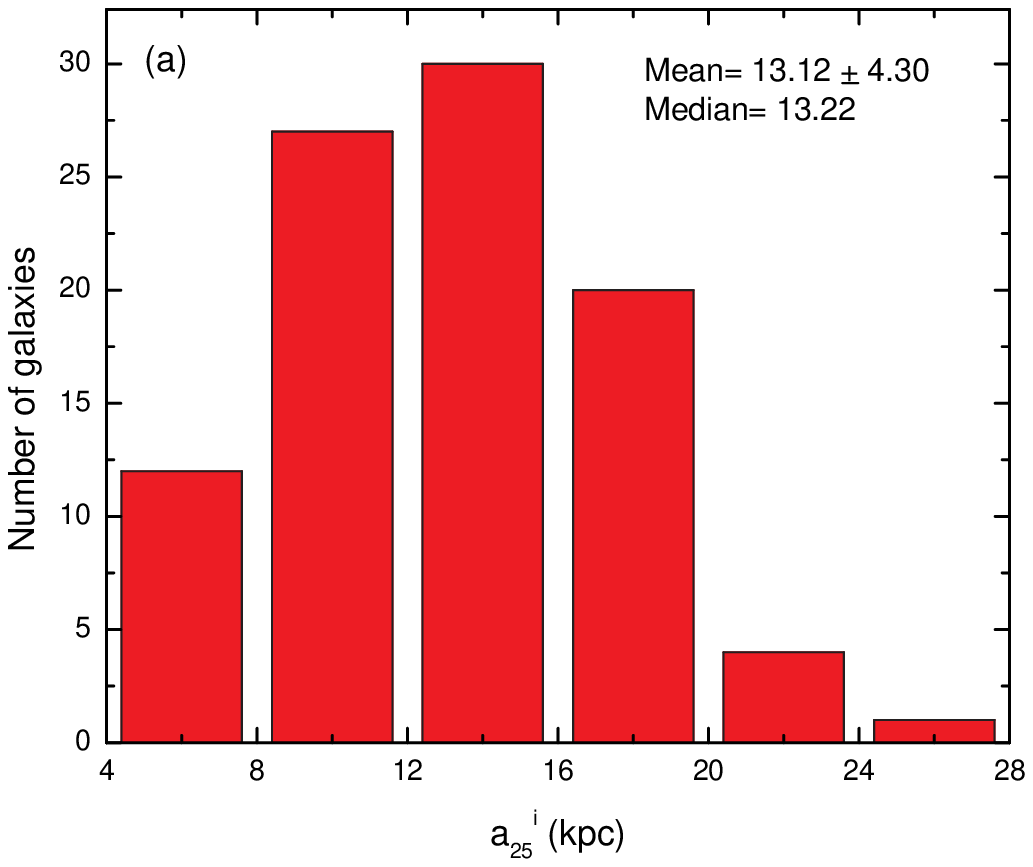}
\includegraphics[width=\columnwidth,clip=true]{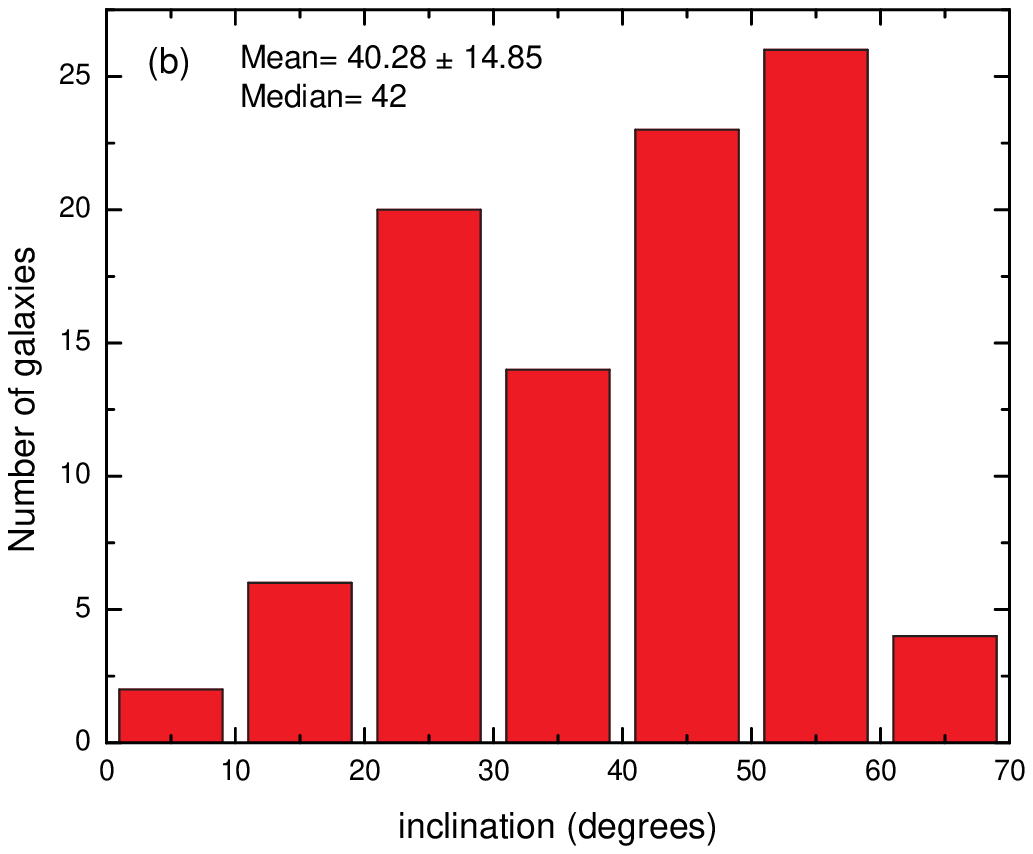}
\includegraphics[width=\columnwidth,clip=true]{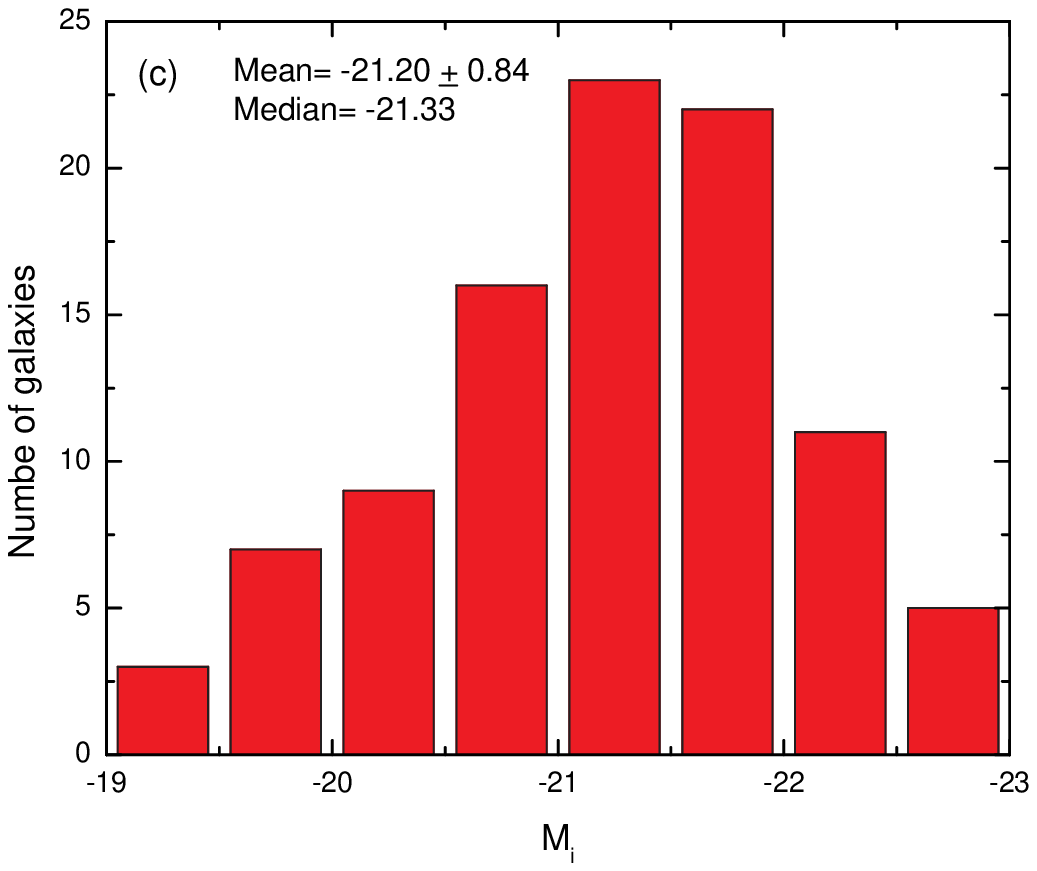}
\includegraphics[width=\columnwidth,clip=true]{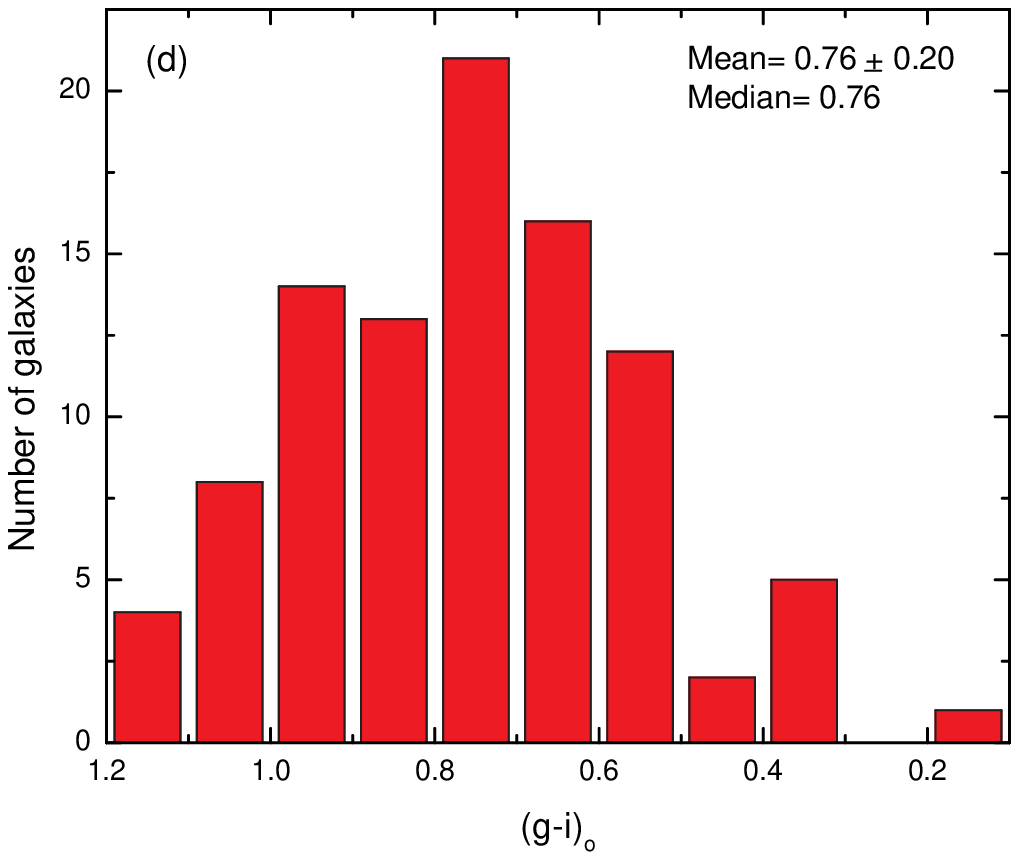}
\caption{Basic properties of the Sb-Sc CIG sample: (a) The
distribution of the disk size $a_{25}^{i}$ in i-band. (b)
Distribution of the inclination. (c) Distribution of the total
absolute magnitude M$_{i}$ in i-band. (d) Distribution of the global
color (g-i)$_{o}$.}\label{fig1}
\end{figure*}

\clearpage

\begin{figure*}
\centering
\includegraphics[width=\columnwidth,clip=true]{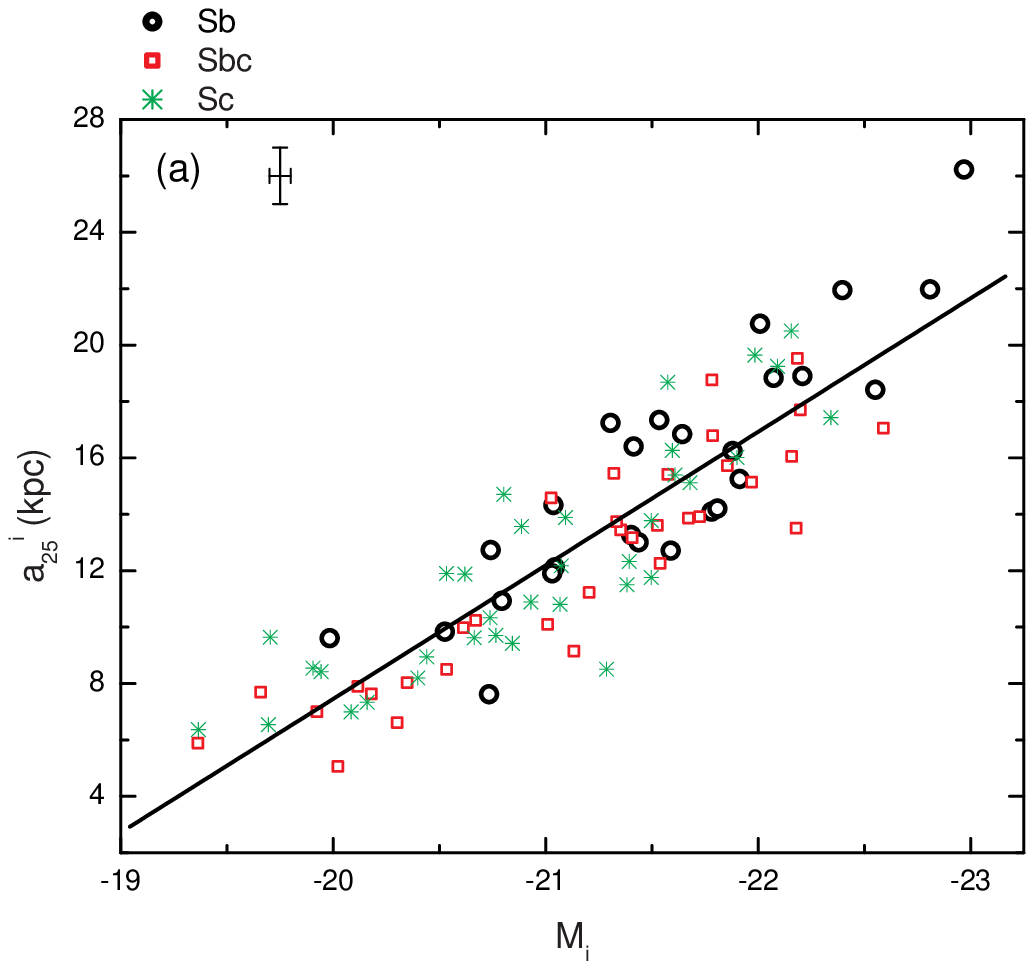}
\includegraphics[width=\columnwidth,clip=true]{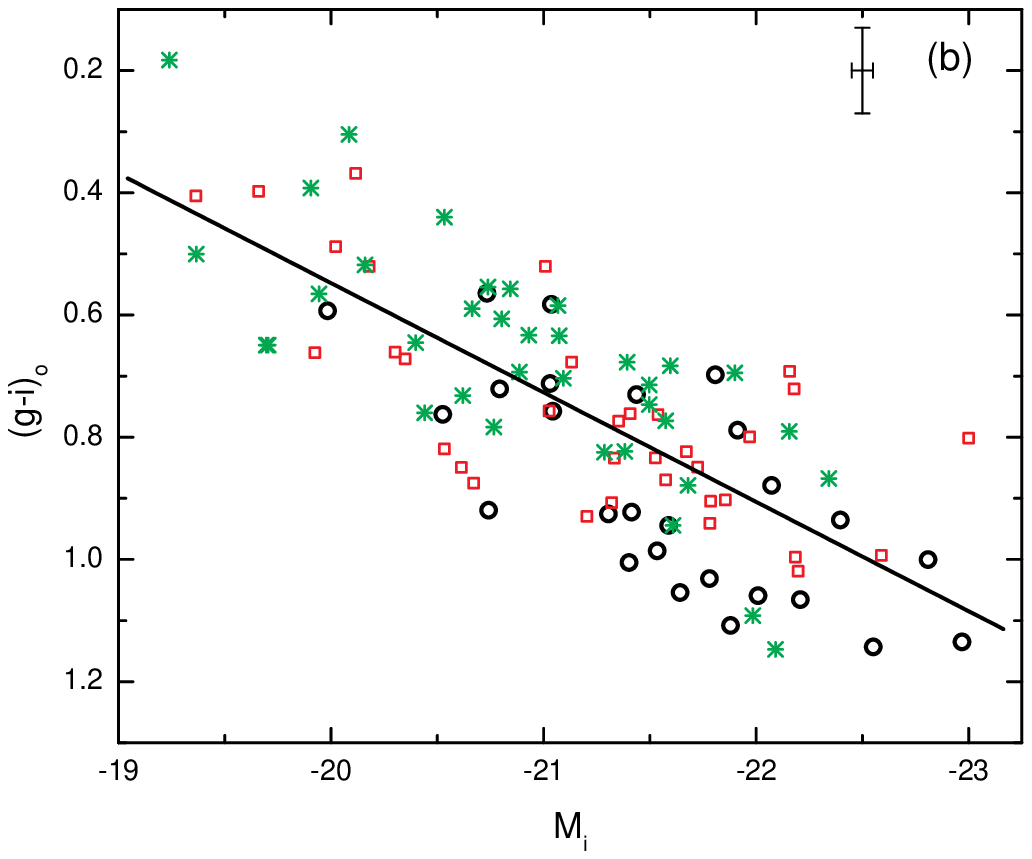}
\includegraphics[width=\columnwidth,clip=true]{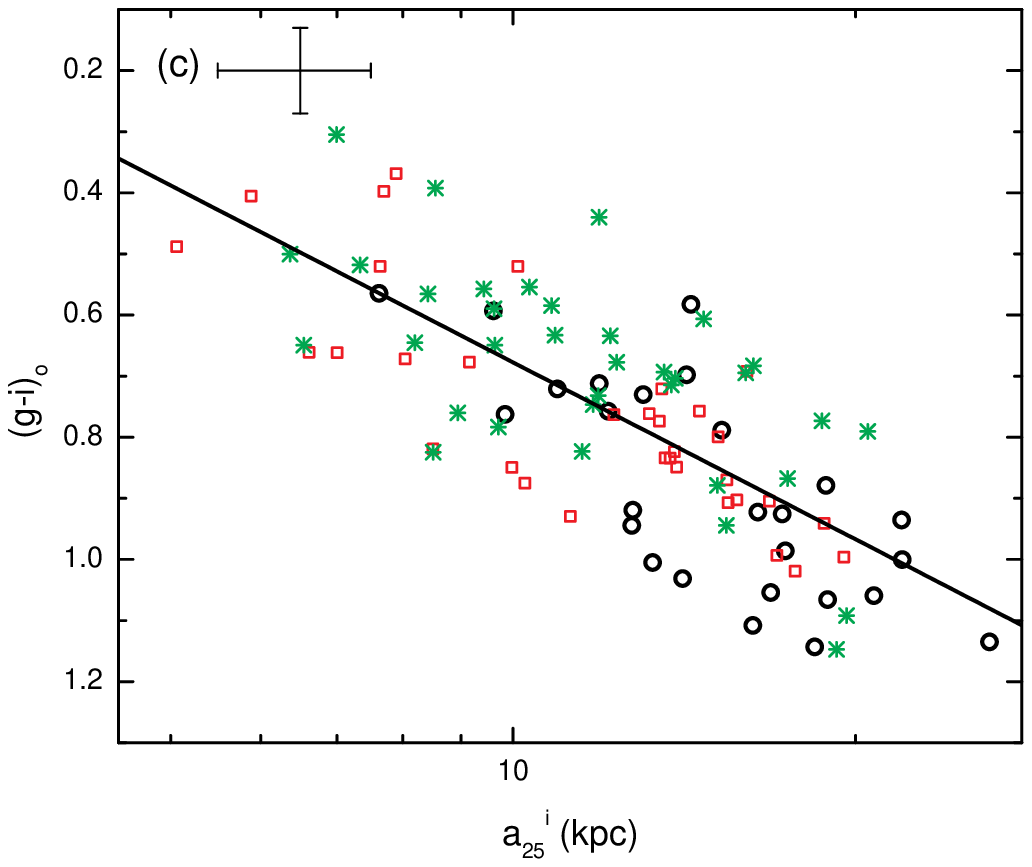}
\caption{(a) Disk size $a_{25}^{i}$ - absolute magnitude M$_{i}$
diagram. (b) Color (g-i)$_{o}$ versus total i-band absolute
magnitude M$_{i}$. (c) Color (g-i)$_{o}$ versus disk size
$a_{25}^{i}$ in i-band. The three morphological types (Sb-Sbc-Sc)
are shown with different symbols. A linear regression fit to the
whole sample is shown in each panel. The typical 2$\sigma$ error
bars are shown in each panel. } \label{fig2}
\end{figure*}

\begin{figure*}
\includegraphics[width=2.0\columnwidth]{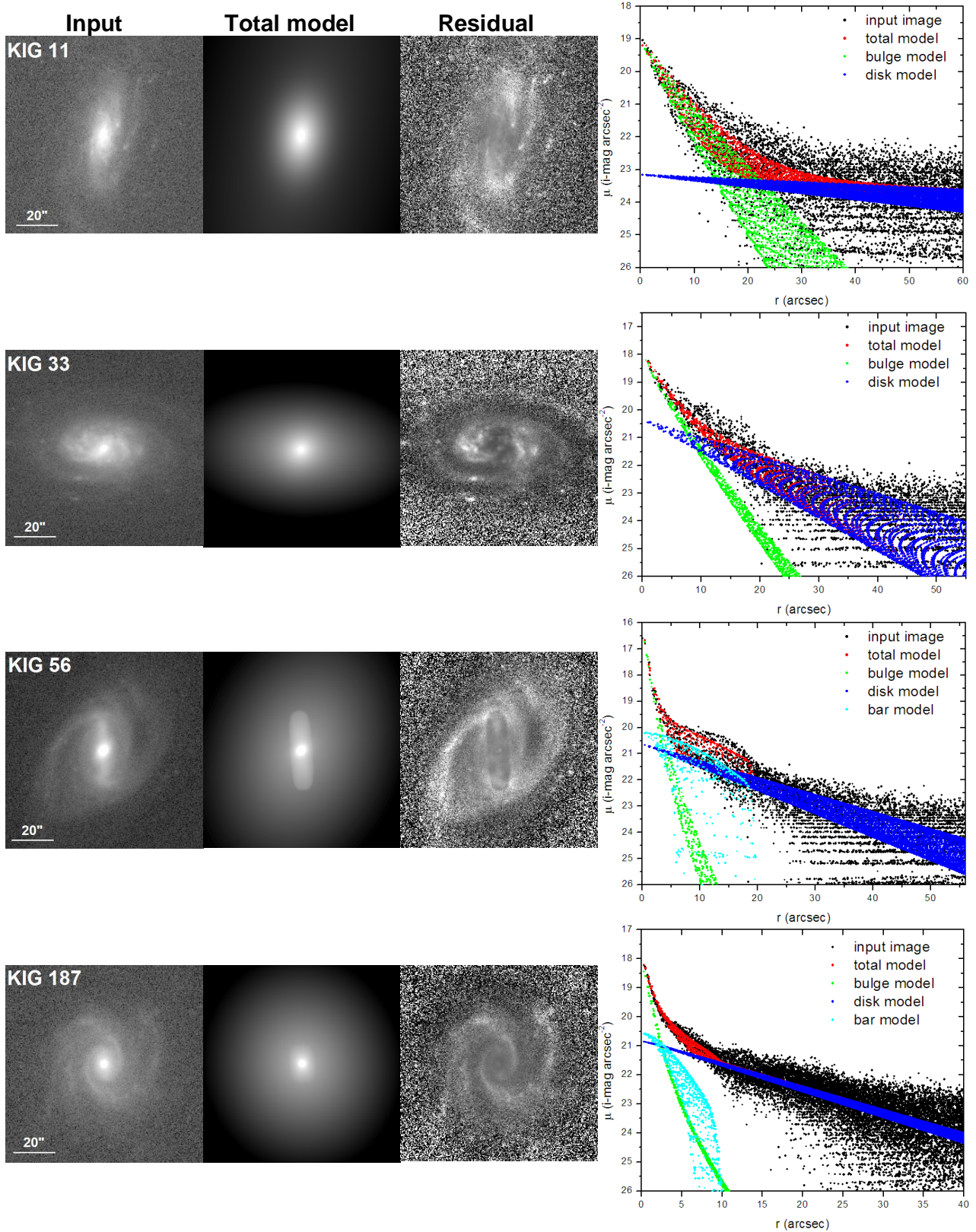} \caption{Examples of BUDDA-decomposition.}\label{fig3}
\end{figure*}

\clearpage

\begin{figure}
\plotone{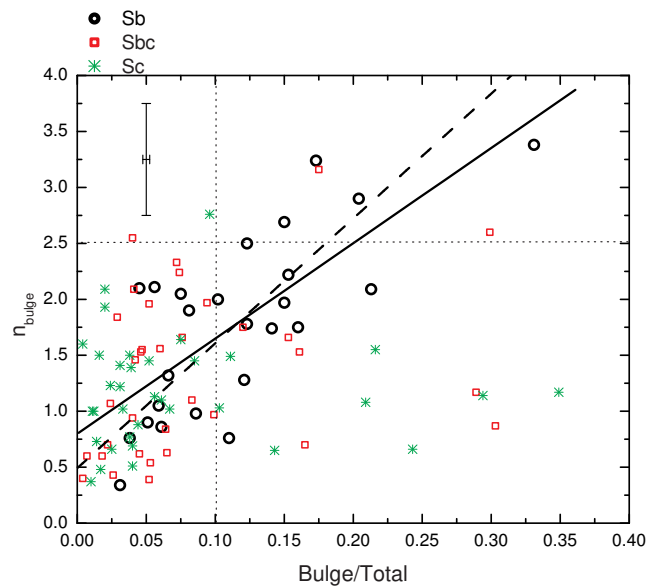} \caption{Bulge S\'{e}rsic index versus Bulge/Total
luminosity ratio. A linear regression fit (solid line) and a
bisector fit (dashed line) are shown for Sb-type only. The typical
2$\sigma$ error bars are shown.}\label{fig4}

\end{figure}

\begin{figure*}
\centering
\includegraphics[width=\columnwidth,clip=true]{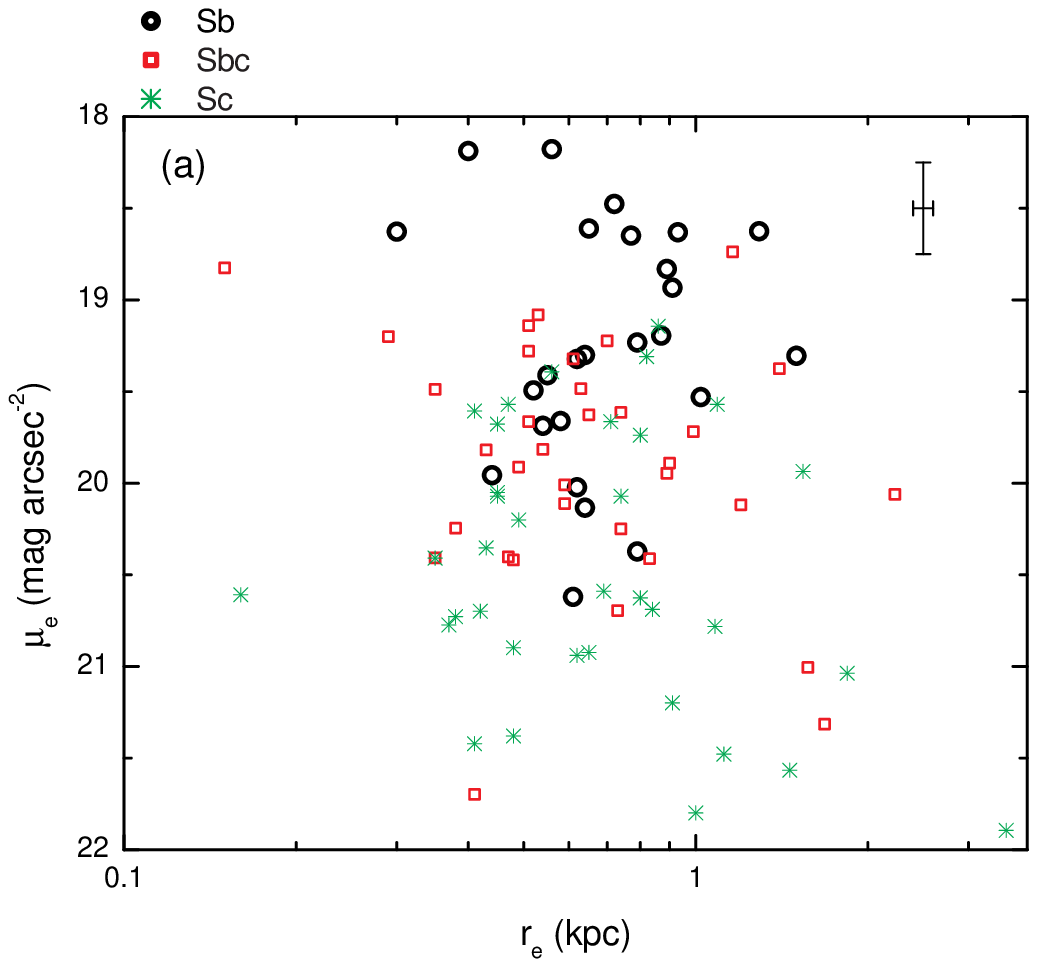}
\includegraphics[width=\columnwidth,clip=true]{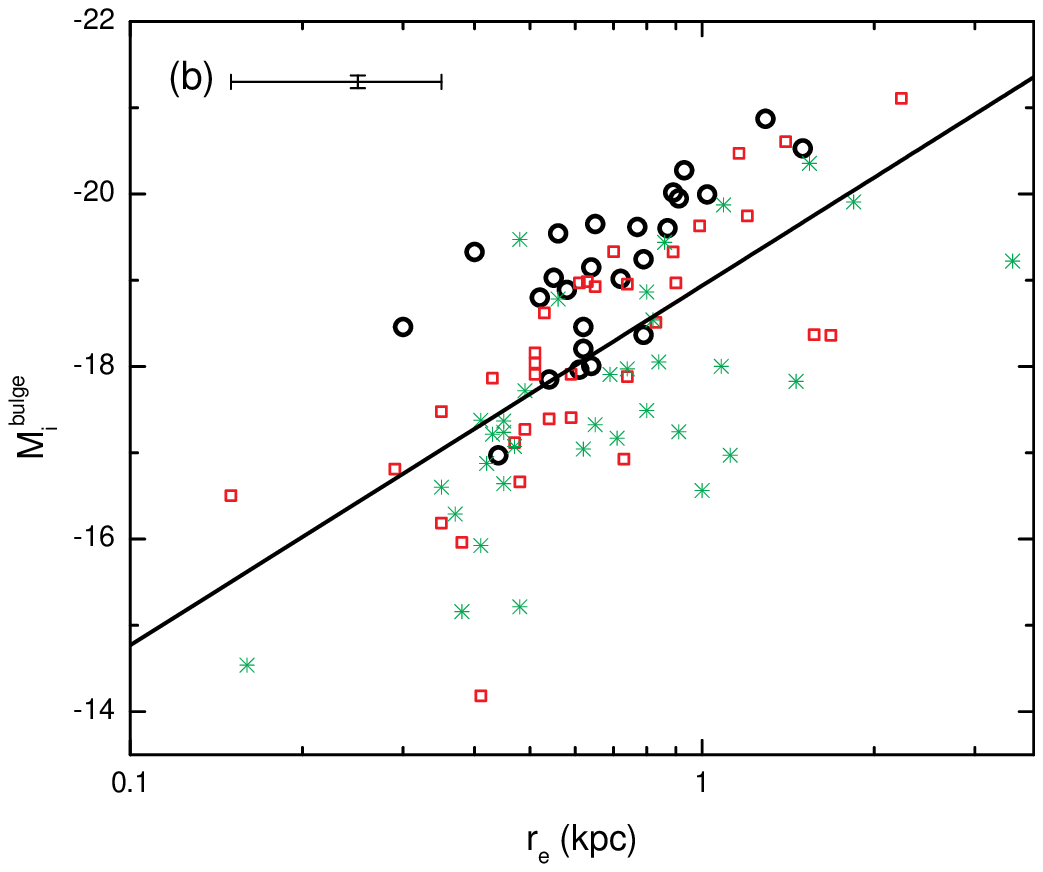}
\includegraphics[width=\columnwidth,clip=true]{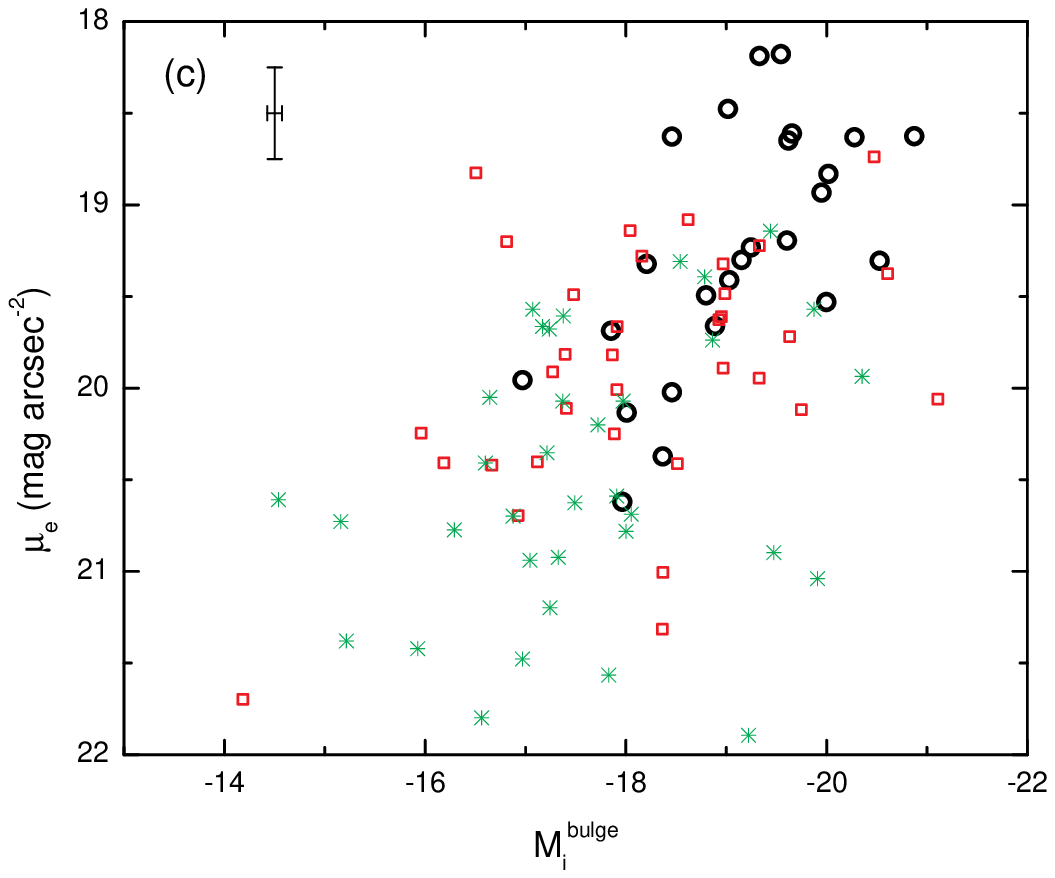}
\includegraphics[width=0.95\columnwidth,clip=true]{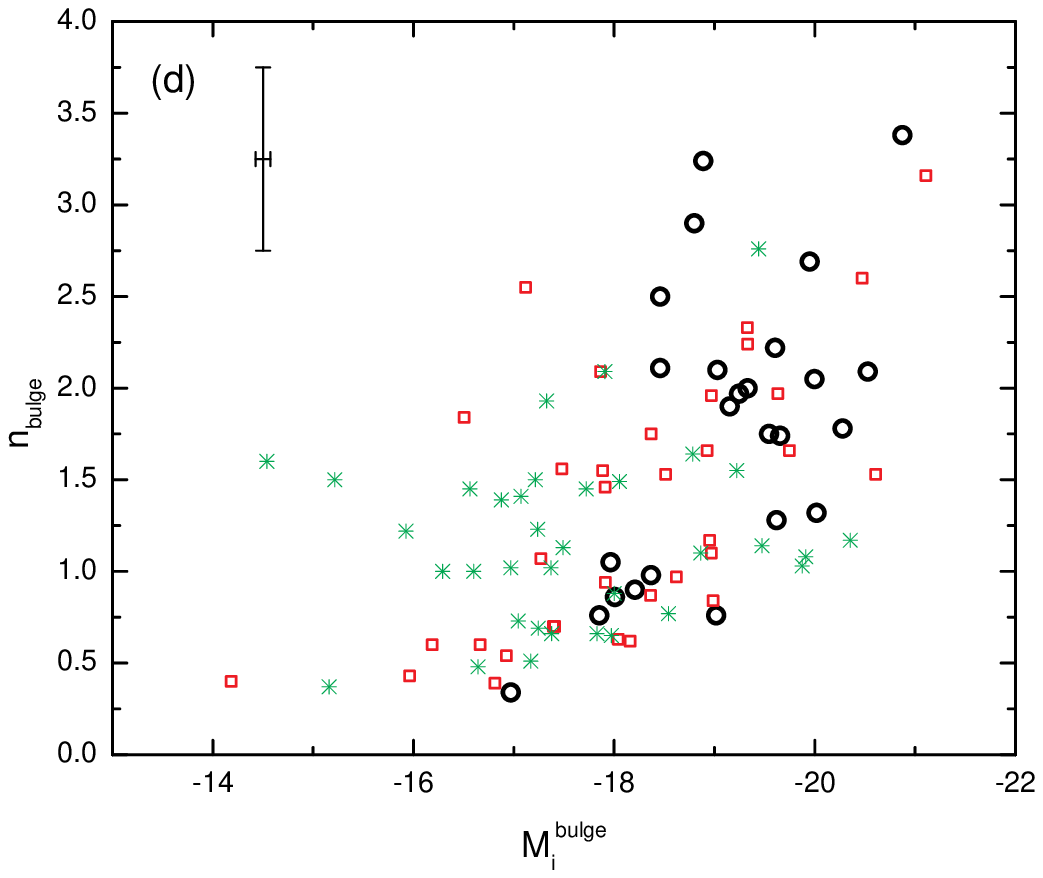}
\caption {(a) Fundamental Plane for Bulges (effective surface
brightness as a function of effective radius); (b-d) Relationship
between the parameters describing the bulge and its absolute i-band
magnitude. The three morphological types are indicated with
different symbols (see figure's legend). A linear regression fit to
the whole sample is shown as a solid line (panel b). The typical
2$\sigma$ error bars are shown in each panel.} \label{fig5}
\end{figure*}

\clearpage

\begin{figure}
\plotone{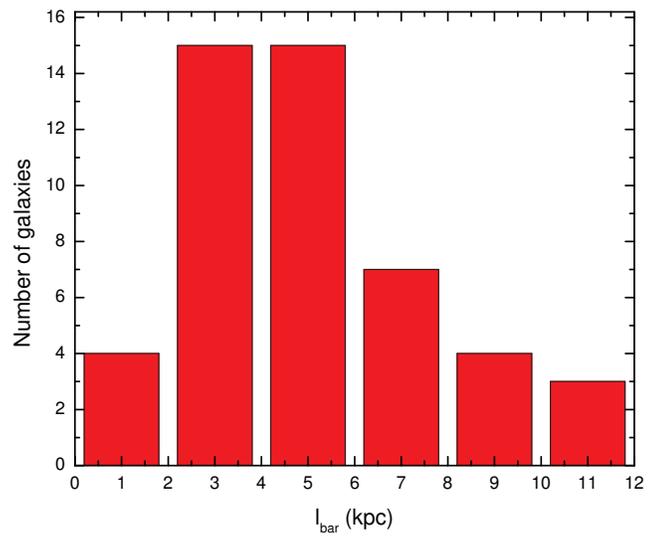} \caption{Bar size (semimajor axis of the bar)
distribution for all barred Sb-Sc galaxies (N=48)} \label{fig6}
\end{figure}

\begin{figure*}
\centering
\includegraphics[width=\columnwidth,clip=true]{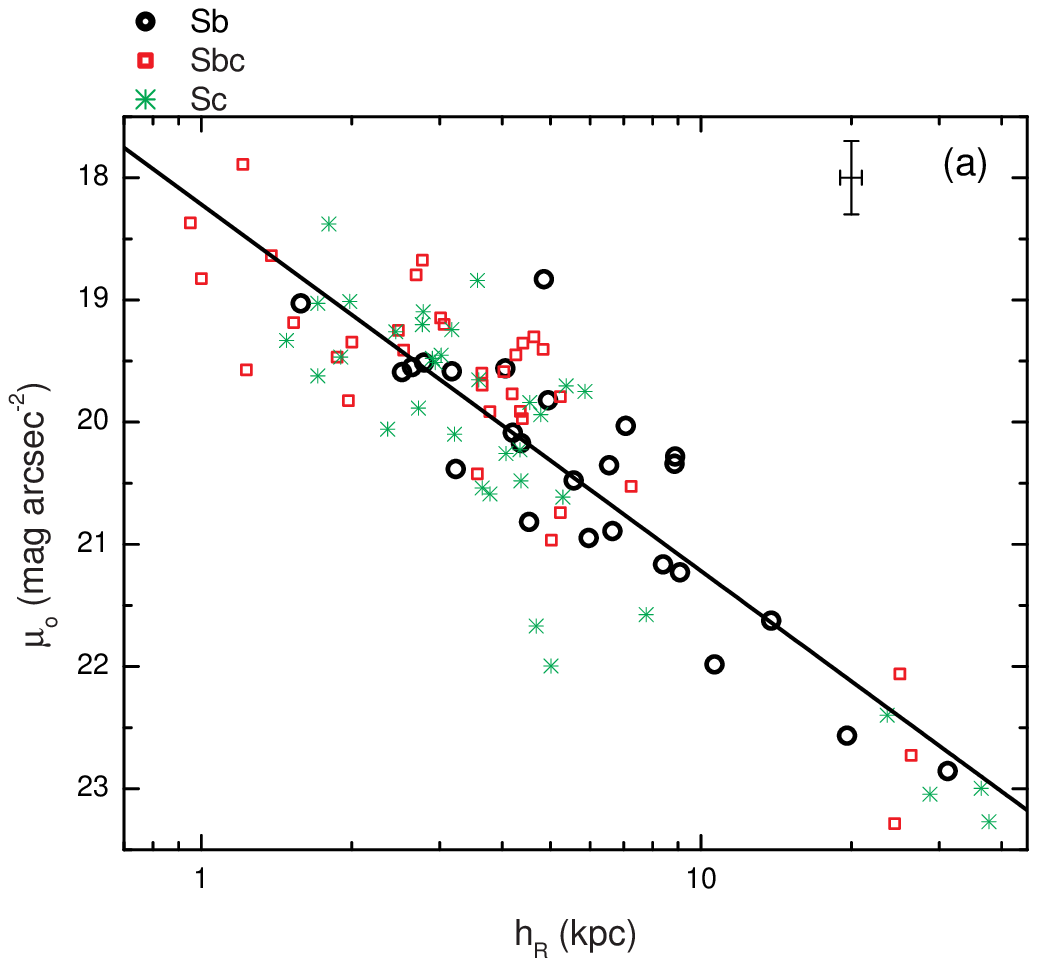}
\includegraphics[width=\columnwidth,clip=true]{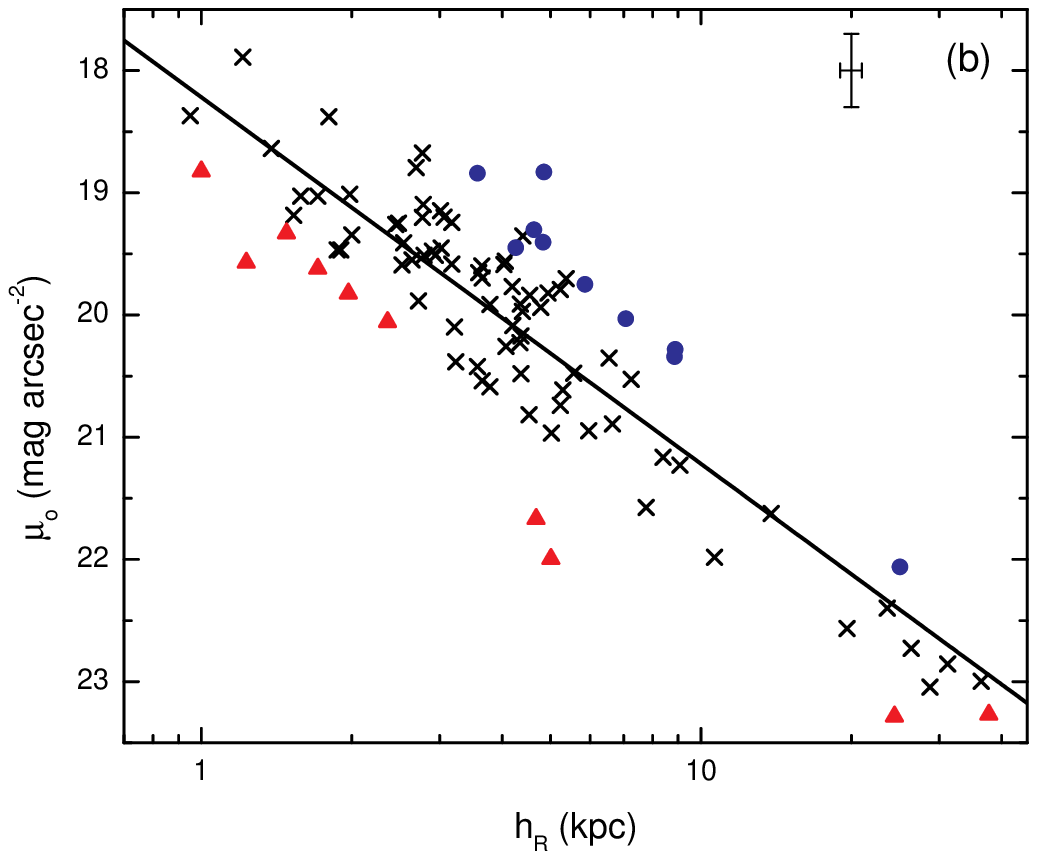}
\caption{Relationship between the two parameters describing the disk
profile, central surface brightness $\mu_{o}$ versus disk
scalelength h$_{R}$. Panels (a) and (b) illustrate the same plot
with the following differences:(a) The three morphological types are
indicated with different symbols (see figure's legend). (b) Solid
circles denote the 10 most luminous galaxies and solid triangles
denote the 10 least luminous galaxies in our sample. In panel (b) we
make no distinction between morphological types. A linear regression
fit to the whole sample is shown as a solid line in both panels. The
typical 2$\sigma$ error bars are shown in each panel.} \label{fig7}
\end{figure*}

\begin{figure*}
\centering
\includegraphics[width=\columnwidth,clip=true]{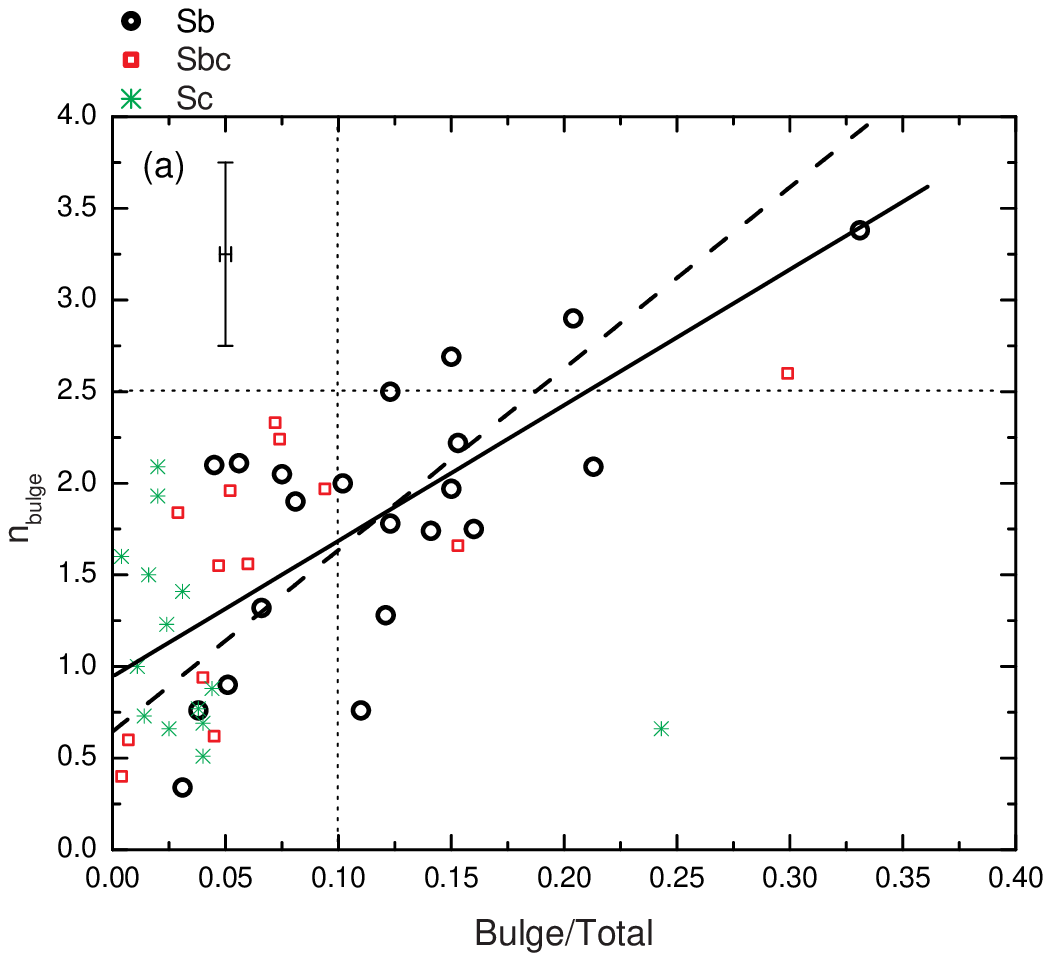}
\includegraphics[width=\columnwidth,clip=true]{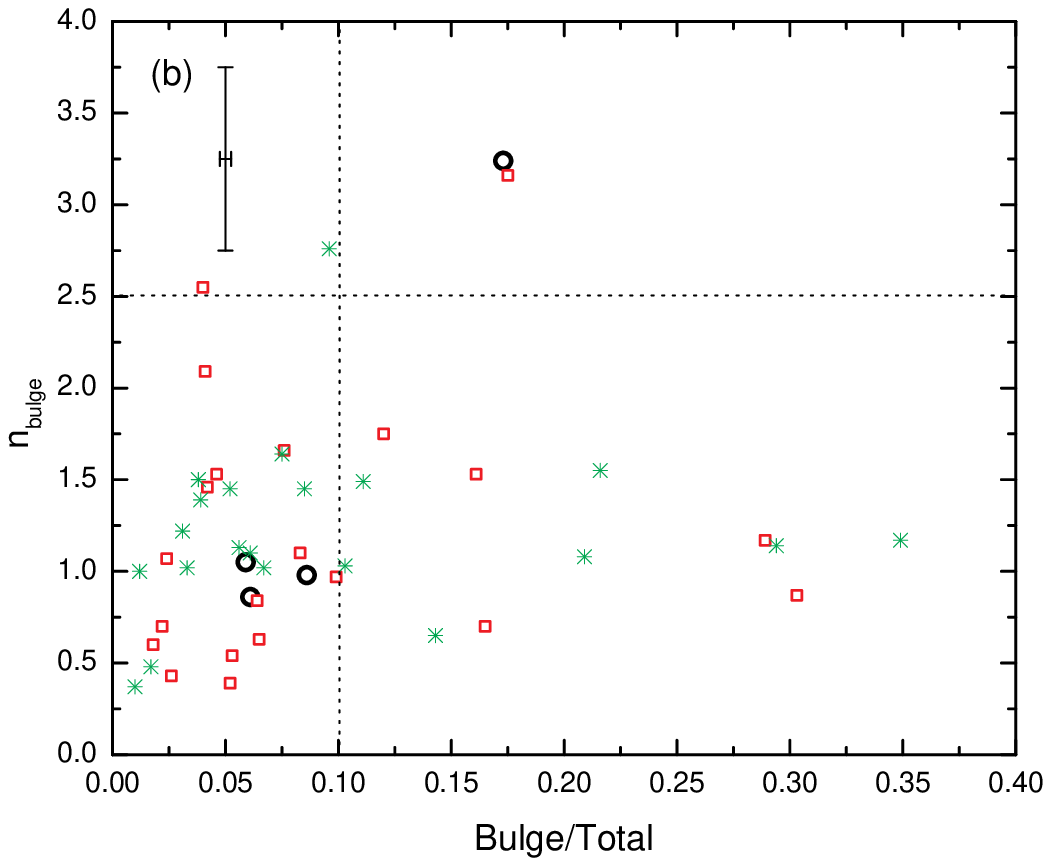}
\caption {(a) Bulge S\'{e}rsic index versus Bulge/Total luminosity
ratio for \textbf{barred} galaxies. (b) Bulge S\'{e}rsic index
versus Bulge/Total luminosity ratio for \textbf{non-barred}
galaxies. A linear regression fit (solid line) and a bisector fit
(dashed line) are shown for only for Sb-type in panel a. The three
morphological types are indicated with different symbols (see
figure's legend). The typical 2$\sigma$ error bars are shown in each
panel.}\label{fig8}
\end{figure*}

\begin{figure*}
\centering
\includegraphics[width=\columnwidth,clip=true]{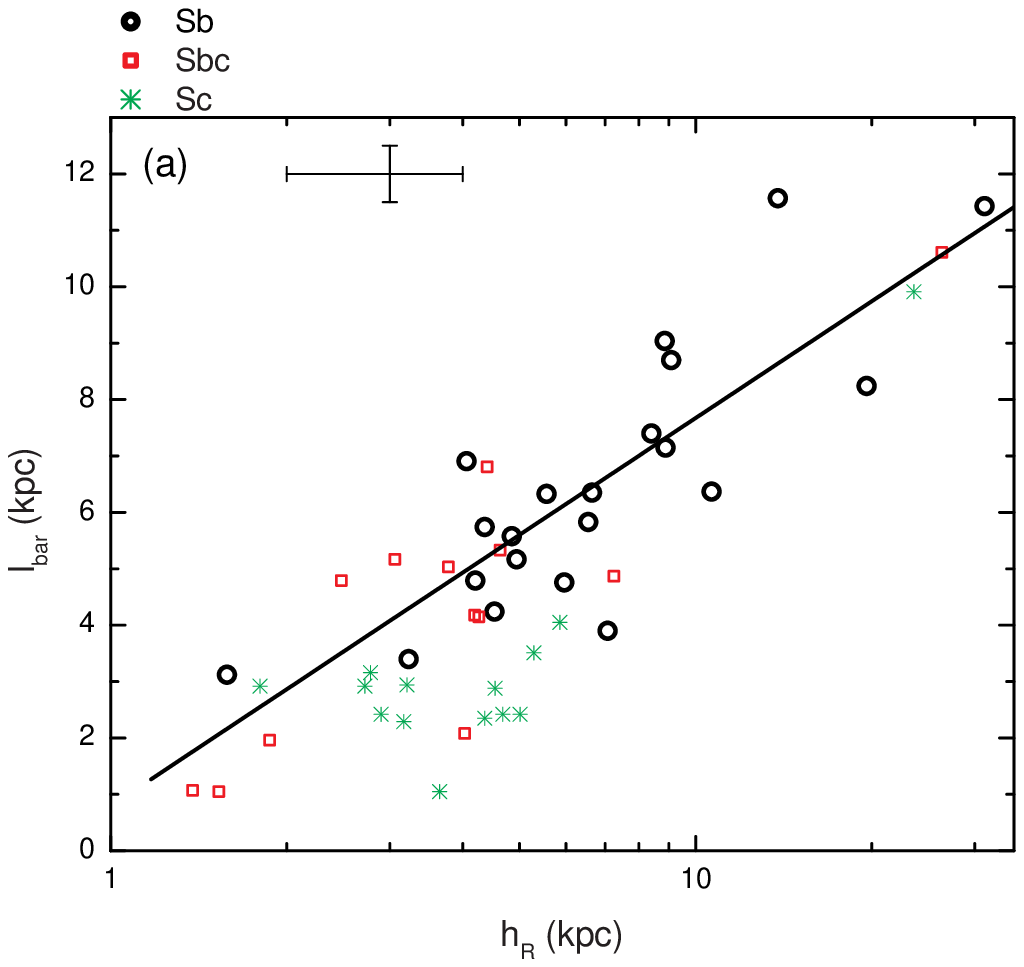}
\includegraphics[width=\columnwidth,clip=true]{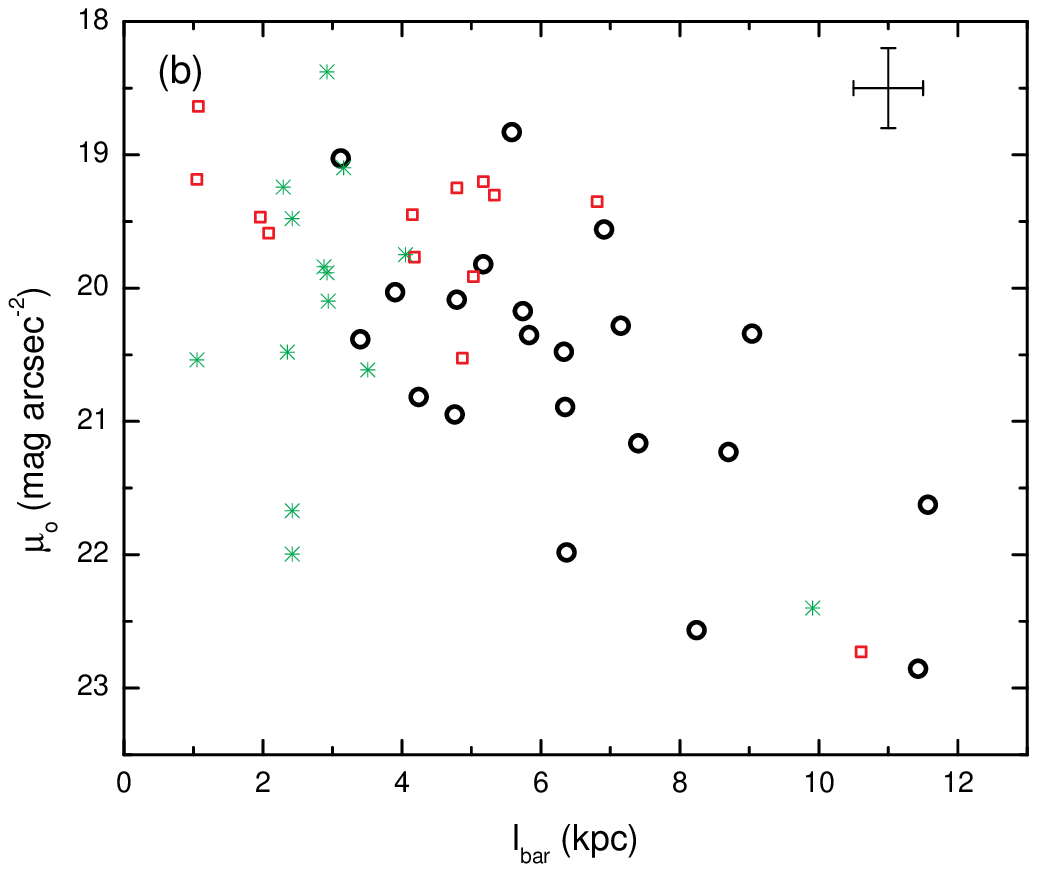}
\includegraphics[width=\columnwidth,clip=true]{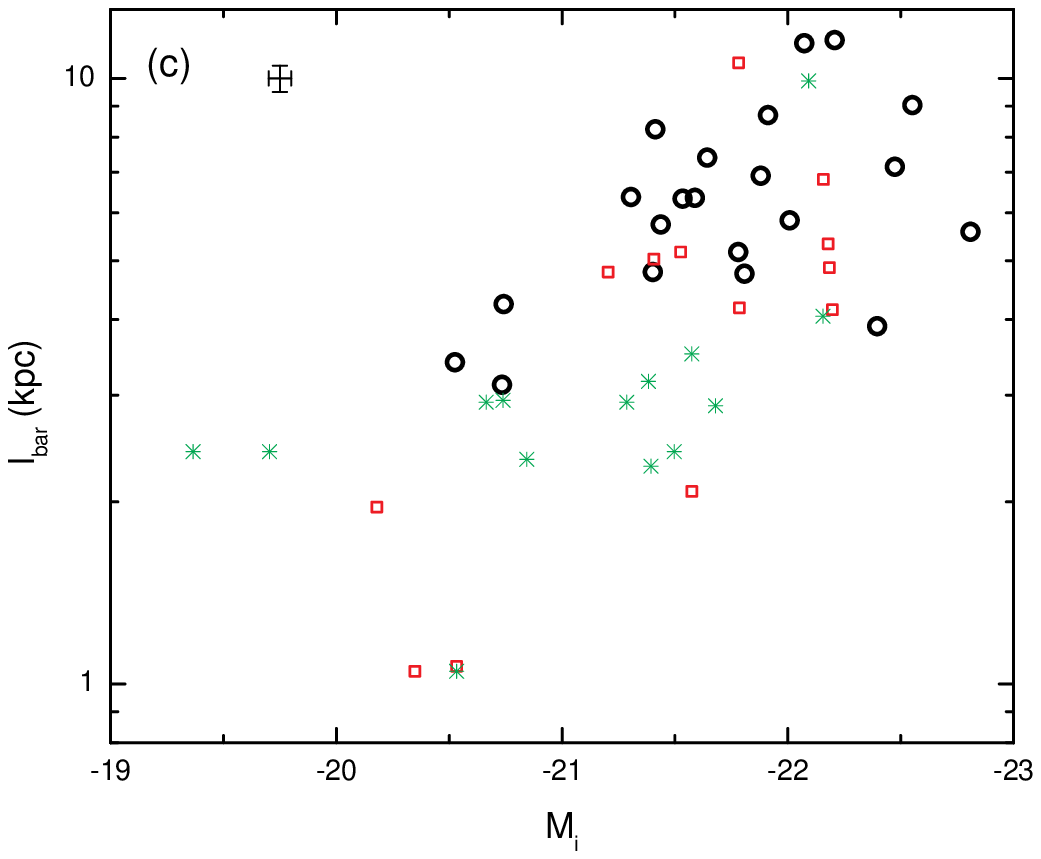}
\includegraphics[width=\columnwidth,clip=true]{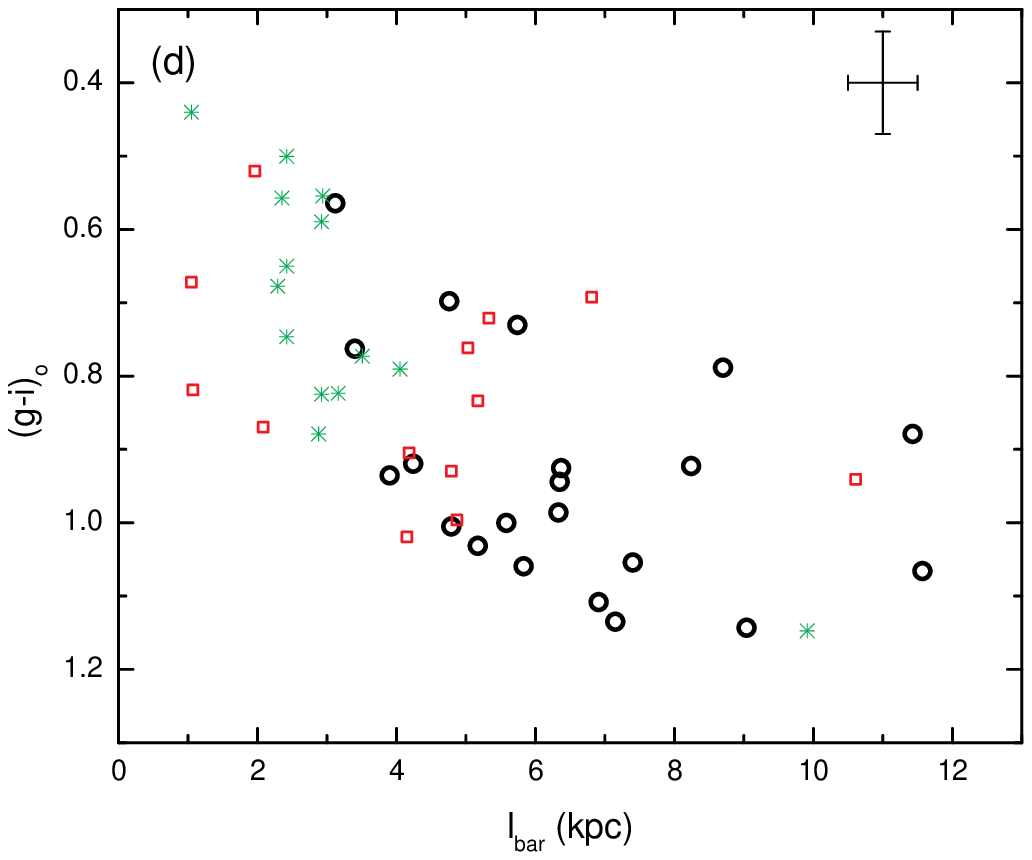}
\includegraphics[width=\columnwidth,clip=true]{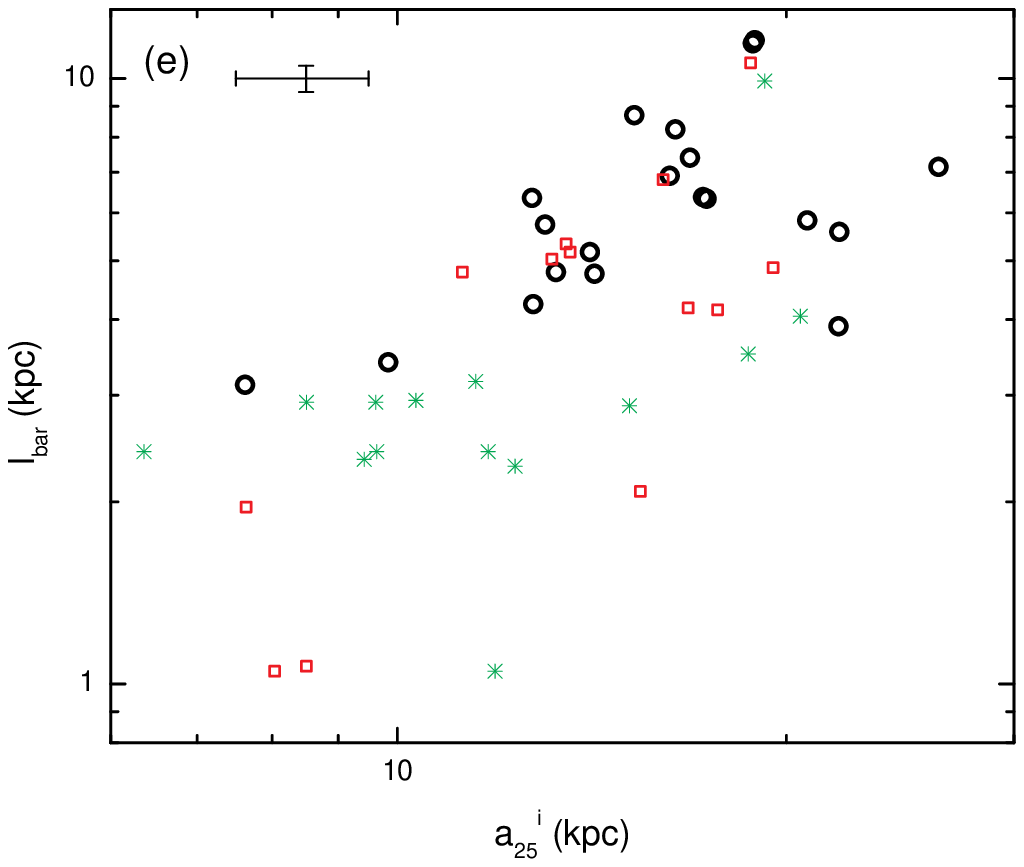}
\caption{Relationship between bar size l$_{bar}$ and the parameters
describing the disk: disk scalelength h$_{R}$ (panel a) and disk
central surface brightness $\mu_{o}$ (panel b). A linear regression
fit for all barred galaxies in our sample (48) is shown as a solid
line in panel a. Panel (c) shows the relation between the size of
the bar and the total absolute magnitude of the galaxy M$_{i}$. (d)
Galaxy color (g-i)$_{o}$ versus bar size l$_{bar}$. (e) Size of the
bar l$_{bar}$ versus the galaxy size $a_{25}^{i}$. The three
morphological types are indicated with different symbols (see
figure's legend). The typical 2$\sigma$ error bars are shown in each
panel.} \label{fig9}
\end{figure*}

\begin{figure*}
\centering
\includegraphics[width=\columnwidth,clip=true]{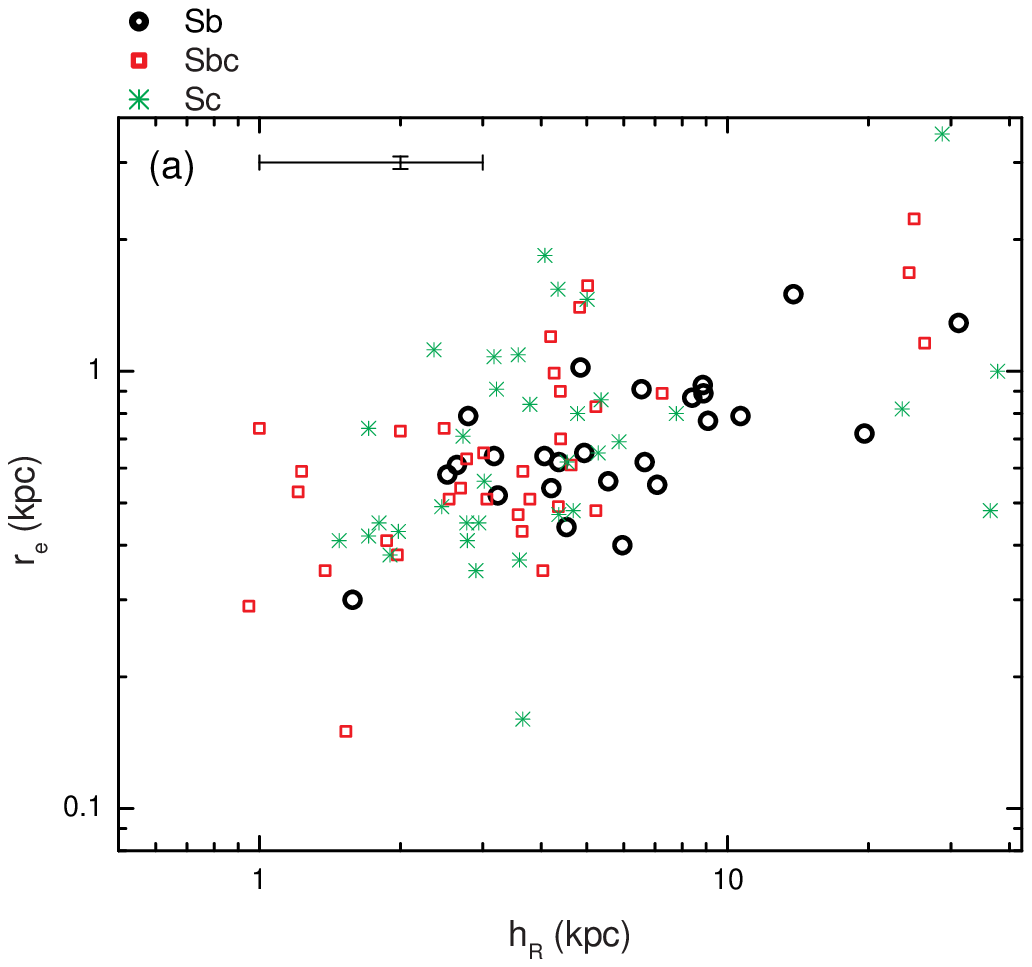}
\includegraphics[width=\columnwidth,clip=true]{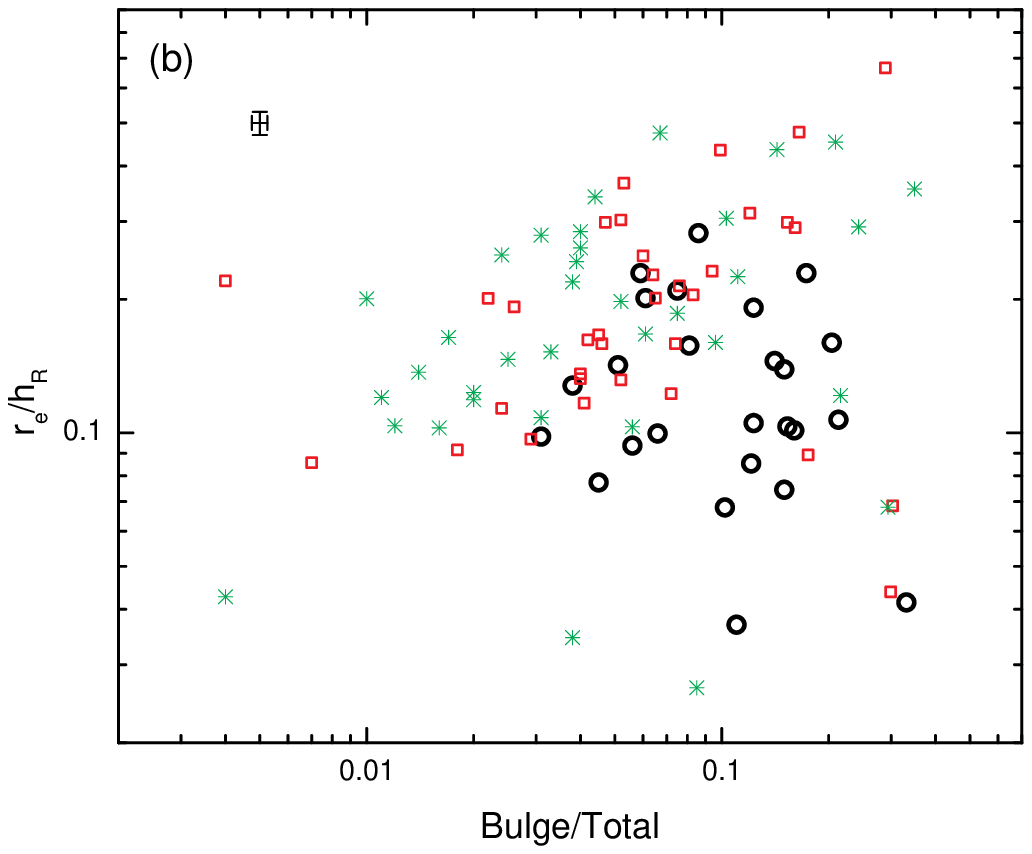}
\includegraphics[width=\columnwidth,clip=true]{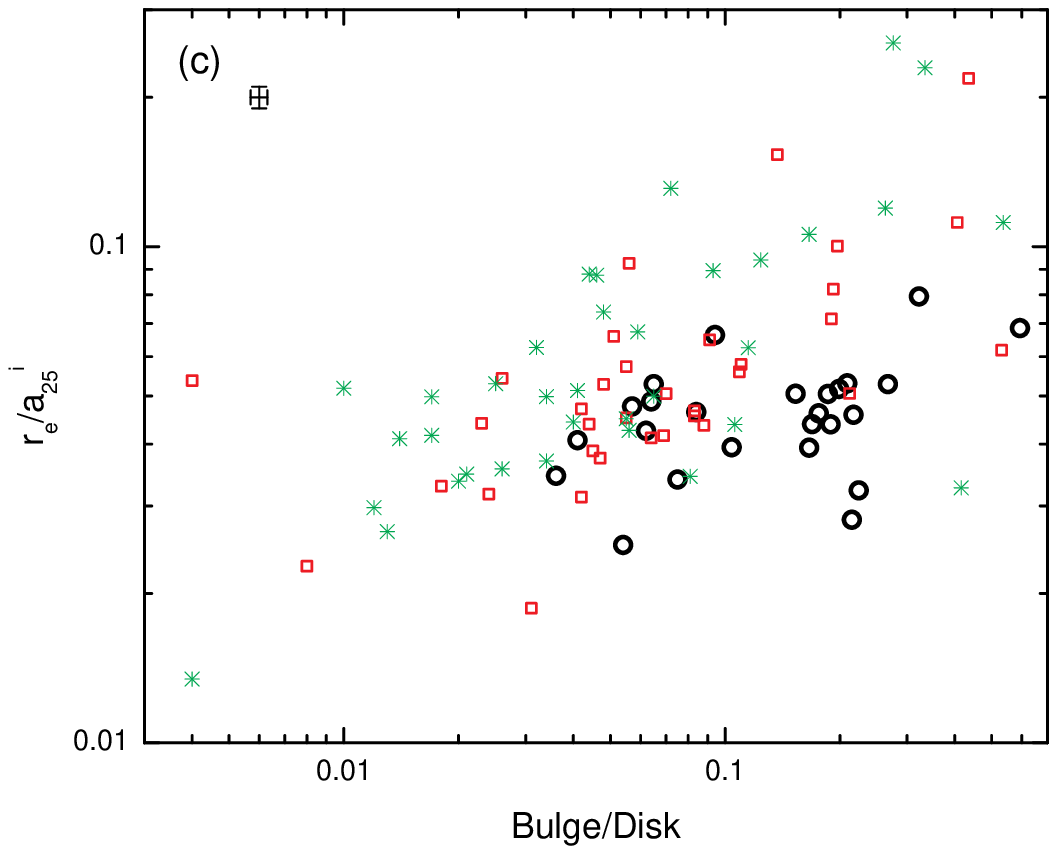}
\caption{(a) Bulge effective radius r$_{e}$ shown in relation to the
scalelength of the disk h$_{R}$. (b) Bulge effective radius r$_{e}$
normalized to disk scalelength h$_{R}$ as a function of Bulge/Total
luminosity ratio. (c) Bulge effective radius r$_{e}$ normalized to
the disk size $a_{25}^{i}$ versus Bulge/Disk luminosity ratio. The
three morphological types are indicated with different symbols (see
figure's legend). The typical 2$\sigma$ error bars are shown in each
panel.} \label{fig10}
\end{figure*}

\clearpage

\begin{figure*}
\centering
\includegraphics[width=\columnwidth,clip=true]{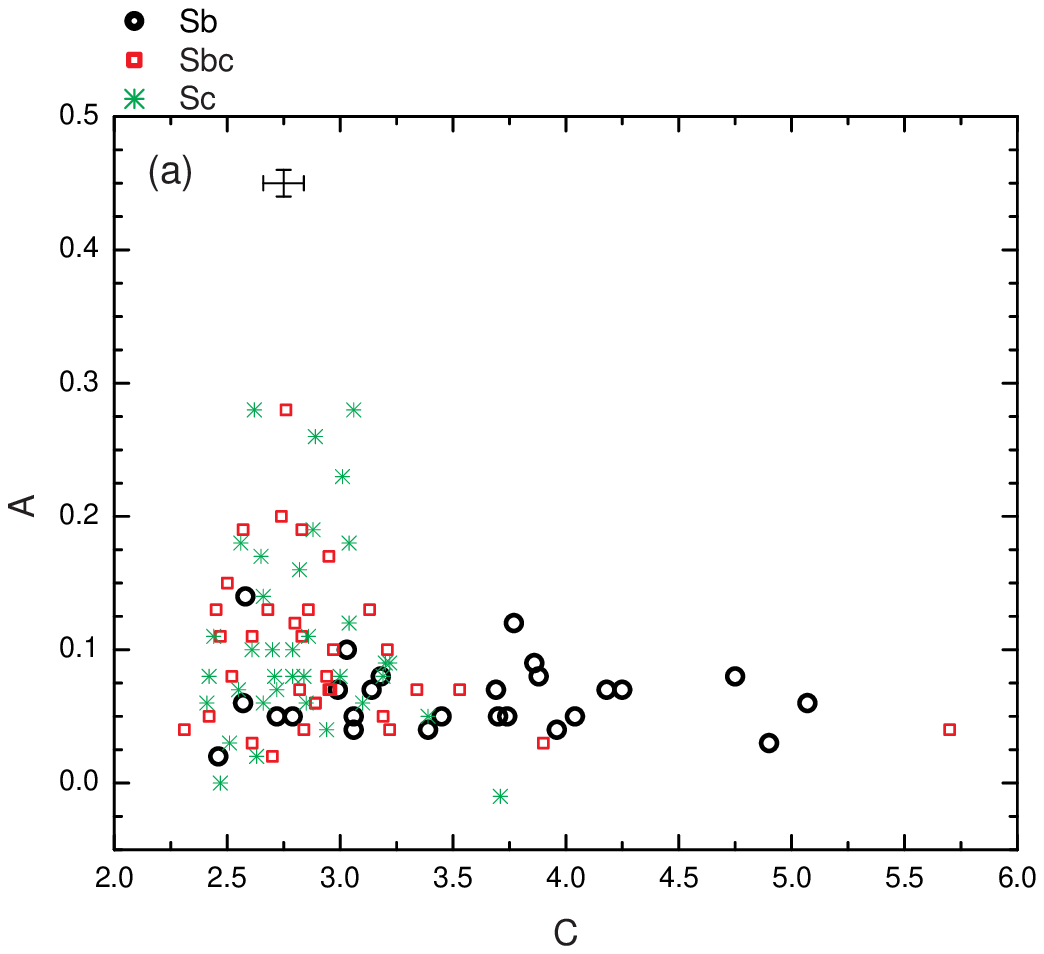}
\includegraphics[width=\columnwidth,clip=true]{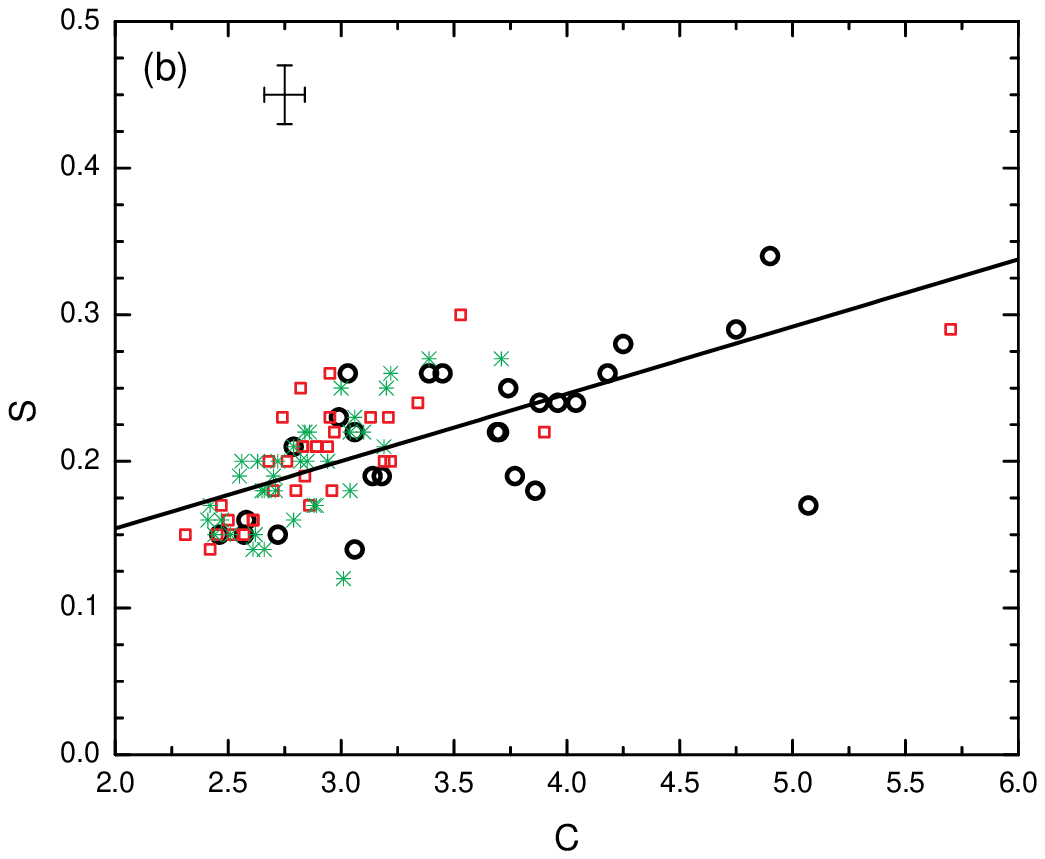}
\includegraphics[width=\columnwidth,clip=true]{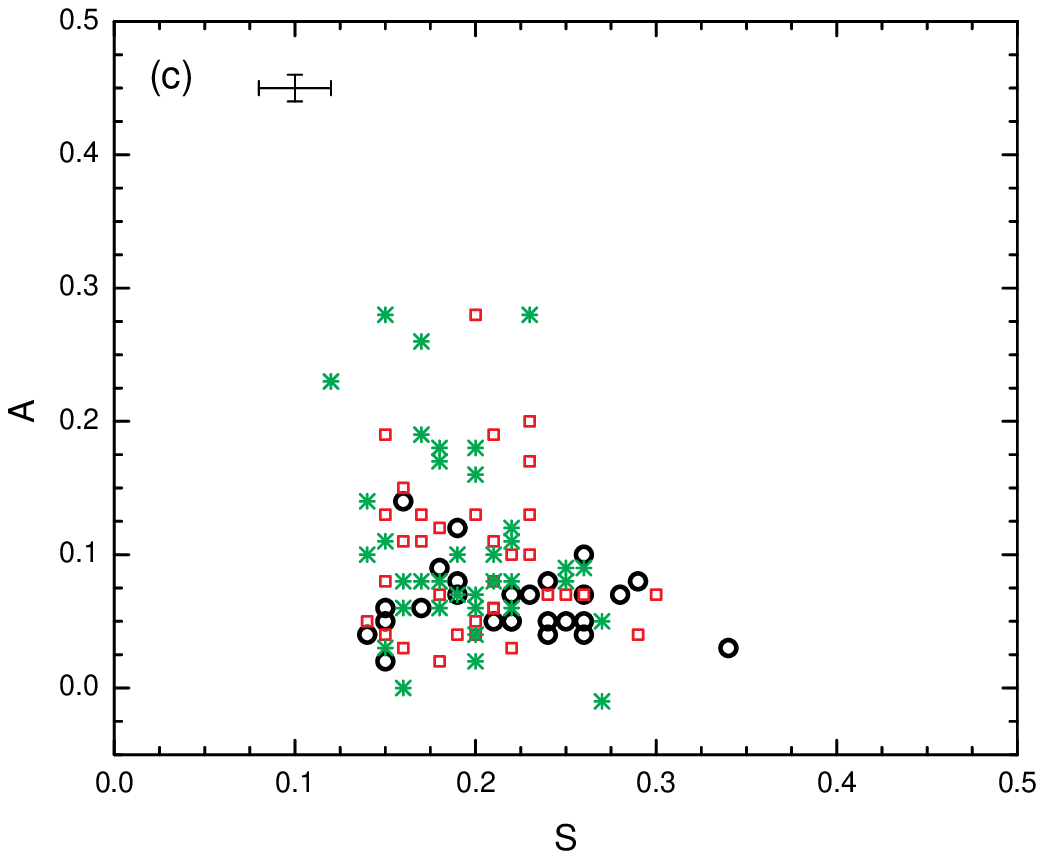}
\caption{CAS (Concentration-Asymmetry-Clumpiness) Parameters paired
in AC-SC-AS planes in (a), (b), (c) respectively. The three
morphological types are indicated with different symbols (see
figure's legend). A linear regression fit for the whole sample is
show as a solid line in panel (b). The typical 2$\sigma$ error bars
are shown in each panel.} \label{fig11}
\end{figure*}

\clearpage

\begin{figure*}
\centering
\includegraphics[width=\columnwidth,clip=true]{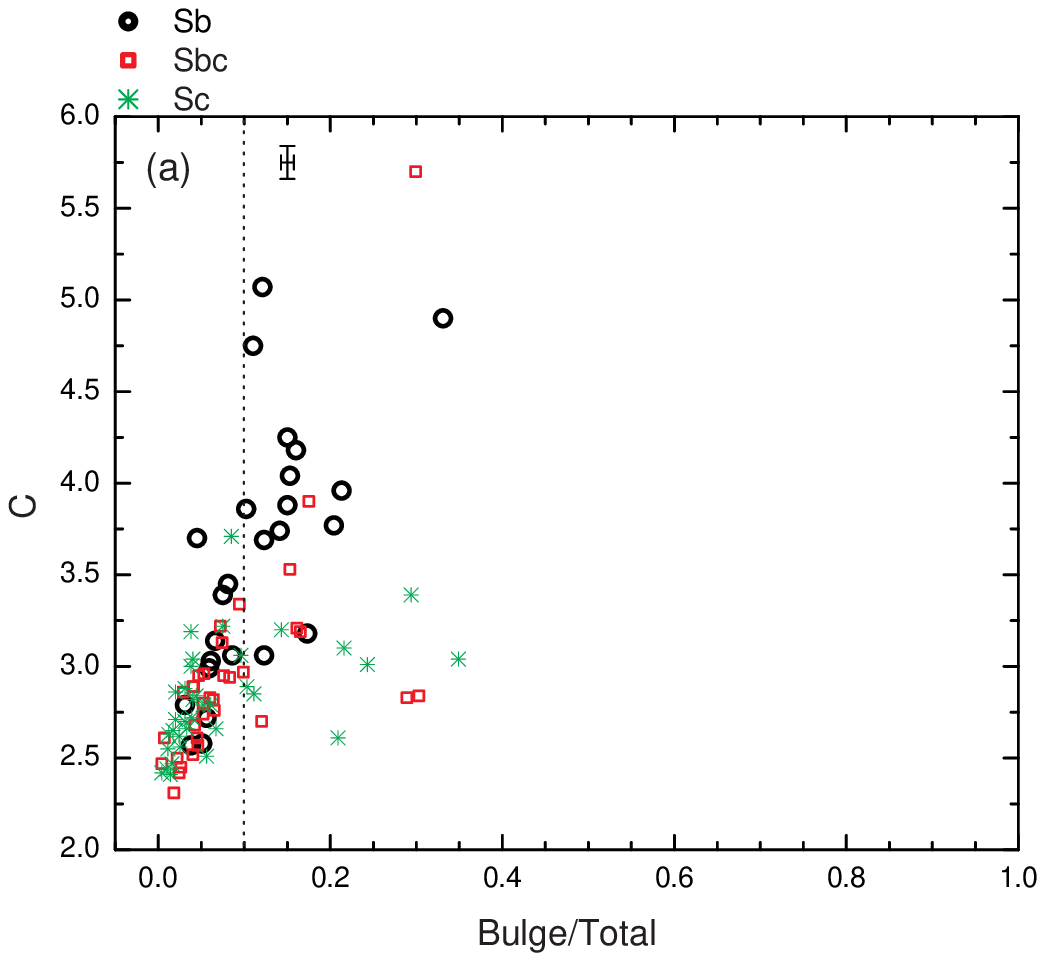}
\includegraphics[width=\columnwidth,clip=true]{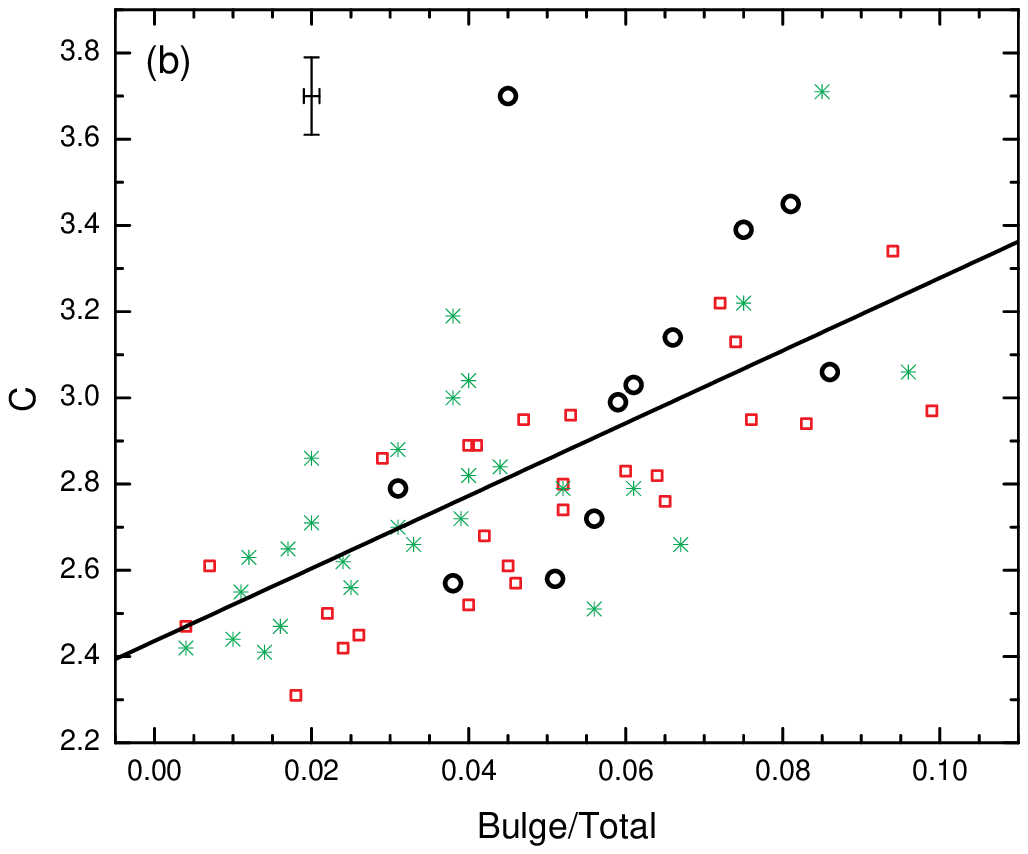}
\includegraphics[width=0.95\columnwidth,clip=true]{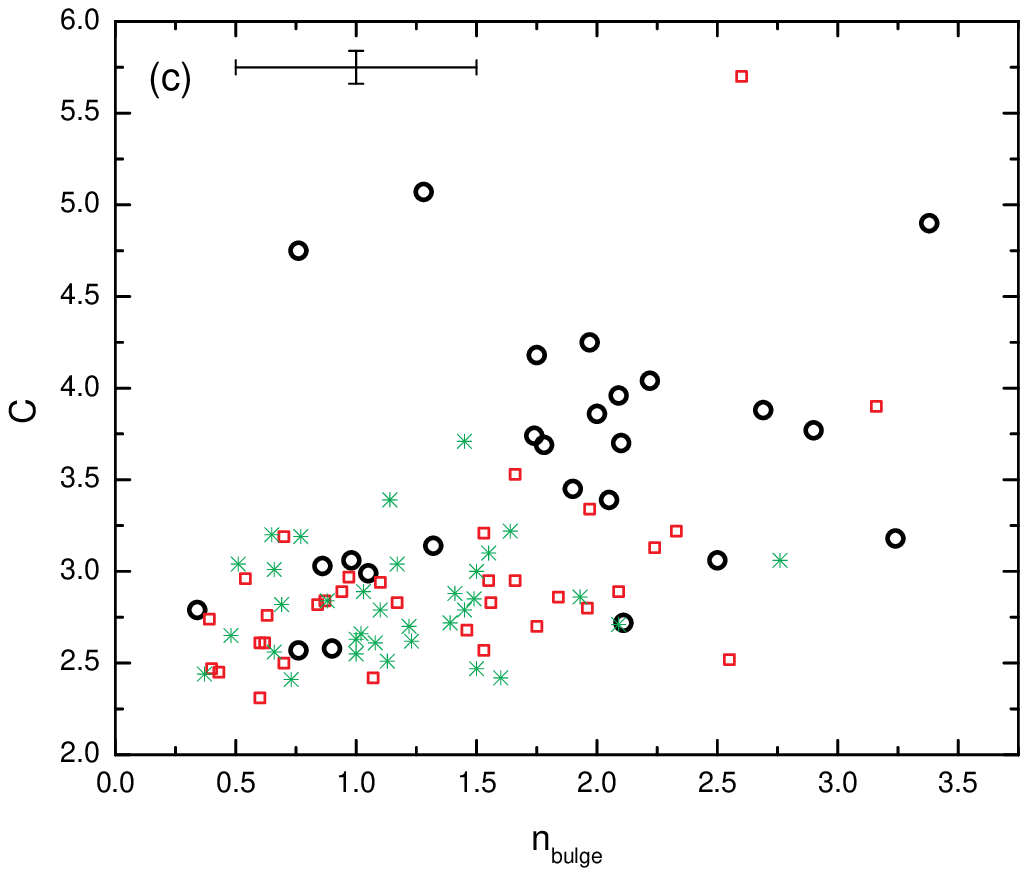}
\includegraphics[width=\columnwidth,clip=true]{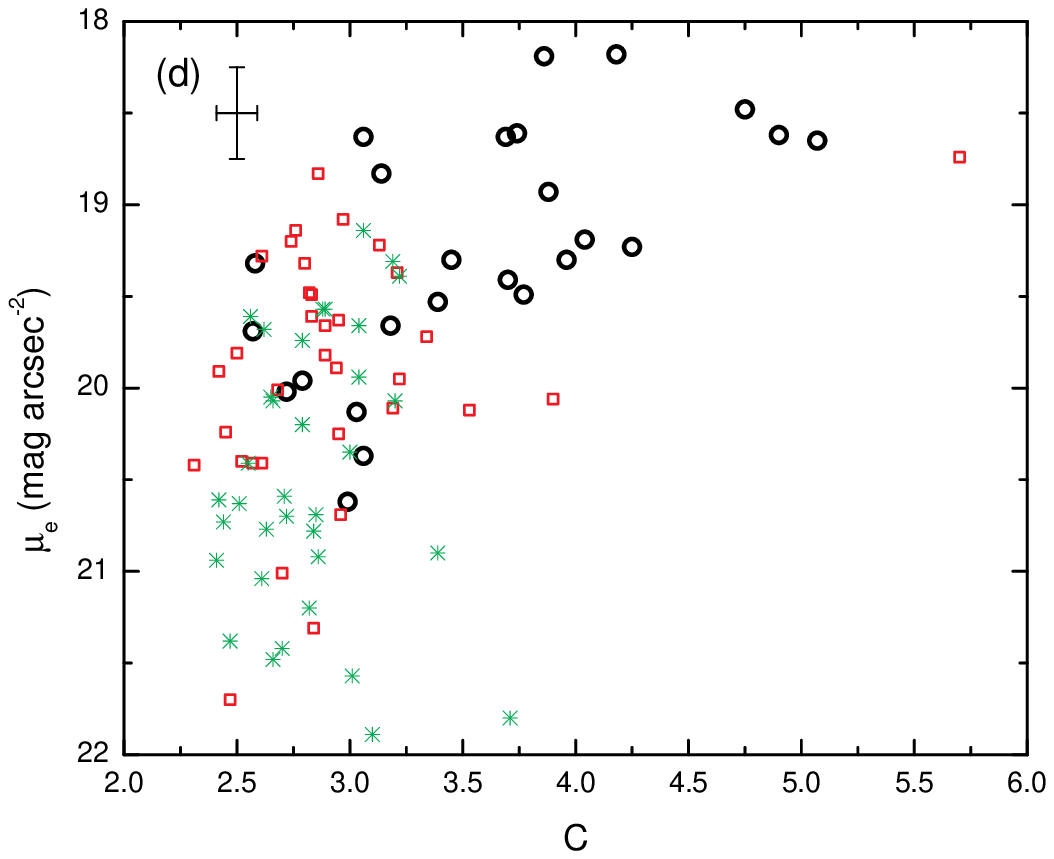}
\caption{Concentration index C shown in relation to parameters
describing the bulge: Bulge/Total luminosity ratio (a-b), S\'{e}rsic
index n$_{bulge}$ (c) and bulge effective surface brightness
$\mu_{e}$ (d). Panel (b) offers a detailed look at C versus
Bulge/Total from panel (a), with an emphasis on the region to the
left of Bulge/Total = 0.1, denoted by the vertical dotted line in
(a). The three morphological types are indicated with different
symbols (see figure's legend). The typical 2$\sigma$ error bars are
shown in each panel.} \label{fig12}
\end{figure*}

\clearpage

\begin{figure}
\plotone {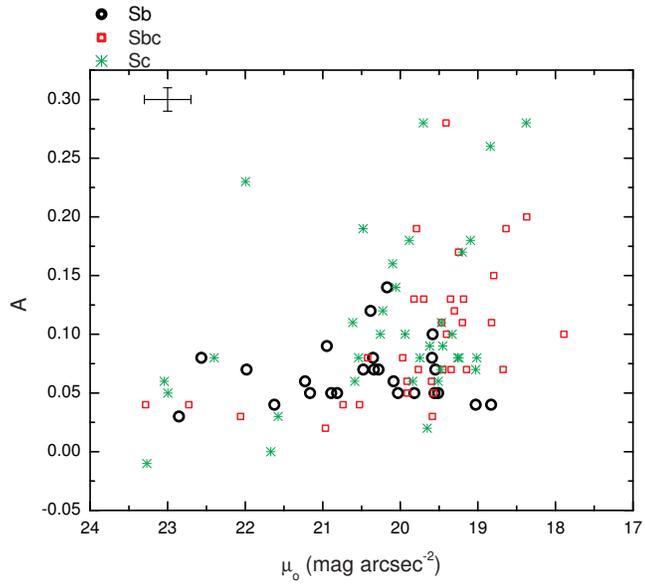} \caption{Asymmetry index A in relation to disk
central surface brightness $\mu_{o}$. The three morphological types
are indicated with different symbols (see figure's legend). The
typical 2$\sigma$ error bars are shown.} \label{fig13}
\end{figure}

\clearpage

\end{document}